\documentclass[12pt,letterpaper]{umd-thesis}    

\usepackage{amstex}	
\usepackage{amssymb}	
\usepackage{footnpag}	
\usepackage{psfig}		

 \spacing{1}	

\newcommand{\rmmcaption}[1]{\spacing{1.0}\caption{#1}}

\setboolean{masters}{false}     
\title{
ON THE DYNAMICS OF PHASE TRANSITIONS  \\
AND THE NONEQUILIBRIUM FORMATION \\
OF TOPOLOGICAL DEFECTS  \\ 
}
\author{Gregory James Stephens}
\department{Department of Physics}
\advisor{Professor Bei-Lok Hu}
\chairtitle{Chairman/Advisor}
\committee      {      
                        Professor Robert Dorfman \\
                        Professor Emeritus Richard Ferrell \\
                        Professor Theodore Jacobson     \\
                        Professor John Weeks  \\
                }
\setboolean{hasfigures}{true}
\setboolean{hastables}{false}
\setboolean{hascopyright}{true}
\abstractfile{abstract.tex}
\acknowledgements{ack.tex}
\date{2000}
\begin{document}
%


%
%
%

\makefrontmatter

\chapter{Introduction} 
\section{Phase transitions and topological defects}
Phase transitions are common in the world around us and in the laboratory.  For example, a ferromagnet exhibits 
spontaneous magnetization below the Curie temperature and Helium-4 exhibits superfluidity below
$T_c \approx 2$K (at ordinary pressures).  In the energetic
environment of the early universe phase transitions played a decisive role in shaping the cosmos we observe today.  
These include transitions at the Grand Unification (GUT) scale, $T_c \approx 10^{16}$ GeV, 
and at the electroweak scale, $T_c \approx 10^2$ GeV. In addition, Quantum Chromodynamics (QCD) is expected
to display a color deconfinement and/or chiral phase transition at temperatures $T_c \approx 100$ MeV.  
An inflationary phase transition, a possible explanation for the flatness, horizon
and monopole problems of the standard Friedmann-Robertson-Walker (FRW) cosmology, also 
may have occurred at temperatures near the GUT scale.  At still earlier epochs near 
the Planck energy, $T_c \approx 10^{19}$ GeV, the classical properties of spacetime described by general relativity 
may emerge from a phase transition in candidate theories of quantum gravity.
Understanding the dynamic processes by which 
phase transitions occur is of universal physical importance.

Phase transitions are often accompanied by broken symmetry. 
For example, crystalline ice formed when water freezes does not share the rotational and 
translational symmetry of liquid water.  If the broken-symmetry phase also has a degenerate ground state,
then the phase transition can produce topological defects. Relics of the high-temperature symmetric 
phase of the system, topological defects are topologically stable field configurations that are locally 
trapped in an excited state.  Examples of defects include vortices in both type-II superconductors and the superfluids, 
Helium-3 and Helium-4. Vortex topological defects also appear as cosmic strings in various models of the early universe.
Topological defects are classified by the homotopy groups $\Pi_n(M)$ of the vacuum
manifold, $M$. In three spatial dimensions simple defects are domain 
walls if $\Pi_0(M) \neq 1$, strings if $\Pi_1(M) \neq 1$, 
monopoles if $\Pi_2(M) \neq 1$ and textures if  $\Pi_3(M) \neq 1$ \cite{vilenkin:1994}. 

In some condensed matter systems, topological defects are not merely formed during the phase transition, they
are the root {\it cause} of it.  In two spatial dimensions a system with O(2) symmetry undergoes 
a Kosterlitz-Thouless (KT) transition \cite{kosterlitz:1973}.   The KT transition from a quasi-ordered phase to 
a disordered phase occurs when enough thermal energy exists to unbind vortex-antivortex pairs.  A similar transition occurs in
a system with O(2) symmetry in three spatial dimensions \cite{antunes:1998a,antunes:1998b}.  In this case, 
the disordered phase coincides with the appearance of long vortex loops spanning the size of the system.

Kibble was the first to show that topological defects are a generic
feature of nonequilibrium phase transitions \cite{kibble:1976}.  In the course of  
a symmetry-breaking phase transition, a field decays from an unstable, symmetry-restored state to
a stable (degenerate) vacuum.  The correlation length limits the length scale on which the 
field can choose the same vacuum state.  In the
laboratory, if the phase transition proceeds slowly, the correlation length is 
bounded only by the size of the system; the field will effectively choose
a homogeneous vacuum state.  In the early universe, however, the correlation
length is bounded by the size of the particle horizon.  With
correlated regions limited in size by causality, Kibble argued that phase transitions in the
early universe necessarily leave both a domain structure of vacuum states and
a network of topological defects.

Topological defects formed in the early universe can have profound 
consequences on the subsequent spacetime evolution.  In the standard FRW 
cosmology, GUT scale monopoles, even if formed with an initial 
density equal to 
Kibble's lower bound of one defect per horizon volume, contribute 
more than $10^{12}$ times the largest possible density of the universe  
consistent with observations. The overproduction of monopoles  
is a puzzle within the standard FRW cosmology 
and provided part of the original motivation for an early epoch of 
inflation \cite{guth:1981}. If monopoles are formed in a phase transition before or during 
an inflationary phase, their number density is safely diluted by the exponentially expanding scale factor.  
However, it has also recently been observed that in the very energetic process of reheating, as coherent
oscillations of the inflaton decay into other fields, a nonthermal spectrum of topological
defects may be produced, thus again raising the monopole problem \cite{kofman:1996}.  The details of the 
reheating process and the possible formation of a defect distribution at the end of an inflationary 
phase are currently an area of active study (see e.g \cite{felder:2000}). 
Other topological defects such as cosmic strings are viable
candidates for the seeds of structure formation and produce a characteristic 
signature in the angular power spectrum of the cosmic microwave background (for a recent, short  
review see \cite{durrer:2000}).  
Satellites such as MAP and Plank designed specifically to measure fluctuations in the cosmic
microwave background radiation to high accuracy are scheduled to be 
launched in the near future.  The data obtained from these
missions is expected to clarify the precise role of topological defects 
in the origin of structure formation in the universe.

Nonequilibrium phase transitions are observed in many laboratory systems and the study of 
topological defects has helped forge links
between the study of high energy processes in the early universe and condensed matter systems \cite{zurek:1996}.  
In liquid crystals, the formation of a network of topological string defects has been observed 
in the rapid cooling of the system through its critical point, demonstrating not only the viability of the 
Kibble mechanism in producing topological defects, but also, the scaling of the defect network 
\cite{chuang:1991,bowick:1994}.  
Other condensed matter systems, like superfluid $^3$He, have complex symmetry-breaking patterns and offer 
many potential analogs to early universe processes \cite{volovik:2000}.

The proliferation of field theories containing topological defects and their importance  
in both condensed matter and early universe processes provides strong motivation to extract predictions from 
defect models that can be compared with experiments and observations.  These predictions generally contain
three main ingredients: the density of topological defects
immediately following the formation process, such as a phase 
transition; the subsequent evolution of the defect distribution;
the calculation of physical observables, such as cosmic microwave background radiation anisotropies and polarization.  
While both numerical and analytical 
work has been done on the evolution of the defect distribution and the
extraction of physical predictions, less attention has been focused on quantitative predictions of the initial defect
density.

\section{The Kibble and Zurek mechanisms of defect formation}

The first estimate of the initial topological defect density formed in a phase transition was made by 
Kibble in a cosmological context \cite{kibble:1976}.  The basic ingredients of the 
Kibble mechanism are causality and the Ginzburg temperature $T_G$.  The 
Ginzburg temperature is defined as the temperature at which thermal fluctuations contain just enough energy for 
correlated regions of the field to overcome the potential energy barrier
between inequivalent vacua,
 \begin {equation}
kT_G \sim \xi(T_G)^3 \Delta F(T_G),
\end {equation}
\noindent  where $\Delta F$ is the difference in free energy density between the true and false vacua and
 $\xi$ is the equilibrium correlation length.  We count a fluctuation over the barrier as a single degree
of freedom.
In the Kibble mechanism, the length scale characterizing the initial defect network is set by the equilibrium 
correlation length of the field, evaluated at $T_G$.
In a recent series of experiments the Kibble mechanism was tested in the laboratory
\cite{bauerle:1996,ruutu:1996}.  The results, while
confirming the production of defects in a symmetry-breaking phase transition, 
indicate that 
$\xi(T_G)$ does {\it not} set the characteristic length scale of the initial defect distribution.  These
experiments were suggested by Zurek, who criticized Kibble's
use of equilibrium arguments and the Ginzburg temperature \cite {zurek:1985}.
  
Combining equilibrium and nonequilibrium ingredients, Zurek offers a
``freeze-out'' proposal to estimate the initial density of defects. 
In equilibrium, and near the critical temperature,  
the correlation length and the relaxation time of the field grow without bound as  
\begin {eqnarray}
\xi = \xi_0 {\mid \epsilon \mid}^{-\nu}, \\
\tau = \tau_0 {\mid \epsilon \mid}^{-\mu}, 
\end {eqnarray}
\noindent where $\epsilon$ characterizes the proximity to the critical temperature,
\begin {equation}
\label{eq-epsilon}
\epsilon = \frac {T_c-T} {T_c}, 
\end {equation}
and $\mu$ and $\nu$ are critical exponents appropriate for the theory under consideration.  
In the freeze-out scenario, the dynamics of the phase transition occur through the time-dependent linear
temperature quench,
\begin {equation}
\epsilon=\frac {t} {\tau_Q},
\end {equation}
so that for $t<0$, the temperature of the heat bath is above the critical temperature
and the critical temperature is reached at $t=0$. The field starts in thermal equilibrium with the heat 
bath at a temperature above the critical temperature.
Initially, as the temperature of the bath is lowered adiabatically, 
the field remains in local thermal equilibrium.  However, the ability of the field to maintain equilibrium
depends on the relaxation time, which diverges at the critical point.
The divergence of the equilibrium relaxation time as the heat bath approaches the critical temperature
is known as critical slowing down. Critical slowing down results from the combination of a finite speed of propagation 
for perturbations of the order parameter and a diverging correlation length.  As the correlation length diverges, small 
perturbations of the order parameter ({\it e.g.,} lowering of the temperature) take longer to
propagate over correlated 
regions, therefore it takes longer to reach equilibrium.  As the critical temperature is approached from above 
there comes a time $t^*$ during the quench when the time remaining before the 
transition equals the equilibrium relaxation time
\begin {equation}
 \left | t^* \right |=\tau(t*).
\end {equation}
Beyond this point the correlation length can 
no longer adjust to follow the changing temperature of the bath.  At time $t^*$ the dynamics of the
correlation length freezes.   The correlation 
length remains frozen until a time $\left |t^*\right |$ after the critical temperature is reached.  
In the freeze-out scenario the correlation length at the freeze-out time $t^*$ sets the characteristic length scale
for the initial defect network.  
Evaluating the correlation length at $\epsilon(t^*)$, we find the frozen correlation length, 
and therefore the initial defect density, scale with the quench rate as
\begin {equation}
\xi (t^*) \sim {\tau_Q}^{\frac {\nu} {1+\mu}}.
\end {equation}
The power-law scaling of the topological defect density with the quench rate has been verified 
using 1-dimensional, 2-dimensional and 3-dimensional simulations of 
phenomenological time-dependent Landau-Ginzburg equations \cite{laguna:1997,yates:1998,antunes:1999}. 
Experimental efforts to test the Zurek prediction in condensed matter systems are wide-ranging but inconclusive. 
While nonequilibrium phase transitions in $^3He$ appear to show qualitative agreement \cite{bauerle:1996,ruutu:1996}, 
the experiments to date lack reliable error estimates and a mechanism to vary the quench rate.  In addition, 
quenches in $^4He$ \cite{dodd:1999} and high-temperature
superconductors \cite{carmi:1999} do not seem to produce defect densities at the level 
expected from the freeze-out mechanism.

The Kibble prediction for the initial length scale of the defect 
distribution is inconsistent with experiment, yet the alternative  
freeze-out proposal raises many conceptual questions. The freeze-out scenario relies on 
phenomenological ideas derived from experience with classical 
Landau-Ginzburg systems.  It is not clear how much of this picture, if any, is applicable
to {\it quantum} fields that undergo phase transitions in the early universe.
Quantum fields bring new conceptual issues. In particular, consideration of 
decoherence is necessary to understand when and how the quantum system at the late stages of the transition 
can be described by an ensemble of classical defect configurations.  
Even in the classical context, the use of {\it equilibrium} critical scaling
in a fully dynamical setting is, at best, an approximation.  Critical dynamics is generally
much richer and less universal than equilibrium critical behavior \cite{hohenberg:1977}.  For a system 
coupled to a rapidly changing environment, the 
correlation length evolves through the dynamical equations of motion of the field.  
In fact, the relaxation time is not always well-defined. 
In addition, a realistic bath consists of interactions between the system and other fields in the universe.  
Physical interactions 
between the bath and the system produce interesting and nonequilibrium behavior 
such as noise and dissipation \cite{hu:1994}.  The reduction of the complicated
bath-system interaction to one parameter, the quench time scale, and a prescribed linear time-dependence in the 
effective mass of the time-dependent Landau Ginzburg equation neglects these potentially important processes.

\section{Foundational issues}

Despite significant effort, the detailed mechanisms responsible for the 
formation of topological defects remain only partially understood. Throughout this dissertation we explore  
three interwoven and complementary themes that are relevant in a broad theoretical context and 
specifically to the problem of nonequilibrium defect formation.
These themes are (i) equilibrium and nonequilibrium, (ii) quantum and classical and (iii) fundamental and  phenomenological. 
Phase transitions and topological defects occur in both classical and 
quantum systems, under equilibrium and nonequilibrium conditions and are depicted by  
phenomenological and fundamental theories. The study of the nonequilibrium formation of topological
defects therefore provides a broad theoretical vantage point 
from which to further elucidate the connections between these complementary and disparate
concepts.

\subsection{Equilibrium and Nonequilibrium}
The formation of topological defects is a nonequilibrium process.  Yet, both
the Zurek and Kibble mechanisms rely on aspects of equilibrium physics.  This is, at best, inconsistent.
Certainly, neither mechanism applies in a far-from-equilibrium phase transition.  If, as it seems, some
aspects of the phenomenological freeze-out scenario are correct then it is important to grasp why and under
what physical conditions.  Ultimately, a complete understanding requires 
a {\it first-principles} theory of the dynamics of defect formation.

\subsection {Quantum and Classical}
Phase transitions and the formation of topological defects occur in quantum environments such as the hot,
early universe.  However, both the Kibble and Zurek mechanisms provide a description of 
defect formation based entirely on classical reasoning. We expect differences in a fully quantum process. 
In some situations, the differences between a quantum and classical system are relatively small. For example, quantum theory
can modify critical scaling exponents and change the coarsening dynamics from that of a classical system.  
However, the conceptual differences between quantum and classical systems can also be vast: 
even the definition of a topological defect in a 
quantum field theory with arbitrary quantum state is unknown.  In addition, the formation of topological 
defects in a system of quantum fields involves the element of decoherence;  topological defects can only 
be identified as classical (stochastic) field
configurations when a positive-definite probability distribution emerges from the density matrix
of the quantum system.

\subsection{Fundamental and Phenomenological}
Topological defects appear as classical field configurations 
in phenomenological Landau-Ginzburg theories and in the classical limit of symmetry-broken
quantum field theories. The existence and significance of topological defects 
require that we enlarge our notions of fundamental degrees of freedom.  In particle physics and cosmology 
we customarily describe many systems by linear perturbation of interacting field theories. 
Topological defects do not survive this reductionist approach;  they represent an important 
class of nonperturbative fluctuations. 
In special situations there are exact dualities between field and defect descriptions.  
More often, the degrees of freedom we render important and fundamental, defect or field, string or membrane, 
metric or black hole depends on the nature of the questions we ask.

\vspace{5mm}

\noindent 
\section{Organization}
This dissertation is organized as follows.  In Chapter 2 we review the formalism of
nonequilibrium quantum field theory, focusing on the techniques useful in describing the
dynamics of a second-order phase transition.  We derive the quantum
effective action, giving a diagrammatic loop expansion for the both the 1PI and 2PI effective action.
We then review the closed-time-path (CTP) formalism, describing in detail the differences
from the conventional formulation of quantum field theory.  We present the CTP
effective action and give an explicit expression at two-loop order.
As an example, we apply the CTP and effective action formalism to a simple system, 
the quantum dynamics of an anharmonic oscillator.  

In Chapter 3, which describes original work published in Ref.~\cite{stephens:1999}, we apply the techniques of
nonequilibrium quantum field theory to the problem of defect formation during a phase transition in the early universe.
We study the nonequilibrium dynamics leading to the formation of topological defects 
in a symmetry-breaking phase transition of a quantum scalar field with
$\lambda \Phi^4$ self-interaction in a spatially flat, radiation-dominated FRW Universe.  
The quantum field is initially in a finite-temperature symmetry-restored state and the phase transition develops as the
universe expands and cools. We present a 
first-principles, microscopic approach in which the nonperturbative, nonequilibrium dynamics of the quantum field is 
derived from the two-loop, two-particle-irreducible closed-time-path effective action. 
We numerically solve the dynamical equations for the two-point function
and we identify signatures of correlated domains in the infrared portion of the momentum-space power spectrum. 
During the phase transition, the scale factor grows by a factor $a_{final} \sim 13 \: a_{initial}$ for the fastest quench 
and $a_{final} \sim 1.3 \: a_{initial}$ for the slowest quench.
We find that the size of correlated domains formed after the phase transition scales as a power-law with the 
expansion rate of the universe. 
We calculate the
equilibrium critical exponents of the correlation length and relaxation time for this model and show that the
power-law exponent of the domain size, for both overdamped and underdamped evolution (determined by the expansion
rate $\tau$), is in good agreement with 
the freeze-out scenario proposed by Zurek.  We also introduce an analytic dynamical model, valid near the critical 
point, that exhibits 
the same power-law scaling of the domain size with the quench rate.  By incorporating the realistic quench of the 
expanding universe our approach illuminates the {\it dynamical} mechanisms important for topological defect formation 
and provides a preliminary step towards a complete and 
rigorous picture of defect formation in a second-order phase transition of a quantum field. The observed power-law scaling
of the size of correlated domains, calculated here in a nonequilibrium quantum field theory context, provides preliminary
evidence for the freeze-out scenario in three spatial dimensions. 

In Chapter 4, which, in part, describes original work published in Ref.~\cite{stephens:2000},
we study the formation and interaction of topological textures in a nonequilibrium phase transition
of an classical O(3) scalar field theory in $2+1$ dimensions.  We present two models of the nonequilibrium
phase transition, a dissipative quench and a pressure quench.  The analysis of the dissipative quench is complicated
by the presence of large thermal fluctuations and is not yet complete. In the pressure quench, the phase transition 
is triggered through an external, time-dependent effective mass, parameterized by quench timescale $\tau$. 
When measured near the end of the transition ($\langle \vec{\Phi}^2 \rangle=0.9$) the average 
texture separation $L_{sep}$ and the average texture width $L_w$ scale as
 $L_{sep} \sim \tau^{0.39 \pm 0.02}$ and $L_w \sim \tau^{0.46 \pm 0.04}$, 
significantly larger than the single power-law $\xi_{freeze} \sim \tau^{0.25}$
predicted from the Kibble-Zurek mechanism.  We show that Kibble-Zurek scaling is recovered at
very early times but that by the end of the transition $L_{sep}(\tau)$ and $L_w(\tau)$ result instead
from a competition between the length scale determined at freeze-out and the ordering dynamics of a 
textured system.  
We offer a simple proposal for the dynamics of these length scales:  
$L_{sep}(t)=\xi_1(\tau)+L_1(t-t_{freeze})$ and $L_{w}(t)=\xi_2 (\tau)+L_2(t-t_{freeze})$ 
where $\xi_1 \sim \tau^\alpha$ and $\xi_2 \sim \tau^\beta$ are determined by the freeze-out
mechanism.  $L_1(t)=(\xi_1)^{\frac{1}{3}}t^{\frac{1}{3}}$ and $L_2(t)=t^{\frac{1}{2}}$ are dynamical length scales 
previously known from phase ordering dynamics, and $t_{freeze}$ is the freeze-out time.  
We find that $L_{sep}(t)$ and $L_w(t)$ fit closely to the length scales observed at the
end of the transition and yield $\alpha=0.24 \pm 0.02$ and $\beta=0.22 \pm 0.07$, in good
agreement with the Kibble-Zurek mechanism using the (overdamped) mean field critical exponents $\mu=2\nu=1$.
In the context of phase ordering these 
results suggest that the multiple length scales characteristic of the late-time ordering of a 
textured system derive from the critical dynamics of a {\it single} nonequilibrium correlation length. 
In the context of defect formation these results imply that significant evolution of
the defect network can occur before the end of the phase transition.  Therefore a quantitative understanding
of the defect network at the end of the phase transition generally requires an understanding of both critical
dynamics {\it and} the interactions among topological defects. 

In Chapter 5, which describes work in progress \cite{stephensA:2000,stephensB:2000,stephensC:2000}, we discus
black hole phase transitions in semiclassical gravity.  We review the thermodynamics
of the black hole phase transition, focusing on the calculation of the semiclassical free energy of a black hole
in thermal equilibrium.  We determine that the phase transition is entropically driven: the large amount of black hole entropy
makes the transition to a black hole spacetime inevitable. 
We introduce a quantum atomic model of the equilibrium black hole system and show that the phase transition is
realized as the abrupt excitation of a high energy state. 
We then discuss the nonequilibrium dynamics of the black hole phase transition. 
We highlight the inadequacy of the homogeneous nucleation theory of first-order phase transitions and explore  
similar examples from condensed matter and string theory.

\chapter{Nonequilibrium quantum field theory}
\section{Introduction}
In this chapter we present a first-principles, dynamical treatment of quantum field theory.  We review 
the Closed-Time-Path (CTP) formalism of nonequilibrium quantum field theory 
\cite{schwinger:1961,bakshi:1963,keldysh:1964,niemi:1981,niemi:1984,landsman:1987,
chou:1985,su:1988,dewitt:1986,jordan:1986,calzetta:1987,calzetta:1988} with particular emphasis on
the techniques useful in describing the dynamics of a second-order phase transition.
There are a number of rapidly evolving high-energy systems in which
nonequilibrium processes are important.  The description of these systems requires a fully dynamical, 
initial value formulation of quantum field theory.  Examples include the far-from-equilibrium dynamics
of the quark-gluon plasma now being probed in heavy-ion collisions \cite{rajagopal:2000} and the
dynamics of inflation (see e.g.~\cite{ramsey:1997,ramsey:1998} and references therein).  Furthermore,
in the hot early universe it is likely that the broken symmetries of the standard model
of particle physics were partially or totally restored.  As the universe expanded and cooled,
symmetry-breaking phase transitions occurred through the nonequilibrium dynamics of quantum fields 
(see e.g \cite{boyanovsky:1994a}).

Solving the exact dynamics of a system consisting of interacting quantum fields is currently
technically impossible. For example, the Heisenberg equations of motion for the operator $O$ of a quantum field
theory with Hamiltonian $H$, 
\begin{equation}
i\partial_t O= \left [ O,H \right ],
\end{equation}
represent an infinite number of coupled, differential equations at each spacetime point.  The additional complexity 
of quantum theory is evident even in the quantum mechanics of a single point particle.  The evolution of a classical
particle is described in Newtonian mechanics by a single second-order ordinary differential equation while
a quantum particle is described by the
Schroedinger equation, a first-order {\it partial} differential equation.  
However, it may be that we don't need all of the information contained in the wave
function.  In fact, in many situations, it is satisfactory to follow only the degrees of
freedom contained in the expectation values of a few well-chosen operators.  In this work we
focus on obtaining evolution equations for the mean field and the two-point function of a quantum field theory.

\subsubsection{Organization}

The organization of this chapter is as follows. In Sec.~\ref{sec:qea} we derive the quantum
effective action, giving a diagrammatic loop expansion for the both the one-particle-irreducible (1PI) and 
two-particle-irreducible (2PI) effective action.
In Sec.~\ref{sec:ctp} we review the closed-time-path (CTP) formalism, describing in detail the differences
from the conventional formulation of quantum field theory.  We then combine this discussion with
the results of the previous section to derive the closed-time-path effective action.  We also
give an explicit expression for the two-loop truncation of the 2PI-CTP effective action.
In Sec.~\ref{sec:anosc} we apply the CTP formalism to a simple system, the quantum dynamics of an
anharmonic oscillator.  A summary is given in Sec.~\ref{sec:ch1sum}

\section{The quantum effective action}
\label{sec:qea}
The quantum effective action is a functional of correlation functions whose variation yields 
quantum-corrected equations of motion.  Analogous to a free energy in statistical mechanics,
the quantum effective action is useful in understanding the extent to which quantum fluctuations
influence the static and dynamic behavior of correlation functions of a quantum system. 
In particular, suppose we are interested
in the mean field,
\begin{equation}
\phi(x)=\langle 0_+ |\Phi_H(x) | 0_- \rangle,
\end {equation}
and the two-point function,
\begin{equation}
G(x,x')=\langle 0_+ |T[\Phi_H(x)\Phi_H(x')]|0_-\rangle.
\end{equation}
In the presence of a source $J(x)$ we define the vacuum persistence amplitude
\begin{equation}
{\langle 0_+|0_- \rangle}_J \equiv Z[J]=\int D\phi\: e^{\frac{i}{\hbar} \left [ S[\phi]+\int J(y)\phi(y) \right ]}
\end{equation}
and the connected generating functional $W[J]$ via
\begin {equation}
Z[J]=e^{\frac{i}{\hbar}W[J]}.
\end {equation}
The mean field is given by
\begin{equation}
\phi(x)=\left .{\frac {\delta W[J]} {\delta J(x)}} \right|_{J=0},
\end{equation}
while
\begin{equation}
G(x,x')=\left. \frac {\delta^2 W[J]}{\delta J(x) \delta J(x')} \right |_{J=0}.
\end{equation}
If the interaction coupling of the fields is small it is useful to partition
the action into free and interacting pieces
\begin {equation}
S=S_{free}+S_{int},
\end{equation}
so that
\begin{equation}
e^{\frac{i}{\hbar}S}=e^{\frac{i}{\hbar}S_{free}}\left (1+\frac{i}{\hbar}S_{int}-\frac{1}{\hbar^2}(S_{int})^2+...\right).
\end{equation}
Since the exponent in the path integral is now Gaussian, the correlation functions derived from $Z[J]$ can 
be computed as a power series in the interaction coupling.  Such perturbative 
expansions have proven quite successful in 
computing scattering cross sections for weakly-coupled quantum field theories such as quantum electrodynamics. 
However, a number of interesting phenomena in quantum field theory are fundamentally nonperturbative in nature.  In 
a phase transition in particular, the old vacuum is nonperturbatively different from the new vacuum. 
To analyze such situations we seek a formalism that is both nonperturbative in the coupling
of the fields yet tractable in terms of quantum interactions.  The effective action provides one such
avenue.

\subsection{The 1PI effective action}

We define the one-particle irreducible (1PI) effective action $\Gamma[\phi]$ by
means of the Legendre transform
\begin{equation}
\Gamma[\phi]=W[J]-\int d^4y \: J(y)\phi(y).
\end{equation}
The equation of motion satisfied by the mean field is found by variation of the 1PI effective action,
\begin {equation}
\frac {\delta \Gamma[\phi]} {\delta \phi(x)}=\int d^4y\: \frac {\delta W[J]}{ \delta J(y)} \frac {\delta J(y)} 
{\delta \phi(x)} -J(x)-\int d^4y \: \frac {\delta J(y)} {\phi(x)} \phi(y).
\end{equation}
Since 
\begin{equation}
\label{eq-Jphi}
\phi(x)=\langle \Phi_H(x) \rangle= \frac {\delta W[J]} {\delta J[x]},
\end{equation}
then (in the absence of sources) the mean field is determined as a solution to the equation
\begin {equation}
\label{eq-meanfield}
\frac {\delta \Gamma[\phi]} {\delta \phi(x)}=0.
\end {equation}
At this stage, we have done nothing more than rewrite the generating functional in a form appropriate for
generating equations of motion satisfied by the mean field.  We still have the full complexity of the 
quantum field theory. The advantage of this approach is that the effective action $\Gamma[\phi]$ can be computed 
diagrammatically to arbitrary order in a {\it loop} expansion.  The loop expansion is a semiclassical
expansion where the formal expansion parameter is Planck's constant $\hbar$.  The loop expansion provides
a calculational technique to sum an infinite class of perturbative Feynman diagrams and thus satisfies
our nonperturbative criteria.  The computation of the effective action in the loop expansion proceeds in two stages.  First
$W[J]$ is evaluated to the desired loop order.  Then, in order to achieve the Legendre transform, we 
must solve the functional equation Eq.~(\ref{eq-Jphi}) for $J[\phi]$.  This process is illustrated below.     

Expansion of $W[J]$ in powers of $\hbar$ is the loop expansion.  We expand the classical action about a
background field configuration $\phi_0$
\begin{equation}
S[\phi_0+\phi]=S[\phi_0]+\left . \frac{\delta S} {\delta \phi} \right |_{\phi_0} \phi
	      +\frac{1}{2} \phi \left. \frac{\delta^2 S}{\delta \phi^2} \right |_{\phi_0} \phi+...,
\end{equation}
and let the background field satisfy
\begin{equation}
\label{eq-phi0}
\left. \frac{\delta S}{\delta \phi} \right |_{\phi_0}=-J. 
\end{equation}
The connected generating functional is 
\begin{equation}
W[J]=S[\phi_0]+\phi_0J+\frac{1}{2}i\hbar\ln{\det{(iD^{-1}})}+O(\hbar^2),
\end{equation}
where
\begin{equation}
iD^{-1}=\left.\frac {\delta^2 S} {\delta \phi^2} \right |_{\phi_0}.
\end{equation}
The tricky part in the calculation of the effective action is the Legendre transform.  Without quantum corrections,
\begin{equation}
\phi_0=\frac {\delta W[J]} {\delta J}.
\end{equation}
We are thus motivated to define
\begin{equation}
\phi_0=\bar{\phi} + \phi_1
\end{equation}
where
\begin{equation} 
\bar{\phi}=\frac {\delta W[J]} {\delta J}
\end{equation}
is the mean field we are seeking and $\phi_1$ (a functional of $\bar{\phi}$) is 
at least of order $\hbar$.  Inverting Eq.~(\ref{eq-phi0}) to view $J$ as a function of $\phi_0$
we have
\begin{equation}
\Gamma[\bar{\phi}]=W[J]-\bar{\phi}J[\bar{\phi}]=S[\bar{\phi}+\phi_1]+(\bar{\phi}
                   +\phi_1)J[\bar{\phi}+\phi_1]+\frac{1}{2}i\hbar\ln{\det{(iD^{-1}})}
\end{equation}
Since $\phi_1$ is at least first order in $\hbar$ we can functionally expand about $\bar{\phi}$
and keep terms only to first order.  The final result is
\begin{eqnarray}
\label{eq-1loopEFA}
\Gamma_{1-loop}[\bar{\phi}]=S[\bar{\phi}]+\frac{1}{2}i\hbar\ln{\det{(iD^{-1}})} \\
iD^{-1}=\left.\frac {\delta^2 S} {\delta \phi^2} \right |_{\bar{\phi}} 
\end{eqnarray}
The generalization of this result to higher loop orders is algebraically involved but straightforward.  
First derived in \cite{dolan:1974} the effective
action to arbitrary loop orders is given by
\begin{equation}
\Gamma[\bar{\phi}]=S[\bar{\phi}]+\frac{1}{2}i\hbar\ln{\det{(iD^{-1}})}+\Gamma_1[\bar{\phi}]
\end{equation}
where $\Gamma_1[\bar{\phi}]$ is the sum of all one-particle irreducible vacuum graphs for a
theory described by the shifted action 
\begin{equation}
S_{shift}[\phi, \bar{\phi}]=S[\phi+\bar{\phi}]-S[\bar{\phi}]-\phi\frac{\delta S[\bar{\phi}]}{\delta \bar{\phi}}.
\end{equation}
As a concrete example we consider the case of a scalar field theory with potential
$\mathcal{V}[\phi]= \frac{1}{2}m^2\phi^2+\frac{\lambda}{4!}\phi^4$. The shifted action is given by
\begin{equation}
\label{eq-shiftact}
S_{shift}[\phi, \bar{\phi}]=
-\int d^4x \left[\frac{1}{2}\:\phi\left( \Box+m^2+\frac{\lambda}{2}\bar{\phi}^2\right)\phi+\frac{\lambda}{3!}\bar{\phi}\phi^3+
			      \frac{\lambda}{4!}\phi^4\right].
\end{equation}
\begin{figure} 
\begin{center}
\strut\psfig{figure=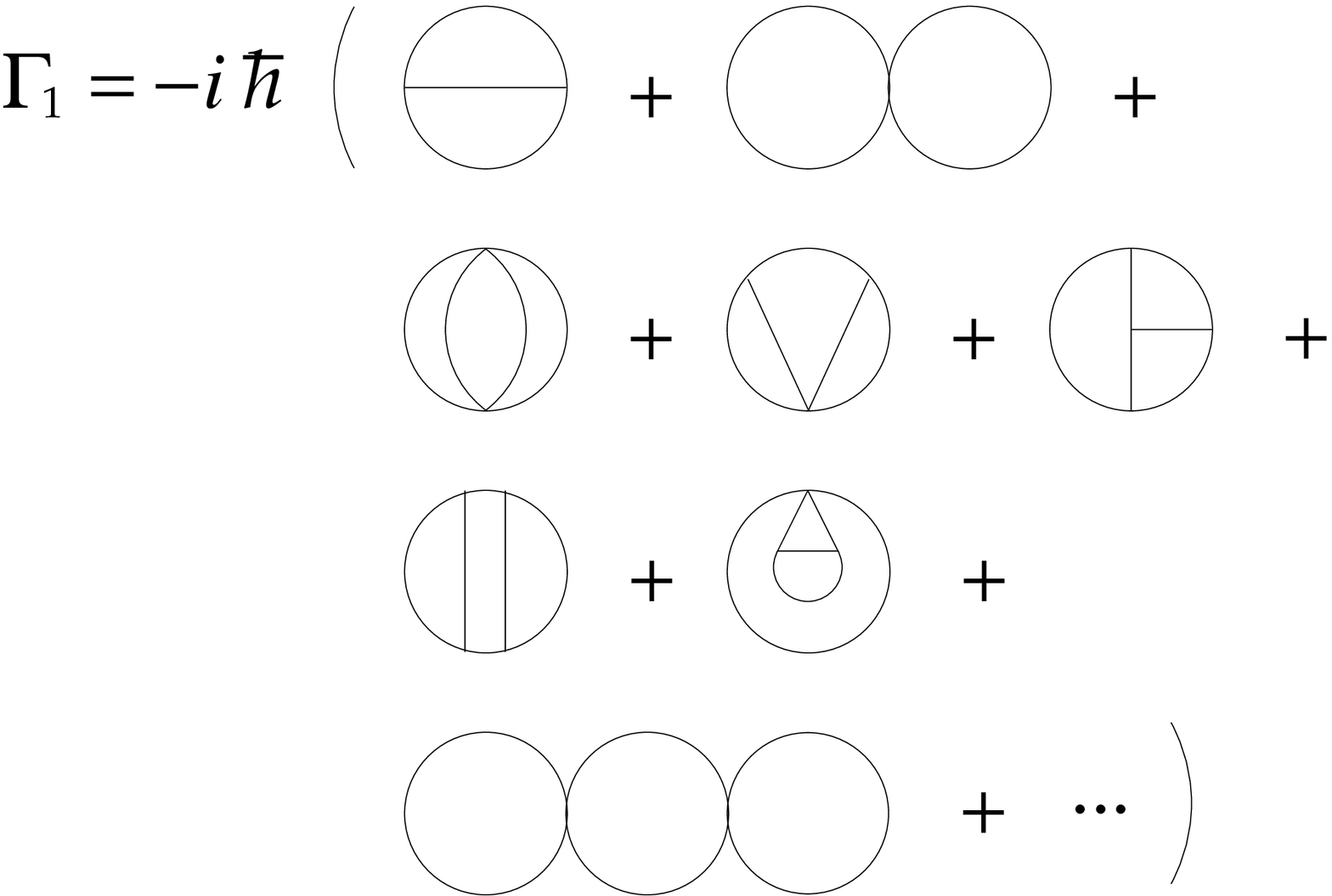,width=3.5in,angle=0}
\rmmcaption{Diagrammatic expansion of $\Gamma_1[{\bar{\phi}}]$ to 3-loop order.}
\label{fig:gamma1}
\end{center}
\end{figure}
$\Gamma_1[\bar{\phi}]$ is shown in Fig.~\ref{fig:gamma1}, in which
lines and vertices are given respectively by the propagator and interactions of the shifted action,
Eq.~(\ref{eq-shiftact}).  In this case the effect of
quantum fluctuations is to change the effective mass of the $\phi$ particles and to add a new cubic interaction vertex
with a strength proportional to the mean field.

\subsection{The 2PI effective action}

In situations where fluctuations about the mean field are important (as for example in second-order phase transitions) the 
1PI effective action and its diagrammatic expansion need to be generalized.  Inthe 1PI effective action the two-point
correlation function $G(x,x')$ is {\it slaved} to the mean field:  $G(x,x')$ itself has no independent dynamics.
Instead, we wish to consider the case where both $\phi(x)$ and $G(x,x')$ are treated on equal footing.  We seek
an effective action whose variation yields independent dynamical equations for $\phi(x)$ and $G(x,x')$.

\begin{figure} 
\begin{center}
\strut\psfig{figure=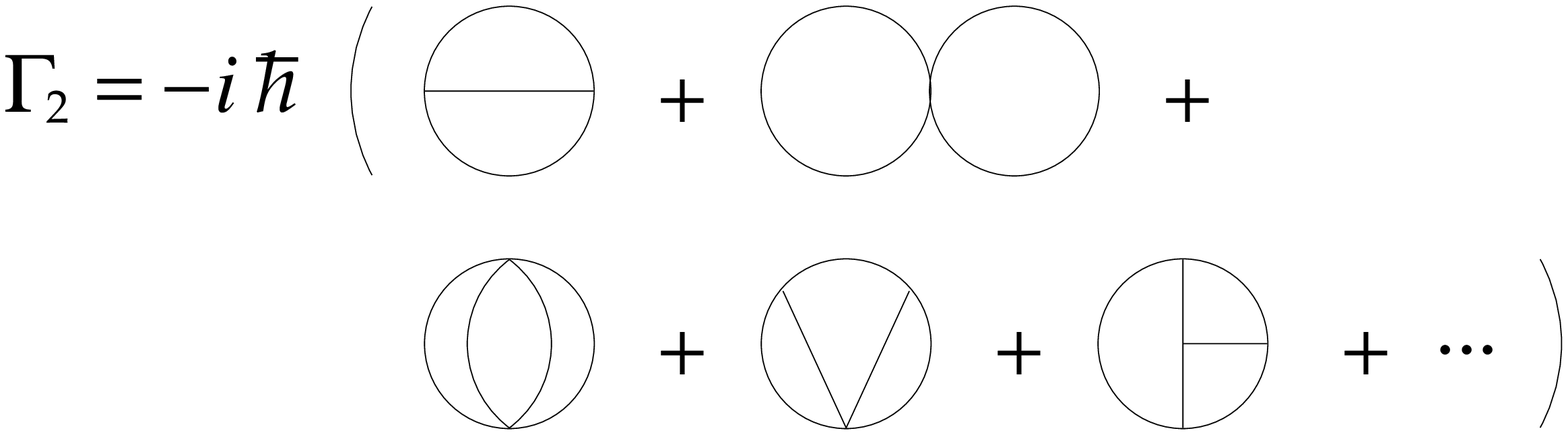,width=3.5in,angle=0}
\rmmcaption{Diagrammatic expansion of $\Gamma_2[\bar{\phi},G]$ to 3-loop order.   }
\label{fig:gamma2}
\end{center}
\end{figure}

The generating functional in the presence of sources coupled to both the mean field and the two-point function is
\begin{equation}
Z[J,K]=e^{\frac{i}{\hbar}W[J,K]}=\int D\phi\: e^{\frac{i}{\hbar} \left [S[\phi]+\int\phi(y) J(y)
 + \frac{1}{2} \int \int \phi(y)K(y,z)\phi(z) \right ]}.
\end{equation}
As before, we make the Legendre transformation to define the two-particle irreducible (2PI) effective action,
\begin{eqnarray}
\Gamma[\phi, G]&=&W[J,K]-\int  \phi(y)J(y) -\frac{1}{2} \int  \int  \phi(y) K(y,z) \phi(z) \nonumber \\
  &-&\frac{\hbar}{2} \int\int  G(y,z)K(z,y).
\end{eqnarray}
The equations of motion for the mean field and the two-point function (in the absence of sources) are
\begin{eqnarray}
\frac{\delta \Gamma[\phi,G]}{\delta \phi(x)}=0, \\
\frac{\delta \Gamma[\phi,G]}{\delta G(x,y)}=0. 
\end{eqnarray}
The calculation of the 2PI effective action in a diagrammatic loop expansion proceeds along similar lines 
as the 1PI case.  Here we
simply quote the result \cite{cornwall:1974}
\begin{equation}
\label{eq-2PIea}
\Gamma[\bar{\phi},G]=S[\bar{\phi}]+\frac{i\hbar}{2}  
Tr \ln{G^{-1}}+\frac{i\hbar}{2}Tr D^{-1}(\bar{\phi})G +\Gamma_2[\bar{\phi},G], 
\end{equation}
where 
\begin{equation}
iD^{-1}= \left. \frac{\delta^2 S[\phi]}{ \delta \phi(x) \delta \phi(y)} \right |_{\phi=\bar{\phi}}.
\end{equation} 
$\Gamma_2(\bar{\phi},G)$ is the sum of two-particle irreducible vacuum graphs for a theory with propagator $G(x,y)$
and interactions determined from the cubic and higher terms in the shifted action $S_{shift}[\bar{\phi}+\phi]$, 
Eq.~(\ref{eq-shiftact}).  
Fig.~\ref{fig:gamma2} shows $\Gamma_2[\bar{\phi},G]$ for a scalar field theory with $\phi^4$ self-interaction.

\section{Closed-Time-Path formulation}
\label{sec:ctp}
In this section we review the closed-time-path (CTP) formalism of nonequilibrium quantum
field theory.  Our discussion follows closely that of Ref.~\cite{calzetta:1987}.
We consider the quantum amplitude in the presence of a source $J$
\begin{equation}
Z[J]={\langle \phi',t' | \phi,t \rangle}_J=\langle \phi',t'|T e^{\frac{i}{\hbar}\int_t^{t'}J\Phi_H} |\phi,t \rangle
\end{equation}
where $|\phi,t\rangle$ is an eigenstate of the Heisenberg field operator at time t
\begin{equation} 
\Phi_H|\phi,t\rangle=\phi|\phi,t\rangle
\end{equation}
and $T$ denotes time ordering.  With vacuum boundary conditions on the fields $\phi$ in the far past and the
far future, $Z[J]$ is the 
generator for all vacuum to vacuum transition amplitudes.
Explicitly,
\begin{equation}
\langle 0_+|T[\Phi_H(x_1)...\Phi_H(x_n)]|0_-\rangle=-(i\hbar)^n \frac {\delta^n Z[J]}{\delta J(x_1)...\delta J(x_n)}.
\end{equation}
The cross section for a scattering process can be derived from $Z[J]$ using the LSZ reduction formula 
(see for example \cite{ryder}). As explained in the beginning of the chapter there are many situations in 
quantum field theory where scattering cross sections do not provide an adequate physical description. 
In particular, if we do not know the vacuum in the far future, $Z[J]$ is not useful.

To seek a generator for causal processes we evolve the vacuum state with two {\it different} sources
\begin{equation}
\label{eq-}
Z[J_+,J_-]=\langle 0_- | 0_- \rangle_{J_-,J_+}=\langle 0_- |\tilde{T}[e^{-\frac{i}{\hbar}\int_{-\infty}^{\infty}J_-\Phi_H}]
T[e^{\frac{i}{\hbar}\int_{-\infty}^{\infty}J_+\Phi_H}]]0_- \rangle,
\end{equation}
where $\tilde{T}$ denotes anti-temporal ordering.  This is the generating functional for 
vacuum {\it expectation values}.  For example,
\begin{equation}
\langle 0_-|T[\Phi_H(x_1)...\Phi_H(x_n)]|0_-\rangle=(i\hbar)^n\frac{\delta^n Z[J_+,J_-]}{\delta J_+(x_1)...\delta J_+(x_n)}.
\end{equation}
All other expectation values can be constructed by taking suitable combinations of derivatives 
and sources.
To obtain a path integral representation of the generating functional $Z[J_+,J_-]$ we 
insert a complete set of states at an arbitrary time in the far future,
\begin{equation}
\langle 0_-|0_-\rangle_{J_-,J_+}=\int D\psi \: \langle 0_-|\tilde{T}[e^{-\frac{i}{\hbar}\int_{-\infty}^{\infty} J_-(x)\Phi_H(x)}]|
\psi\rangle \langle \psi |T[e^{\frac{i}{\hbar}\int_{-\infty}^{\infty}J_+(x)\Phi_H(x)}]0_- \rangle
\end{equation}
This form immeadiately suggests the path integral representation,
\begin{equation}
\label{eq-Z}
Z[J_+,J_-]=\int D\phi_+ D\phi_-e^{\frac{i}{\hbar} \left [S[\phi_+]-S[\phi_-]+\int J_+(x)\phi_(x)+J_-(x)\phi_-(x) \right ]}.
\end{equation}
The functional integration is taken over fields with boundary conditions appropriate for vacuum in the far past 
but equal and otherwise {\it unspecified} in the future, 
\begin{eqnarray}
\label{eq-CTPbc}
\phi_{\pm}(t\rightarrow -\infty) \rightarrow e^{\pm iwt}, \\
\phi_+(t\rightarrow \infty)=\phi_-(t\rightarrow \infty). \nonumber
\end{eqnarray}
Due to these boundary
conditions the path integral does simply factorize and propagators beyond the Feynman propagator need to be considered.
We illustrate this for the case of a free field theory,
\begin{equation}
S[\phi]=-\frac{1}{2}\int d^4x \:\phi(x)(\Box+m^2)\phi(x)
\end{equation}
To evaluate the generating functional, Eq.~(\ref{eq-Z}), we make a change of field variables,
\begin{equation}
\phi \rightarrow \phi^c+\phi,
\end{equation}
where $\phi_c$ satisfies the classical equations of motion,
\begin{equation}
(\Box +m^2)\phi^c_{\pm}(x)=J_{\pm}(x),
\end{equation} 
with the correct boundary conditions, Eq.~(\ref{eq-CTPbc}).
The classical solutions are
\begin{eqnarray}
\label{eq-classsol}
\phi^c_+(x)=-\int d^4x'\triangle_F(x,x')J_+(x')+\triangle^+(x,x')J_-(x'), \\
\phi^c_-(x)=- \int d^4x'\triangle_D(x,x')J_+(x')+\triangle^-(x,x')J_-(x'),
\end{eqnarray}
where $\triangle_F$ and $\triangle_D$ are Greens functions and $\triangle^+$ and $\triangle^-$ are
homogeneous solutions whose properties we review below.  The action is
\begin{equation}
S[\phi^c+\phi]=\int d^4x \frac{1}{2} [\phi^c(x)J(x)-\phi(x)(\Box+m^2)\phi(x)].
\end{equation}
Since we have disentangled the boundary conditions, the Gaussian path integral can be done exactly,
and the free-field CTP generating functional is
\begin{eqnarray}
Z_{free}[J_+,J_-]=\exp\frac{-i}{\hbar}\int\int \frac{1}{2}\left [ J_+(x) \triangle_F(x-x')J_+(x')+J_+(x)\triangle^+(x-x')J_-(x') \right. \nonumber \\
\left.\: -J_-(x)\triangle^-(x-x')J_+(x')-J_-(x)\triangle_D(x-x')J_-(x') \right ]. \nonumber 
\end{eqnarray}
The Closed-Time-Path boundary conditions require the consideration of four two-point functions,
the Feynman and Dyson propagators and the positive and negative frequency 
Wightman functions.  We review the properties of these functions.
The Feynman propogator is
\begin{equation}
\triangle_F(x-x')=\int \frac {d^4p} {(2\pi)^4} \frac {e^{ip(x-x')}} {p^2-m^2+i\epsilon}=
\int \frac {dp_0d^3\vec{p}} {(2\pi)^4} \frac {e^{i(p_0 (t-t')-\vec{p}\cdot (\vec{x}-\vec{x'}))}} 
{p_0^2-(\vec{p}^2+m^2-i\epsilon)}.
\end{equation}
$\triangle_F(x-x')$ has simple poles in the complex $p_0$ plane and the $p_0$ integration can 
easily be done using contour integration. We find
\begin{eqnarray}
\triangle_F(x-x')&=&i\theta(t-t')\int \frac {d^3\vec{p}} {(2\pi)^3} \frac {1} {2w_{\vec{p}}} 
e^{i[\vec{p}\dot(\vec{x}-\vec{x'})-w(t-t')]} \\
                 &+&i\theta(t'-t) )\int \frac{d^3\vec{p}} {(2\pi)^3} \frac {1} {2w_{\vec{p}}} 
e^{-i[\vec{p}\dot(\vec{x}-\vec{x'})-w(t-t')]}, \nonumber
\end{eqnarray}
where $w_{\vec{p}}^2=\vec{p}^2+m^2$.  When $t \rightarrow -\infty$, $\triangle_F \rightarrow e^{iwt}$.  
The Dyson propogator is simply the Feynman propagator with the poles reversed,
\begin{eqnarray}
\triangle_D(x-x')&=&i\theta(t-t')\int \frac {d^3\vec{p}} {(2\pi)^3} \frac {1} {2w_{\vec{p}}} 
e^{-i[\vec{p}\dot(\vec{x}-\vec{x'})-w(t-t')]} \\
                 &+&i\theta(t'-t) )\int \frac{d^3\vec{p}} {(2\pi)^3} \frac {1} {2w_{\vec{p}}} 
e^{+i[\vec{p}\dot(\vec{x}-\vec{x'})-w(t-t')]}. \nonumber
\end{eqnarray}
In addition to these propagators there are the positive and negative frequency Wightman functions
\begin{eqnarray}
\triangle^+(x-x')&=&\int \frac {d^4p} {(2\pi)^4} e^{ip(x-x')} 2\pi i \delta(p^2-m^2)\theta(p_0) \\
                &=&i\int \frac{d^3\vec{p}}{(2\pi)^3} \frac {1} {2w_{\vec{p}}} 
                e^{i[w(t-t')-\vec{p}\cdot(\vec{x}-\vec{x'})]}, \nonumber
\end{eqnarray}
and
\begin{equation}
\triangle^-(x-x')=-i \int \frac{d^3\vec{p}}{(2\pi)^3} \frac {1} {2w_{\vec{p}}} e^{-i[w(t-t')-\vec{p}\cdot(\vec{x}-\vec{x'})]}.
\end{equation}
Since the free field generating functional is exact these four functions are the four quantum
expectation values,
\begin{eqnarray}
\label{eq-ctpG}
i\triangle_F(x-x') & \equiv &\hbar G_{++}(x,x') = \langle 0_- | T\left [\Phi_H(x)\Phi_H(x')\right] |0_-\rangle, \\
-i\triangle_D(x-x') & \equiv & \hbar G_{--}(x,x') = \langle 0_- | \tilde{T}\left [\Phi_H(x)\Phi_H(x')\right] |0_-\rangle, \nonumber \\
-i\triangle^+(x-x') & \equiv & \hbar G_{+-}(x,x') = \langle 0_- |\Phi_H(x)\Phi_H(x') |0_-\rangle, \nonumber \\
i\triangle^-(x-x') & \equiv & \hbar G_{-+}(x,x') = \langle 0_- |\Phi_H(x')\Phi_H(x) |0_-\rangle. \nonumber
\end{eqnarray}
As with the in-out formalism the CTP generating functional of an interacting field theory is built
from $Z_{free}[J_+,J_-]$.  For a scalar field theory with quadratic self-interaction,
\begin{equation}
\label{eq-ctpZint}
Z_{int}[J_+,J_-]=\exp {\left [ \frac {-i\lambda}{4!} \int d^4x \left [ \frac {\delta^4}{\delta J_+(x)^4}-
\frac {\delta^4}{\delta J_-(x)^4} \right] \right ]}Z_{free}[J_+,J_-].
\end{equation}
The diagrammatic rules of the CTP formalism follow from Eqns.~(\ref{eq-ctpZint}) and (\ref{eq-ctpG}).  There are two external legs, 
one corresponding to $J_+$ the other to $J_-$, and two interaction vertices with coupling $\pm \lambda$.  
Connecting the external legs and the vertices are the four Greens functions, Eq.~(\ref{eq-ctpG}).  Each diagram in the
conventional (in-out) formalism of quantum field theory is
expanded in CTP language by writing all possible ($\pm$) combinations of the external legs, vertices and Greens functions.

\subsection{The 2PI-CTP effective action}
\label{subsec:2PICTP}
The CTP effective action combines the nonperturbative nature of the 
quantum effective action considered in Sec.~\ref{sec:qea} with the dynamical content of the CTP formalism.
The 2PI-CTP effective action generates real and causal equations of motion for the mean field $\langle \Phi_H(\vec{x},t) \rangle$
and the four two-point functions $G_{ab}(x,x')$ of the quantum field theory.
The 2PI-CTP equations of motion are
\begin{eqnarray}
\left. \frac{\delta \Gamma_{CTP}[\phi,G_{ab}]}{\delta \phi_+(x)} \right |_{(\phi_+=\phi_-)}=0, \\
\left.\frac{\delta \Gamma_{CTP}[\phi,G_{ab}]}{\delta G_{ab}(x,y)}\right |_{(\phi_+=\phi_-)}=0. 
\end{eqnarray}
The 2PI-CTP effective action is constructed from Eq.~(\ref{eq-2PIea}) and Fig.~\ref{fig:gamma2} by expanding each 
diagram in all possible ($\pm$) combinations of vertices and Greens functions.  For definiteness and since
we will use this expression in the following chapter we give the analytic expression for a two-loop truncation
of the 2PI-CTP effective action for a scalar field theory with quartic self-interaction,
\begin{eqnarray}
\Gamma_{CTP}[\phi_a,G_{ab}]&=&S[\phi_+]-S[\phi_-]-\frac{i\hbar}{2} \ln{\det{G_{ab}}} \nonumber \\
&+&\frac{i\hbar}{2}\int d^4x \int d^4x'  A^{ab}(x,x')G_{ab}(x',x)  \nonumber \\
&+& \Gamma_2[\phi_a,G_{ab}],
\end{eqnarray}
where        
\begin{equation}
iA^{ab}(x,x')=\frac{\delta^2 (S[\phi_+]-S[\phi_-])}{\delta \phi_a(x) \delta \phi_b(x')},
\end{equation}
and $a,b=\{+,-\}$.
The two-loop contribution is
\begin{eqnarray}
\Gamma_2[\phi_a,G_{ab}] &=&\frac{\lambda \hbar^2}{8} \int d^4x \left[ G_{++}^2(x,x)-G_{--}^2(x,x)\right] \nonumber \\
 &+&\frac{i\lambda^2 \hbar^2}{4} \int d^4x \int d^4x' \left[ \phi_+(x)G_{++}^3(x,x')\phi_+(x')\right] \nonumber \\
 &+&\frac{i\lambda^2 \hbar^2}{4} \int d^4x \int d^4x' \left[\phi_-(x)G_{--}^3(x,x')\phi_-(x') \right]\nonumber \\
 &-&\frac{i\lambda^2 \hbar^2}{2}  \int d^4x \int d^4x' \left[\phi_+(x)G_{+-}^3(x,x')\phi_-(x') \right].
\end{eqnarray}

\section{An example: The quantum anharmonic oscillator}
\label{sec:anosc}
The preceding discussion was somewhat formal. To illustrate CTP and effective action techniques
we derive the one-loop equations of motion for the mean position of a quantum anharmonic oscillator with classical action
\begin{equation}
S[x]=-\int \frac{dt}{2}\left[x(t) \left(\frac{d^2}{dt^2}+w^2\right)x(t)+\frac{\lambda}{2}x^4\right].
\end{equation}
The 1PI-CTP effective action truncated at one loop is
\begin{equation}
\Gamma[x_+,x_-]=S[x_+]-S[x_-]-\frac{i\hbar}{2}Tr \ln{A^{ab}},
\end{equation}
where
\begin{equation}
iA^{ab}=\frac{\delta^2 S}{\delta x^a \delta x^b}.
\end{equation}
The nonzero components of $A^{ab}$ are
\begin{equation}
iA^{++}(t,t')=-\left(\frac{d^2}{dt^2}+w^2+3\lambda x_+^2\right)\delta(t-t'),
\end{equation}
and
\begin{equation}
iA^{--}(t,t')=\left(\frac{d^2}{dt^2}+w^2+3\lambda x_-^2\right)\delta(t-t').
\end{equation}
The equation of motion for the mean position of the oscillator is given by
\begin{equation}
\left (\frac{\delta \Gamma[x_+,x_-]}{\delta x_+}\right )_{x_+=x_-}=0.
\end{equation}
Using $\delta\ln A= A^{-1}\delta A$, variation of the log term gives $(A^{-1})^{++}(t,t) 6i\lambda x(t)$ 
and the equation of motion for the mean position is
\begin{equation}
\label{eq-osc1}
\left(\frac{d^2}{dt^2}+w^2+\lambda x^2+3\lambda\hbar (A^{-1})^{++}(t,t)\right)x(t)=0.
\end{equation}
The inverse of the operator $(A^{-1})^{++}$,
\begin{equation}
G^{++}(A^{-1})^{++}=1,
\end{equation}
is the Feynman propagator for fluctuations about the mean field $x(t)$,
\begin{equation}
\hbar G^{++}(t,t')=\langle 0 | T\:[\delta \hat{x}_H(t), \delta \hat{x}_H(t')]|0\rangle,
\end{equation}
where $\delta \hat{x}_H(t)=\hat{x}_H(t)-x(t)$ is the fluctuation operator.
We can rewrite Eq.~(\ref{eq-osc1}) as a pair of coupled equations
\begin{eqnarray}
\label{eq-osc2}
\left(\frac{d^2}{dt^2}+w^2+\lambda x^2+3\lambda \hbar G^{++}(t,t) \right) x(t)=0\\
\left(\frac{d^2}{dt^2}+w^2+\lambda x^2\right)G^{++}(t,t')=-i\delta(t-t')
\end{eqnarray}
As they stand these equations are not in a convenient form for analysis (in particular for numerical solution).
We expand the Heisenberg fluctuation operator $\delta \hat{x}_H(t)$ in terms of time-dependent 
complex mode functions $f(t)$ and time-independent creation and annihilation operators,
\begin{equation}
\label{eq-fieldmode}
\delta \hat{x}_H(t)=f(t)\hat{a}+f^*(t)\hat{a}^{\dagger}.
\end{equation}
The creation and annihilation operators are defined with respect to the initial vacuum 
\begin{equation}
\hat{a}|0\rangle=0,
\end{equation}
and obey the usual canonical commutation relations 
\begin{equation}
[\hat{a},\hat{a}^\dagger]=1.
\end{equation} 
Since $[\hat{x},\hat{p}]=i\hbar$, the mode functions must satisfy the Wronskian condition
\begin{equation}
\label{eq-wronk}
f(t)\dot{f}^*(t)-f^*(t)\dot{f}(t)=i\hbar.
\end{equation}
Using Eq.~(\ref{eq-fieldmode}) the Feynman propagator
is simply expressed in terms of the mode functions
\begin{equation}
\hbar G^{++}(t,t')=\theta(t-t')f(t)f^*(t')+\theta(t'-t)f^*(t)f(t').
\end{equation}
The 1PI-CTP equations of motion for the mean field $x(t)=\langle \hat{x}_H \rangle $
and the fluctuation mode functions are
\begin{eqnarray}
\label{eq-1PICTPeom}
\left(\frac{d^2}{dt^2}+w^2+\lambda x^2+3\lambda |f(t)|^2 \right)x(t)=0 \\
\left(\frac{d^2}{dt^2}+w^2+3\lambda x^2 \right ) f(t)=0 \nonumber
\end{eqnarray}
The equations of motion preserve the Wronskian condition, Eq.~(\ref{eq-wronk}) so that we only
require it to be satisfied initially.  The nonlinear (ordinary) differential equations, Eq.~(\ref{eq-1PICTPeom}), are now 
easy to treat numerically.  

\section{Summary}
\label{sec:ch1sum}
In this chapter, we reviewed the Closed-Time-Path (CTP) formulation of nonequilibrium
quantum field theory.  We introduced the quantum effective action and presented a
diagrammatic loop expansion of both the one-particle-irreducible (1PI) and two-particle-irreducible (2PI) 
effective action.  We then combined the CTP formalism with effective action techniques and
gave an explicit expression for the 2-loop truncation of the 2PI-CTP effective action.  
In Chapter 3 we use this expression
to derive a dynamical equation for the two-point function of a quantum scalar field undergoing a
symmetry-breaking phase transition in the early universe.
Finally, as a simple example, we used the 1PI-CTP effective action to derive evolution
equations for the mean position of a quantum anharmonic oscillator.

\chapter{Quantum critical dynamics:  Formation of domains in the early universe}
\section{Introduction} 
A complete understanding of the physical issues involved 
in the formation of topological defects in a quantum field theoretic phase transition requires a
first-principle approach to the nonequilibrium dynamics of quantum fields, 
a realistic treatment of the interaction of
the quantum field with other fields that constitute an environment, and the 
identification of classical defect configurations from the quantum field system.
In this chapter, which represents work published in Ref.~\cite{stephens:1999},
we analyze the nonequilibrium quantum dynamics of the phase transition, focusing on
the formation of correlated domains of true vacuum. We choose as our model 
a quantum scalar field evolving through a second-order phase transition,
initiated by the cooling of a radiation-dominated Friedmann-Robertson-Walker (FRW) Universe.  We leave to
future research a thorough analysis of the field-bath interaction and the quantum-to-classical 
transition (for recent, related work on decoherence see \cite{lombardo:2000}). 

While symmetry restoration in finite-temperature quantum field theory has been known 
since the early work of Kirzhnitz and Linde \cite{kirzhnitz:1972}, most previous efforts have
focused on the equilibrium aspects of the transition.  
Techniques such as the finite-temperature effective potential \cite{dolan:1974,weinberg:1974} and the 
renormalization group \cite{ma:1976,zinn-justin:1996} 
have been developed to deduce 
equilibrium critical properties such as the order of the phase transition and its
critical temperature.  However, equilibrium techniques 
are inadequate to study the {\it dynamics} of the phase transition.  The use of equations of 
motion for the mean field derived from the finite-temperature
effective potential was criticized in \cite{mazenko:1985}.   In general the use of equations generated 
from the finite-temperature effective potential in a dynamical
setting results in unphysical (acausal and complex) solutions.   Although equilibrium techniques
are clearly inappropriate, solving the full equations of motion
for an interacting quantum field theory is generally impossible, even numerically.  
As discussed in the previous chapter, this difficulty is partially overcome by the development of 
approximation schemes which allow for the evolution of a restricted set of correlation functions of 
the quantum field theory.  These methods have been applied to a number 
of dynamical problems in quantum field theory including the dynamics of second-order phase transitions
\cite{calzetta:1989,boyanovsky:1993,cooper:1997,boyanovsky:2000}.

The problem of defect formation in a nonequilibrium second-order phase transition of a quantum field has 
also recently received attention \cite{bray:1994,gill:1995,kara:1997,bowick:1998}.  The results 
are promising but the studies are incomplete.  In these previous approaches the phase transition is incorporated through 
an {\it ad hoc} time dependence of the effective mass of a {\it free} field theory:  an instability in the theory is
induced when the mass becomes tachyonic.  The use of a prescribed time dependence of the effective mass, 
while providing a convenient analytic model, lacks 
physical justification.  The neglect of interactions confines the applicability
of these approaches to very early times before the field amplitude grows substantially. They are therefore 
unable to account for the back reaction which is necessary to  stabilize domain growth and shut off the spinodal 
instabilities of the phase transition. In addition, defects formed during
the linear stages of the phase transition are transient and not likely to survive to late times.

It is useful to contrast our first-principles approach to the dynamics of a phase transition in quantum field
theory with an approach common to condensed matter.   In condensed matter systems it is 
common to model critical dynamics with a classical, phenomenological, time-dependent Landau-Ginzburg equation 
for an order parameter $\Psi$,
\begin{equation}
\partial_t \Psi(\vec{x},t) = -\Gamma \frac {\delta F} {\delta \Psi } +\xi,
\end{equation}
\noindent where $F[\Psi]$ is a phenomenological free energy density for the order parameter, $\Gamma$ is a
phenomenological dissipative coefficient and $\xi$ is a stochastic term
incorporating thermal fluctuations of the environment \cite{goldenfeld:1995}. 
Even if the order parameter is of quantum origin, as {\it e.g.} in the phase of the 
wavefunction for liquid Helium-4, the Landau-Ginzburg equation is rarely 
derived logically from the underlying quantum dynamics of the system.  
In condensed matter systems, to compensate for insufficient microscopic information, the order parameter and
its equation of motion are chosen with great care and physical intuition.
In experimentally inaccessible environments, such as the early universe, it is
not {\it a priori} obvious what the order parameter, or its dynamics, should be.  
In situations where phenomenological approaches are inadequate, it is 
necessary to work with the fundamental quantum dynamics of the fields. 
A first-principles approach to the quantum dynamics of phase 
transitions avoids {\it ad hoc} assumptions about the dynamics of the correlation length and
the effect of the quench.  The system simply evolves under the true 
microscopic equations of motion.  We can therefore explore many 
details of critical dynamics that are inaccessible in phenomenological theories.  

This chapter is organized as follows.  In Sec.~\ref{sec:critdyn}   
we present the model, a derivation of the equations of motion for the two-point function of the theory, 
and discuss renormalization, initial conditions and numerical parameters used in the numerical simulation.
We then provide a dynamical description of the phase transition using results of the numerical simulations.  
Section \ref{sec:domains} begins the discussion of domains and presents the argument that
domains are determined by a peak in the Fourier space structure function $k^2G(k,t)$.  Section \ref{sec:ch2critscal} 
discusses the power-law scaling of the size of domains with the quench rate.  
The equilibrium critical exponents of the correlation length and
relaxation time are calculated in both the underdamped and overdamped cases and the power-law scaling of the 
defect density with the quench rate predicted by the
freeze-out proposal is shown to be in good agreement with the numerical simulations.  An analytical model valid for 
slow quenches and near the 
onset of the instability is introduced and the power-law exponent in the analytic model is found to be the same as 
the numerical simulations. 
Section \ref{sec:ch2sum} provides a summary and discussion of 
these results and presents possible directions for further research.

\section{Nonequilibrium symmetry-breaking through correlation dynamics}
\label{sec:critdyn}
We consider a quantum scalar field in a FRW spacetime. 
The field has the symmetry-breaking classical action
\begin {equation}
S=\int  d^4x \sqrt{-g}\left (\partial_{\mu} \Phi \partial^{\mu} \Phi +m^2 \Phi^2 -\frac {\lambda} {4!} \Phi^4 \right),
\end {equation}
where $g$ is the determinant of the metric of the classical 
background  spacetime.  We assume 
that the stress-energy tensor is dominated by other radiation fields present 
in the early Universe.  These fields maintain the overall 
homogeneity and isotropy of the universe. Small deviations in 
homogeneity and isotropy that may eventually be responsible for the fluctuations observed in the
cosmic microwave background radiation are produced in our model by the topological 
defects of the system and appear only at the end of the phase transition.
We therefore work in the semiclassical test-field approximation (ignoring 
back reaction of the $\Phi$ field on the spacetime), and we assume 
that the scale factor has the time-dependence of a homogeneous and isotropic, spatially flat,
radiation-dominated universe,
\begin {equation}
\label {eq-scale}
a(t)= \left[ \frac {t+\tau} {\tau} \right ] ^{\frac {1} {2}}.
\end {equation}
The expansion of the universe and the resulting redshifting of the modes act here as a physical
quench allowing the dynamics of the phase transition to unfold naturally.  This is in distinction to 
work which uses an 
instantaneous change in the sign of the square of the mass \cite{boyanovsky:1993,gill:1995}.  

The nonequilibrium dynamics of the quantum field is derived through the the use of the 2PI-CTP effective
action, discussed in the previous chapter.  The equations of motion for the mean field and the two-point function 
derived from the 2PI-CTP effective action respect the $\Phi \rightarrow -\Phi$ symmetry of the classical action. 
Since the field starts in a symmetry-restored state 
above the critical point where
\begin{equation}
{\langle \Phi \rangle}_{initial} =0,
\end{equation}
\noindent the mean field remains identically zero throughout the phase transition; the dynamics of the phase transition
unfold through the dynamics of the two-point correlation functions.  
The equations of motion for the two-point functions are obtained through the variation of the two-loop truncation of the 
2PI-CTP effective action presented in Sec.~\ref{subsec:2PICTP}.  The two-loop truncation of the 2PI-CTP effective
action is equivalent to the time-dependent Hartree-Fock approximation.
The equation of motion for the Feynman propagator 
$G(x,x')$ is  
\begin{equation}
\label{eq-xeom}
\left (\frac{\partial ^2}{\partial t^2}-\frac {\nabla_x^2}{a^2}+3\frac {\dot{a}}{a}\frac {\partial} {\partial t} 
-m^2+\frac {\lambda \hbar}{2} G(x,x) \right )G(x,x')=-i\delta(x-x').
\end{equation}
Instead of directly solving Eq.~(\ref{eq-xeom}) we derive the
homogeneous equation for the (complex) mode functions of the quantum field.
We define the spatial Fourier transform,
\begin{equation}
G(x,x')=\int \frac{d^3 \vec{k}}{(2\pi)^3} e^{-i\vec{k}\cdot (\vec{x}-\vec{x}')}G_k(t,t'),
\end{equation}
where $k=|\vec{k}|$, and expand the Heisenberg Field operator $\Phi_H(\vec{x},t)$ as
\begin{equation}
\hat{\Phi}_H(\vec{x},t)=\int \frac{d^3\vec{k}} {(2\pi)^3} \left [ e^{i\vec{k}\cdot\vec{x}}f_k(t)\hat{a}_{\vec{k}} +
 e^{-i\vec{k}\cdot\vec{x}}f^*_k(t)\hat{a}^{\dagger}_{\vec{ k}}\right ].
\end{equation}
The creation operators and annihilation operators $\hat{a}_{\vec{k}}$ and $\hat{a}^{\dagger}_{\vec{k}}$ 
are defined with respect to the initial vacuum,
\begin{equation}
\hat{a}_{\vec{k}}|0\rangle=0,
\end{equation}
and obey the commutation relations 
\begin{equation}
\label{eq-accr}
[\hat{a}_{\vec{k}},\hat{a}^{\dagger}_{\vec{k}'}]=\delta_{\vec{k},\vec{k}'}.
\end{equation} 
The field-momentum canonical commutation relations and Eq.~(\ref{eq-accr}) imply the Wronskian condition on the mode functions,
\begin{equation}
f_k(t)\dot{f}_k^*(t)-f_k^*(t)\dot{f}_k(t)=i\hbar.
\end{equation}
The momentum space two-point function $G_k(t,t')$ expressed in terms of the mode functions is
\begin{eqnarray}
\label{eq-modeG}
\hbar G_k(t,t')&=& i \theta(t-t') \left ( f_k(t)f^{*}_k(t')(N_k+1)+f^{*}_k(t)f_k(t')N_k \right )  \nonumber \\
          &+& i \theta(t'-t) \left ( f^{*}_k(t)f_k(t')(N_k+1)+f_k(t)f^{*}_k(t')N_k \right ),
\end{eqnarray}
where
\begin{equation}
N_k=\langle a_k{^\dagger} a_k \rangle
\end{equation}
is the number of particles of momentum $k$ in the initial state.
Equations (\ref{eq-xeom}) and (\ref{eq-modeG}) lead to the mode function equation,
\begin {equation}
\label {eq-mode}
\left (\frac {d^2} {dt^2} + 3\frac {\dot{a}(t)} {a(t)} \frac {d} {dt} + \frac {k^2} {a^2(t)}-m^2
+ \frac {\hbar \lambda} {2} G(t,t)\right )f_k(t) = 0,
\end {equation}
 
\noindent where
\begin {equation}
\label{eq-equaltime}
\hbar G(t,t) = \int \frac {d^3k} {(2\pi)^3} f_k(t)f^*_k(t)\sigma_k(T)
\end {equation}
is the equal-time limit of the two-point correlation function and
\begin {equation}
2N_k+1\equiv \sigma_k(T)=\coth \left (\frac { \hbar w_k(0)}{2T} \right )
\end {equation}
is a constant factor incorporating the thermal initial conditions with temperature $T$.  The
effective mass of the system is 
\begin{equation}
\label{eq-meffa}
m^2_{eff}(t)=-m^2+\frac {\lambda \hbar} {2} G(t,t)
\end{equation}
\noindent and the initial frequency is
\begin {equation}
w^2_k(0)=k^2+m^2_{eff}(0),
\end {equation}
where $m^2_{eff}(0)$ is the initial finite-temperature effective 
mass in the Hartree-Fock approximation. Since we work in a radiation-dominated FRW universe, the scalar curvature, 
$R$, is zero and the conformal coupling constant $\xi$ may be ignored. 
The dynamics of the phase transition is analyzed by solving the coupled, nonlinear mode function 
equations, Eq.~(\ref{eq-mode}).  Before proceeding to this discussion we first treat the issues of 
renormalization, initial conditions and the parameters of the model.

\subsection{Renormalization}
The equal-time limit of the two-point function, Eq.~(\ref{eq-equaltime}), appearing in the equation
for the effective mass, Eq.~(\ref{eq-meffa}), is divergent and must be regularized.  A simple regularization 
method, amenable to a numerical simulation, is to implement an ultraviolet cutoff in physical spatial momentum.  
After suitable renormalization, the expression for the two-point function must be independent of this cutoff. 
The cutoff dependence of the bare variance is isolated by
considering a WKB-type solution to the mode function equation. 
Identifying the second order adiabatic mode functions from the 
WKB solution and in the limit of large cutoff $\Lambda$ we find
\begin{equation}
\hbar G_{\Lambda}(t,t)=\frac {1} {2\pi^2}\left ( \frac {\Lambda^2}{2a(t)^2}
-\frac{1}{4}\ln{\left (\frac{\Lambda}{\kappa}\right )} m^2_{eff} \right ) + \mathcal{O}\left(\frac{1}{\Lambda}\right ).
\end{equation}
The removal of these divergent pieces in the effective mass is implemented through mass and coupling
constant renormalization.  The quadratic divergence is associated with the divergence of the propagator self-energy
which we cancel with renormalization of the bare mass.  We remove the remaining logarithmic divergence
by renormalization of the bare coupling constant.  Explicitly,
\begin{eqnarray}
m_b^2 & = & -\frac{\hbar \lambda_b}{16\pi^2}\frac{\Lambda^2}{a^2(t)} +
m_r^2 \left[1+\frac{\hbar \lambda_b}{16\pi^2}\ln(\Lambda/\kappa)\right], \\
\lambda_b & = & \frac{\lambda_r}{1-\frac {\hbar \lambda_r} {16\pi^2} \ln(\Lambda/\kappa)}.
\end{eqnarray}
The shift in bare parameters is time independent as long as the cutoff $\Lambda$ 
and the renormalization scale $\kappa$ are implemented in terms of the physical
momentum,
\begin {eqnarray}
\Lambda=\Lambda_0 a(t), \\
\kappa=\kappa_0 a(t),
\end {eqnarray}
where $\Lambda$ and $\kappa$ are comoving and $\Lambda_0$ and
$\kappa_0$ are physical quantities.
The renormalized mode function equation is
\begin{equation}
\label {eq-rmode}
\left (\frac {d^2} {dt^2} + 3\frac {\dot{a}(t)} {a(t)} \frac {d} {dt} + \frac {k^2} {a^2(t)}-m^2_r
+ \frac {\hbar \lambda_r} {2} G_S(t,t)\right )f_k(t) = 0.
\end{equation}
where $G_S(t,t)$ is the subtracted two-point function,
\begin{equation}
\label{eq-Gs}
\hbar G_S(t,t)=\frac {1} {2\pi^2} \int_{0}^{\Lambda} k^2 dk \left ( f_k f^{*}_k \sigma_k (\beta)
 -\frac {1} {ka^2(t)} + \frac {\theta(k-\kappa)} {4k^3} m^2_{eff} \right ).
\end{equation}
Our renormalization procedure simply removes the (divergent) contribution of short-wavelength modes 
while leaving the long-wavelength modes important to the dynamics of phase transition unaffected.
In following sections we will drop the renormalization subscripts for clarity and it is to be 
understood that we are working with renormalized quantities.

\subsection{Initial conditions}
The quantum field is assumed to be in an initial state of thermal equilibrium. 
In this model with a tachyonic tree-level mass, symmetry is restored by finite 
temperature corrections.  The initial effective mass $m^2_{eff}$ is the solution of the 
(renormalized) equation
\begin{equation}
\label {eq-minit}
m_{eff}^2=-m^2+\frac{\hbar \lambda} {4\pi^2} \int_{0}^{\Lambda} k^2 dk \left (\frac{\coth{
\frac{\hbar \sqrt{k^2+m_{eff}^2}}{2T} }}
{2\sqrt{k^2+m_{eff}^2}} -\frac {1} {2k} + \theta(k-\kappa)\frac{ m^2_{eff}} {4k^3} \right ).
\end{equation}
In the high temperature and small $\lambda$ limit this yields  
\begin{equation}
\label {eq-highTm}
m_{eff}^2=-m^2+\frac {\lambda T^2} {24},
\end{equation}
a result familiar from finite-temperature field theory \cite{kirzhnitz:1972}

In an expanding FRW Universe, exact thermal 
equilibrium will persist only for conformally invariant fields.  
If the expansion rate is small relative to internal collisional processes of the field 
then there is an approximate notion of equilibrium \cite{hu:1982,hu:1983}.  
This is evidenced by transforming to conformal time, $\eta$, defined by 
\begin{equation}
dt=a(\eta)d\eta,
\end{equation}
and performing a mode redefinition
\begin{equation}
\tilde{f}_k(\eta)= f_k(\eta)a(\eta)    
\end {equation}
The rescaled conformal mode function equation is 
\begin {equation}
\label {eq-conmode}
\left (\frac {d^2} {d\eta^2} + k^2 + a^2(\eta)(m^2+\frac {\hbar \lambda} {2} G(\eta,\eta))\right )\tilde{f}_k(\eta)=0.
\end {equation}
If the Universe is slowly expanding,  a WKB-type solution 
is appropriate and a low-adiabaticity truncation of the instantaneous
WKB frequency is sufficient. The zeroth-adiabatic order solution to
equation (\ref {eq-conmode}) is given by 
\begin{eqnarray}
\tilde{f}(\eta)=\frac {1} {\sqrt{2w_k}} e^{-i\int^{\eta}w_kd\eta'}, \\
w_k^2=k^2+a(\eta)^2m_{eff}^2.
\end {eqnarray}
This leads to the following initial conditions for the 
mode functions in cosmic time t
\begin{eqnarray}
f_k(0) & = &\frac {1} {a(0)\sqrt{2w_k(0)}}, \\
\dot {f_k}(0) & = &\left(-\frac {\dot{a(0)}} {a(0)} -iw_k(0)\right)f_k(0).  
\end{eqnarray}
Slow expansion assumes that the 
natural frequency of the kth mode is faster than 
the expansion rate of the Universe,
\begin {equation}
w_k(0) \gg \frac {1} {2\tau}.
\end {equation}
The adiabatic equilibrium approximation fails for the 
lowest $k$ modes.  However with high temperature initial conditions the low $k$ modes are not a 
dominant part of the spectrum. 

\subsection{Numerical parameters}
The model described by Eqns.~(\ref{eq-scale}), (\ref{eq-rmode}) and (\ref{eq-Gs}) is characterized by 6 parameters,
$(m,\, \tau, \, \lambda, \, T, \, \Lambda, \, \kappa)$.  In the following we set $\hbar=1$.
Spacetime scales are measured in units of $1/m$, effectively setting $m^2=1$.
The initial rate of expansion or the initial Hubble constant is controlled 
by $\tau$,
\begin {equation}
H(0)=\frac {\dot{a}(0)} {a(0)}=\frac {1} {2\tau}.
\end {equation}
In our simulations the range of the $\tau$ parameter is
\begin {equation}
0.01\leq \tau \leq 100.
\end {equation}
With this range, the scale factor grows during the phase transition by a factor $a_{final} \sim 13 \: a_{initial}$ 
for the fastest quench and $a_{final} \sim 1.3 \: a_{initial}$ for the slowest quench.
The self-coupling is (relatively) strong,
\begin {equation}
\lambda=0.1.
\end {equation}
The initial temperature $T$ is chosen
so that the value of the initial effective mass $m^2_{eff}(0)$ is of order unity.  Specifically
we choose
\begin {equation}
T =20.0,
\end{equation}
so that
\begin{equation}
m^2_{eff}(0)=-1.0+\frac {\lambda}{24}T^2=0.607.
\end {equation}
The values of the initial effective mass $m^2_{eff}(0)$, the coupling $\lambda$ and expansion rate $\tau$, 
were chosen so that the simulations of the phase transition completed on numerically accessible timescales. Due to the 
renormalization scheme, both the comoving cutoff $\Lambda$ and the comoving renormalization scale $\kappa$ 
increase with the scale factor.  The initial 
value 
\begin {equation}
\Lambda_0 = 340
\end {equation}
was chosen as the lowest value that maintained cutoff 
independence of the initial effective mass.  Insensitivity to the cut-off in the dynamical simulations
was verified by doubling $\Lambda_0$ and observing no change in the output
plots of the time-dependent effective mass.  The initial value of the 
renormalization scale 
\begin {equation}
\kappa_0=1.0
\end {equation}
was chosen so that the renormalization scale was always above 
the maximum momentum of the unstable modes.  The coupled, nonlinear system of mode function equations with the 
given initial conditions was solved numerically using an adaptive stepsize,
fifth-order Runge-Kutta code.  Mode integrals were performed using a simple Simpson rule 
with a uniform momentum binning
\begin {eqnarray}
k=nk_{bin}, \\
k_{bin}=\frac {2\pi} {L_0},
\end {eqnarray}
where $L_0=100.0$ is the effective size of the system and n is the total number of modes.  
Insensitivity to the momentum binning was verified by reducing $k_{bin}$ and 
observing no change in output. The number of
modes $n$ varied from $10^4$ to $10^5$. Run times varied from hours to 
days on a DEC 500 MHz workstation which corresponds to dynamical time scales of $t=2$ to $t=100$.      

\section {Domain formation}
\label{sec:domains}
\begin{figure}
\begin{center}
\strut\psfig{figure=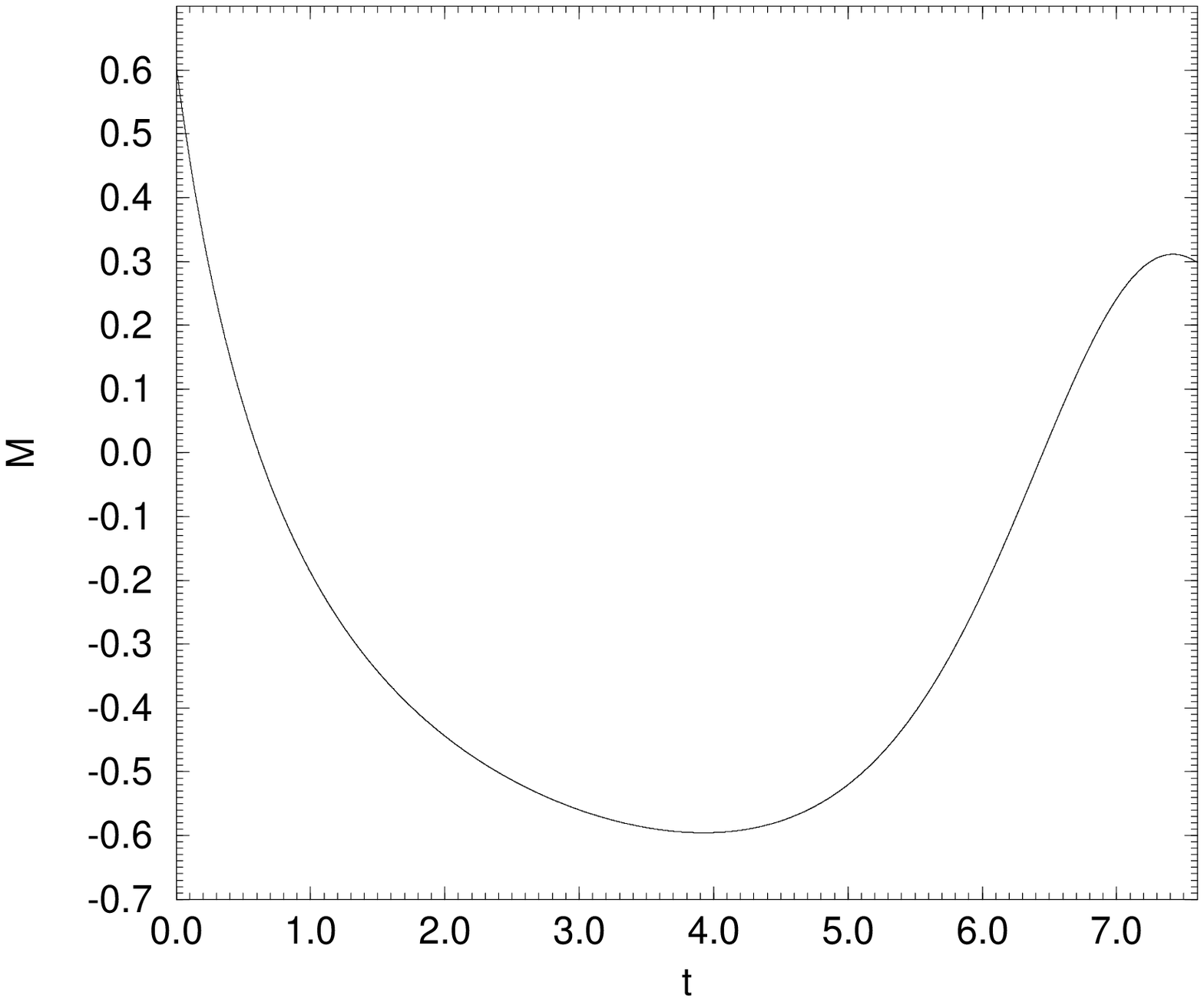,width=3.5in,angle=0}
\rmmcaption{Plot of the square of the (renormalized) effective mass 
$M=m^2_{eff}$ vs. cosmological time $t$ for quench parameter 
$\tau=1$.}
\label{fig:meff}
\end{center}
\end{figure}

\begin{figure} 
\begin{center}
\strut\psfig{figure=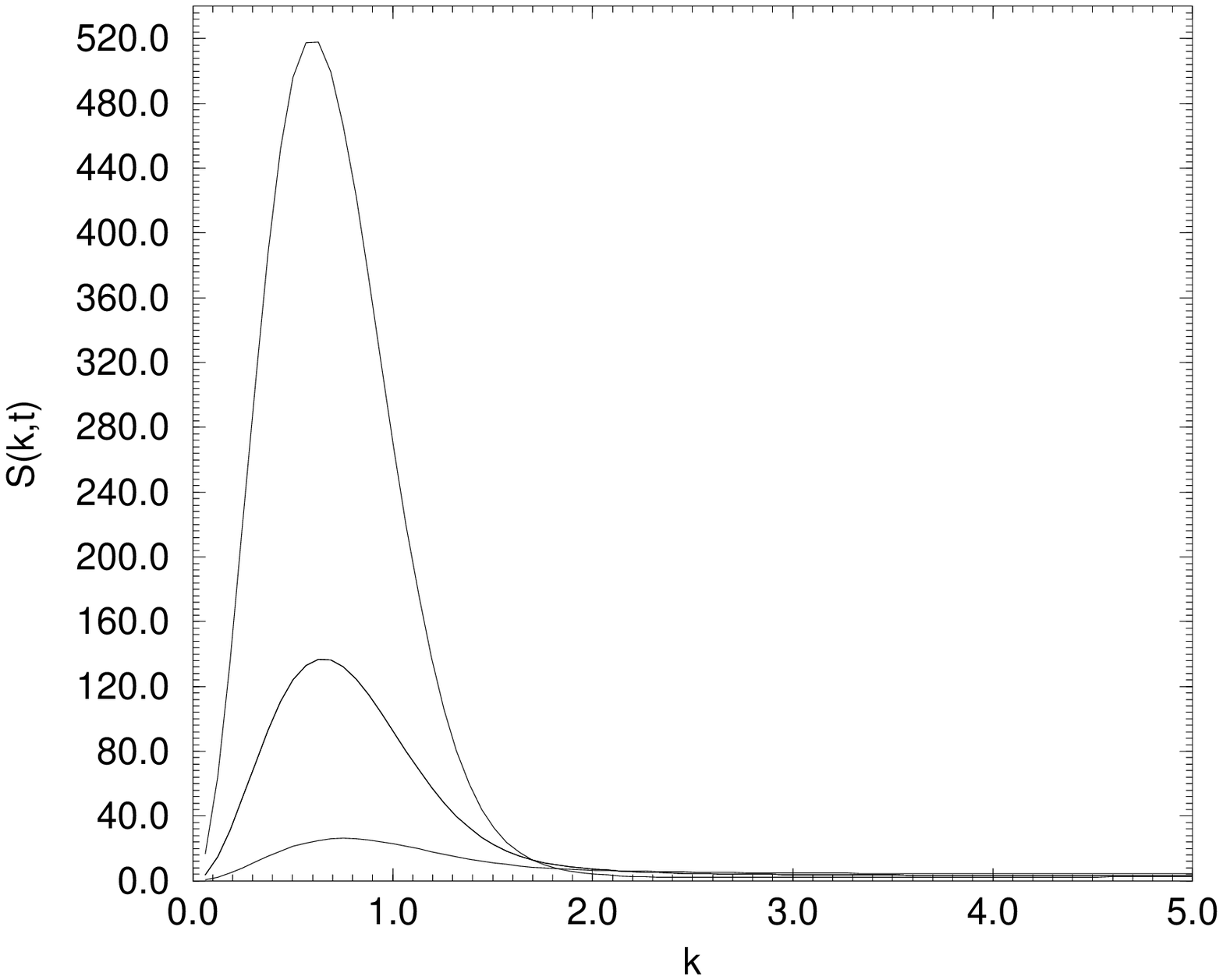,width=3.5in,angle=0}
\rmmcaption{Plot of the Fourier-space structure factor, $S(k,t)=k^2G(k,t)$, vs. 
comoving momentum $k$ at various times during the evolution for quench parameter $\tau=1$.}
\label{fig:structure}
\end{center}
\end{figure}

The results of a typical simulation are shown in Figs.~\ref{fig:meff} and \ref{fig:structure}.  
In Fig.~\ref{fig:meff} 
we plot the renormalized effective mass as a function of 
cosmological time for quench parameter $\tau=1$.  The phase 
transition begins when $m_{eff}^2$ first becomes negative. When $m_{eff}^2$ is negative,
modes with physical momentum $\frac {k} {a} \leq m_{eff}$ 
have imaginary frequencies and begin to grow.  This indicates the onset of 
the spinodal instability which is characteristic of a second-order phase 
transition. In the early stages of the phase transition, the evolution of the 
effective mass is dominated by the redshifting of stable modes 
and the effective mass decreases. As the phase 
transition proceeds, more modes redshift into the unstable momentum band and 
the amplitude 
of unstable modes continues to grow.  Eventually, the redshift of stable 
modes balances the growth of unstable modes and the effective 
mass increases.  As the effective mass passes through zero from below, 
the field reaches the spinodal point. Beyond the spinodal point 
all modes are stable. As the effective mass continues to grow, 
non-linear thermal and other collisional processes are expected to be
important and the Hartree-Fock approximation of the dynamics of the two-point
correlation function breaks down (see for example \cite{calzetta:1989}).  
In Fig.\ref{fig:structure} we show the Fourier space structure factor 
\begin{equation}
S(k,t) \equiv k^2G_k(t,t)
\end{equation} 
at various times during the phase transition.  The bottom 
curve is a snapshot of $S(k,t)$ at time $t=3.9$. The middle curve is a snapshot at $t=5.4$.  
The top curve is a snapshot at $t=6.9$  As argued in the
next section, defects and domains can be identified in the low-k structure of 
$S(k,t)$. As the phase transition begins, $S(k,t)$ develops a peak at 
low k. As the phase transition proceeds, this peak 
grows in amplitude and redshifts until at late times it completely dominates the 
infrared portion of the spectrum.

\subsection {Defect density}

We have argued that the dynamics of a phase transition of a quantum field 
may be approximated by the dynamics of the two-point function 
at least for times less than the spinodal time. Using the two-loop 2PI-CTP
equations of motion we obtained a numerical solution for the evolution 
of the two-point function. To observe the formation of correlated domains 
and topological defects it is necessary to identify these structures 
from the form of the two-point function.

The identification of topological defects from the
underlying quantum dynamics of the two-point function is a complicated
problem.  Intuitively, the existence of well-defined 
correlated domains of true vacuum at the 
completion of the phase transition is similar to the domains that form 
in a condensed matter system like a ferromagnet.  However, our system is described by a 
{\it quantum} field theory and the existence of a classical configuration of 
domains requires a quantum-to-classical transition, of which decoherence is an essential condition. 
In a decohered system we could, in principle, extract a positive-definite probability distribution of field
configurations, $P[\Phi]$, from the density matrix.
Classical defect solutions would then appear as particular
field configurations drawn from $P[\Phi]$ \cite{cooper:1997}.   

In this work our focus is {\it not} on the extraction of classical defects from the quantum system but 
instead on the long wavelength modes that determine the size of correlated domains and, 
therefore, the average defect separation.  The long wavelength modes interact
with an environment of short wavelength and thermal fluctuations that destroys quantum coherence among the long 
wavelength modes and results in a finite correlation length for the system \cite{hu:1993}.  While quantum fluctuations are
indeed important to
the description of phenomena within the scale of one domain, beyond this scale we may hope to treat the modes as classical. 
With
this abbreviated estimate of decoherence we show that the existence and size of correlated domains is indicated by 
an infrared peak in the power spectrum of the equal-time momentum-space two-point function. 

For a free field theory quenched into the unstable region by an 
instantaneous change in the sign of the square of the mass at $t=t_0$ it is 
possible to obtain an analytic expression for the equal-time momentum 
space two-point function \cite {boyanovsky:1993}.  After the quench, the momentum space structure function $k^2G(k,t)$ 
has a strong maximum at the value
\begin {equation}
k_{max} \sim \sqrt{\frac {m_f} {2t}}.
\end {equation}
The real space Fourier transform of this function 
(normalized to unity at $ | \vec{r}-\vec{r}'|=0$) is
\begin {equation}
G( | \vec{r} -\vec{r}'  |,t) \sim e^{-\frac {m_f( |
\vec{r}-\vec{r'}  |)^2} {8t}}.
\end {equation}
This form of the equal-time correlation function is common in condensed matter systems \cite{bray:1994}. 
From the equal-time real space correlation
function we can identify the correlation length
\begin{equation}
\xi(t) \sim \sqrt{ \frac {8t} {m_f}} \sim \frac {1} {k_{max}(t)}.
\end{equation}
In analogy with the free field theory, we assume the 
domain size is proportional to the size of the maximum of the momentum space structure function.

Another estimate of the domain size is motivated by the counting of defects in a 
classical condensed matter system.  While the identification of 
topological defect structures can be a complicated task, when 
defects are well-formed (so that the width of the defect is much smaller 
than the typical defect spacing), it is possible to identify zeros of the 
classical field configurations with topological defects.  In a model 
with a global $O(n)$ symmetry in $n$ spatial dimensions, 
and when the field probability distribution is Gaussian, a formula for the 
ensemble average density of field zeros  
was given by Halperin \cite{halperin:1981} 
and derived explicitly by Liu and  Mazenko \cite {liu:1992}:
\begin{eqnarray}
\label {eq-density}
\rho(t)= C_n\left (\frac {G''(0,t)} {G(0,t)}\right )^{n/2} \\
C_1=\frac{1}{\pi},\: C_2=\frac{1}{2\pi},\: C_3=\frac {1}{\pi^2}\nonumber
\end {eqnarray}
For a single scalar field in 3 spatial dimensions 
Eq.~(\ref{eq-density}) with $n=1$ is
valid as the ensemble-averaged density of zeros along a one-dimensional 
section of the field. The validity of the Gaussian approximation is further discussed
in \cite{calzetta:1999}.  A zero of the classical field configuration does not 
uniquely identify a topological defect.  Thermal 
fluctuations give a large number of zeros of the field configuration 
on small scales \cite{alexander:1993}.   The zeros of the field 
configuration which are not associated with 
different defects lead to an ultraviolet momentum divergence
in the expression for the zero density, Eq.~(\ref{eq-density}).  If we are to make 
physical sense of the zero formula, a coarse graining of the 
field configuration is needed.  To 
effect this coarse-graining in Eq.~(\ref{eq-density}) we impose a 
spatial momentum cut-off at the upper edge of the unstable momentum band. With this choice only the 
unstable modes will contribute to average density of defects, $\rho(t)$.  The length scale set by
the coarse-grained density of defects provides another estimate of the average domain size.

\begin{figure}
\begin{center}
\strut\psfig{figure=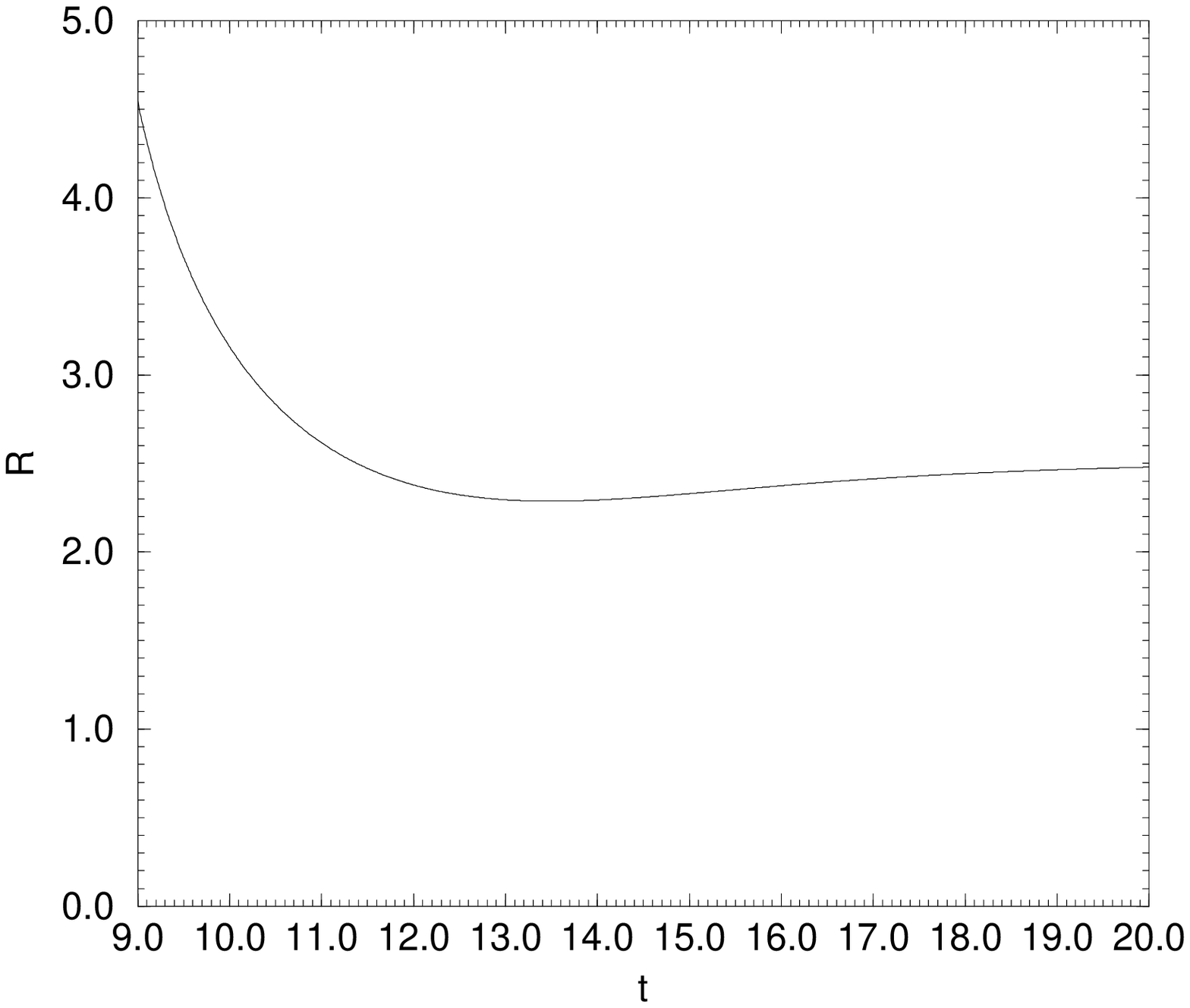,width=3.5in,angle=0}
\rmmcaption{Plot of the ratio $R(t)=\frac{k_{max}(t)}{\rho(t)}$ vs.~cosmological time t for quench parameter $\tau=10$
using the early-time model of Sec.~\ref{subsec:linear}}
\label{fig:singlescale}
\end{center}
\end{figure}

When the structure factor $k^2G(k,t)$ is very strongly peaked about $k_{max}$
the length scales set by $k_{max}(t)$ and $\rho(t)$ differ only by a numerical factor
of order unity. In this case we can approximate the unstable portion of the spectrum as
\begin {equation}
k^2G(k,t) \sim \delta (k-k_{max}).
\end {equation}
The zero density for a single scalar field in 3 dimensions is then
\begin {equation}
\rho = \frac {1} {\pi} \left( \frac {\int k^4 G(k,t)} {\int k^2 G(k,t)}\right)^{3/2} \sim (k_{max})^3.    
\end {equation}
Figure \ref{fig:singlescale} affirms this single-scale relation using the linear model introduced in Sec.~\ref{subsec:linear}.
Additional evidence for a one-scale model was given in 
\cite{antunes:1997} in the context of the dynamics of a classical field phase transition in $1+1$ dimensions.   
For very weak coupling not considered here and when a peak is no 
longer evident in $S(k,t)$, the one-scale approximation  will fail and it is possible in principle 
for the defect density as measured by $\rho(t)$ to depart significantly from
one per correlation volume \cite{karra:1997}.

\section {Critical scaling}
\label{sec:ch2critscal}
\begin{figure}
\begin{center}
\strut\psfig{figure=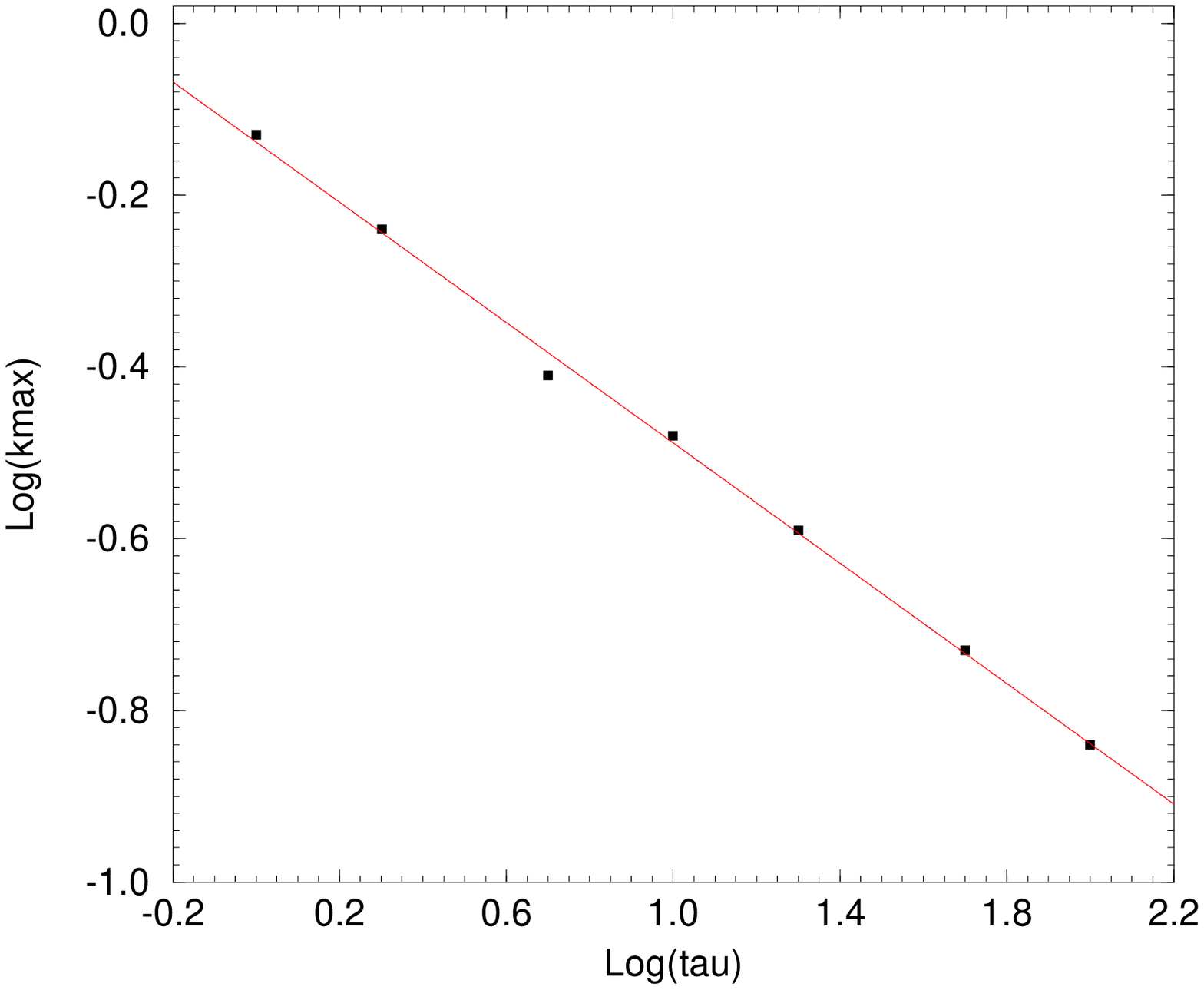,width=3.5in,angle=0}
\rmmcaption{Log-Log plot of $k_{max}$ vs.~$\tau$ in the case of
slow, underdamped quenches ($\tau \geq 1.0$) for domains formed early in the phase transition, at the time when the square
of the effective mass reaches a minimum value.}
\label{fig:eslowpowerlaw}
\end{center}
\end{figure}

\begin{figure} 
\begin{center}
\strut\psfig{figure=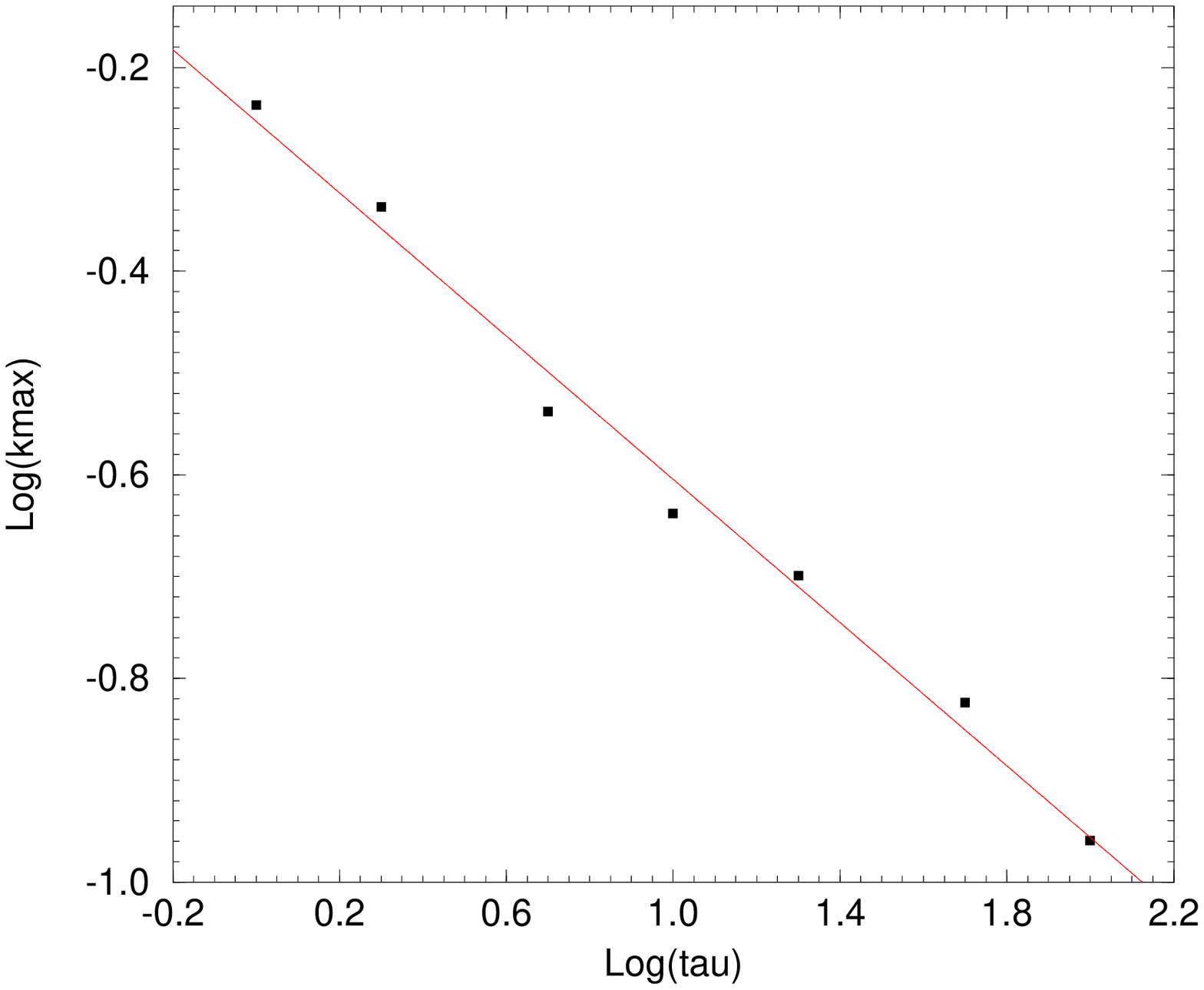,width=3.5in,angle=0}
\rmmcaption{Log-Log plot of $k_{max}$ vs.~$\tau$ in the case of slow,
underdamped quenches ($\tau \geq 1.0$) for
domains formed late in the phase transition, at the time when the square
of the effective mass reaches a local maximum value.}
\label{fig:lslowpowerlaw}
\end{center}
\end{figure}

\begin{figure} 
\begin{center}
\strut\psfig{figure=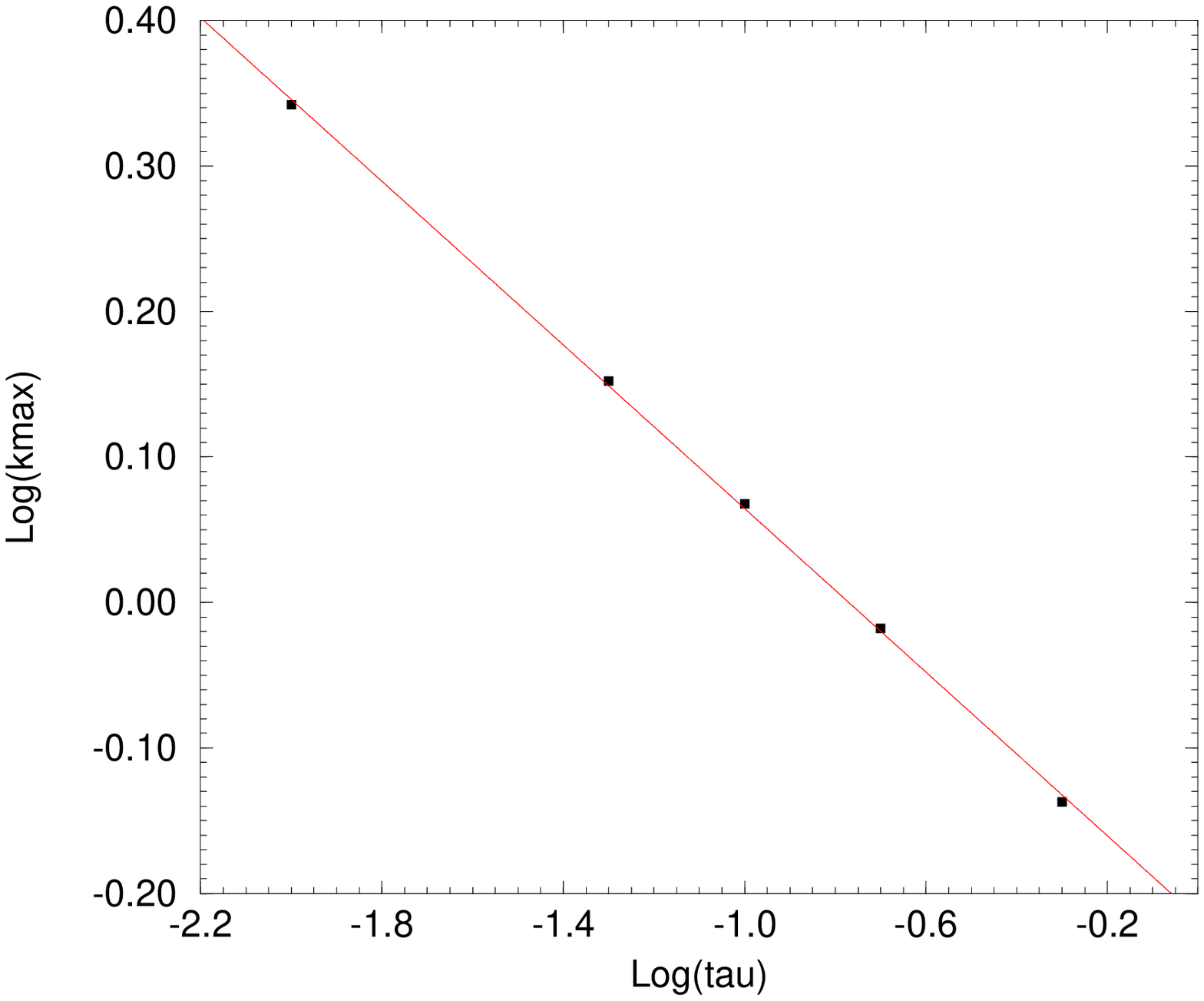,width=3.5in,angle=0}
\rmmcaption{Log-Log plot of $k_{max}$ vs.~$\tau$ in the case of fast,
overdamped quenches ($\tau <1.0$) for
domains formed late in the phase transition, at the time when the square
of the effective mass reaches a local maximum value.}
\label{fig:lfastpowerlaw}
\end{center}
\end{figure}

The evolution of the two-point function and the extraction of 
correlated domains allows a first-principles 
analysis of the mechanisms important for topological
defect formation.  In light of the freeze-out scenario it is 
interesting to compare the initial size of correlated domains for 
scale factors $a_\tau(t)$ with different
expansion rates $\tau$.  To quantify the dependence of the initial size of 
correlated domains on the quench rate $\tau$ of the phase transition, we 
compare the domain size for different values of the parameter $\tau$. As 
discussed above, the average domain size is proportional to the
maximum in the infrared portion of $S(k,t)$. 
We compare the maximum $k_{max}$ for different 
values of $\tau$ and at two distinct sets of times during the phase transition. 
The first set of domains is measured when the square of the effective 
mass reaches a minimum value.
This provides an early-time measure of the size of domains.  The results are 
shown in Fig.~\ref{fig:eslowpowerlaw}. Filled squares are measurements from the numerical simulations. The solid
line is a plot of the best-fit linear function to the data,
\begin{equation} 
\log_{10}(k_{max})=-0.35\log_{10}(\tau)-0.14.
\end{equation} 
The second set of domains is measured when the square of
the effective mass reaches a local maximum value after the phase transition.  
This provides a late-time measure 
of the size of domains.  The results are shown in Fig.~\ref{fig:lslowpowerlaw} and Fig.~\ref{fig:lfastpowerlaw}.  
Filled squares are measurements from the numerical simulations while solid
lines plot the best-fit linear function to the data.  In Fig.~\ref{fig:lslowpowerlaw} the best-fit power-law is
\begin{equation}
\log_{10}(k_{max})=-0.35\log_{10}(\tau)-0.26.
\end{equation}
while in Fig.~\ref{fig:lfastpowerlaw} the best fit is
\begin{equation}
\log_{10}(k_{max})=-0.28\log_{10}(\tau)-0.22.
\end{equation}
\noindent For slow quench rates ($\tau \geq 1.0$), at both early and late times, the dependence of $k_{max}$ on the 
quench rate $\tau$ is well approximated by a power law,
\begin{equation}
k_{max}(\tau) \sim \tau^{-0.35}.
\end{equation} 
For fast quench rates ($\tau<1.0$), the dependence of $k_{max}$ on the quench rate $\tau$ is well approximated by a power law,
\begin{equation}
k_{max}(\tau) \sim \tau^{-0.28}.
\end{equation}
The origin of these exponents and the difference between slow and fast quenches is discussed in the following
section.

\subsection {The freeze-out scenario}
In the Zurek scenario the frozen correlation length and therefore 
the initial size of correlated domains scale as a power law of the 
quench time $\tau$ as
\begin {equation}
\label{eq-powerlaw}
\xi_{freeze}  \sim {\tau}^{\frac {\nu} {1+\mu}}
\end {equation}
where $\mu$ and $\nu$ are the equilibrium critical exponents for 
the correlation length and relaxation time respectively. The correlation length and relaxation time are identified as,
respectively, the length and time scales that characterize the equilibrium
behavior of the propagator near the critical temperature \cite{chaikin:1995}.
Under the scaling hypothesis, the equal-time propagator is written
\begin {equation}
G_k=k^{-(2-\eta)}F[h(\epsilon)k].
\end {equation}
Here $\epsilon$ is the reduced temperature, Eq.~(\ref{eq-epsilon}), $\eta$ is a critical exponent (not to be confused
with conformal time) and $F$ is a dimensionless function.  The finite-temperature equilibrium correlation length is
\begin {equation}
\xi(\epsilon) = h(\epsilon) \sim \epsilon^{-\nu}.
\end {equation}
Similarly the two-time propagator is
\begin {equation}
G_{\omega, k} = \omega^\alpha F[h'(\epsilon) \omega,h(\epsilon)k]
\end {equation}
so that the relaxation time
\begin {equation}
\tau \sim  h' \sim \epsilon^{-\mu}.
\end {equation}
In the Hartree-Fock approximation the equilibrium propagator is
\begin {equation}
G_{\omega, k} \sim \frac {1} {\omega^2-k^2-m^2_{eff}} \coth(\frac {\sqrt{k^2+m_{eff}^2}} {2} \beta)
\end {equation}
where the effective mass is given by the equilibrium value Eq.~(\ref{eq-highTm}).  Near the critical point
\begin {equation}
m^2_{eff}(\epsilon)=A\epsilon,
\end {equation}
\noindent where A is a constant independent of temperature. The scaling
behavior of the propagator is
as $\frac {\omega} {\sqrt{\epsilon}}$ and $\frac {k} {\sqrt{\epsilon}}$ and
therefore the critical exponents for the theory are
\begin{equation}
\label{eq-qcritexp}
\mu=\nu=\frac {1} {2}.
\end{equation}
In the Zurek scenario we therefore expect the power law scaling,
\begin{equation}
k_{max}  \sim {\tau}^{-\frac {1} {3}}.
\end{equation}
Our observation of the power law exponent of $0.35$ in the numerical simulations 
for slow quenches, $\tau \geq 1.0$, (Fig.~\ref{fig:eslowpowerlaw} and Fig.~\ref{fig:lslowpowerlaw}) 
is consistent with this result.

The calculation of the equilibrium critical exponent for the relaxation time depends on the 
dynamics of the mode functions for the the low k modes
near the critical point and in particular whether the dynamics is overdamped or underdamped.   Near the critical point, 
$m^2_{eff} \approx 0$, and the mode function equation is approximately
\begin {equation}
\label {eq-critmode}
\left (\frac {d^2} {dt^2} + 3\frac {\dot{a}(t_c)} {a(t_c)} \frac {d} {dt}
+ \frac {k^2} {a^2(t_c)}
\right )f_k(t) = 0.
\end {equation}
The critical time $t_c$ can be estimated as the time when the equilibrium effective mass, Eq.~(\ref{eq-equilmass}),
goes to zero, 
\begin{equation}
t_c=\frac {2} {3} \tau,
\end{equation}
and the value of the scale factor at $t_c$ is
\begin {equation}
a(t_c)= \left[ \frac {t_c+\tau} {\tau} \right ] ^{\frac {1} {2}} 
\approx 1.3.
\end {equation}
The mode function equation near the critical point Eq.~(\ref{eq-critmode}) is that of a damped 
harmonic oscillator with natural frequency
$\omega_0=\frac {k} {a(t_c)}$ and damping constant $\Gamma$ where
\begin{equation}
\Gamma=3H(t_c) \approx \frac {1} {\tau}.
\end {equation}  
The modes for which $\omega_0 > \Gamma$ or 
\begin {equation}
\label{eq-dampcond}
k> \frac {1.3}{\tau}
\end {equation} 
are underdamped.  The range of wavenumbers $k$ for the modes that determine
the size of correlated domains also depends on $\tau$. Eq.~(\ref{eq-dampcond}) is 
a condition on $\tau$ such that when
\begin{equation}
\tau > \tau_*
\end {equation}
the modes responsible for domain formation are underdamped and $\tau_*$ is to be
determined.  For the underdamped case we expect  
$k_{max} \sim \tau^{-\frac {1} {3}}$.  Using $k_{max}$ as a representative wavevector of the modes
responsible for domains, Eq.~(\ref{eq-dampcond}) implies the underdamped condition
\begin {equation}
\tau_* \sim 1.
\end {equation}   
Slow quenches with $\tau \geq 1.0 \:$ are therefore underdamped. Fast quenches 
are overdamped and the dynamics of the mode functions 
will be dominated by the first time derivative.  In the overdamped case, the critical 
exponent, $\mu$, now assumes the non-relativistic value
\begin{equation}
\mu_{overdamped}=2\mu_{underdamped},
\end{equation}     
and
\begin{equation}
k_{max} \sim \tau^{-\frac{1}{4}}
\end{equation} 
This is consistent with the power law exponent of $0.28$ measured for 
overdamped, fast quenches ($\tau <1.0$) shown in Fig.~\ref{fig:lfastpowerlaw}.
In both the overdamped and underdamped cases, the prediction of the ``freeze-out'' proposal 
for the power law exponent of the scaling of the initial defect density with the quench rate 
appears to be in excellent agreement with the numerical simulations.

\subsection {Early-Time linear approximation}
\label{subsec:linear}
If the initial size of correlated domains and the initial density of 
topological defects are determined by processes occurring very near 
the critical point, as claimed in the freeze-out proposal, then the 
power law scaling of domains with the quench parameter $\tau$ observed 
in the simulation of the Hartree-Fock mode function Eq.~(\ref{eq-rmode}) 
will also appear in a linear approximation.  A linear amplitude approximation to the 
full mode function equation is valid for slow quenches and only near the onset of the 
spinodal instability, before the unstable modes have grown appreciably and 
back reaction is important. 

\begin{figure} 
\begin{center}
\strut\psfig{figure=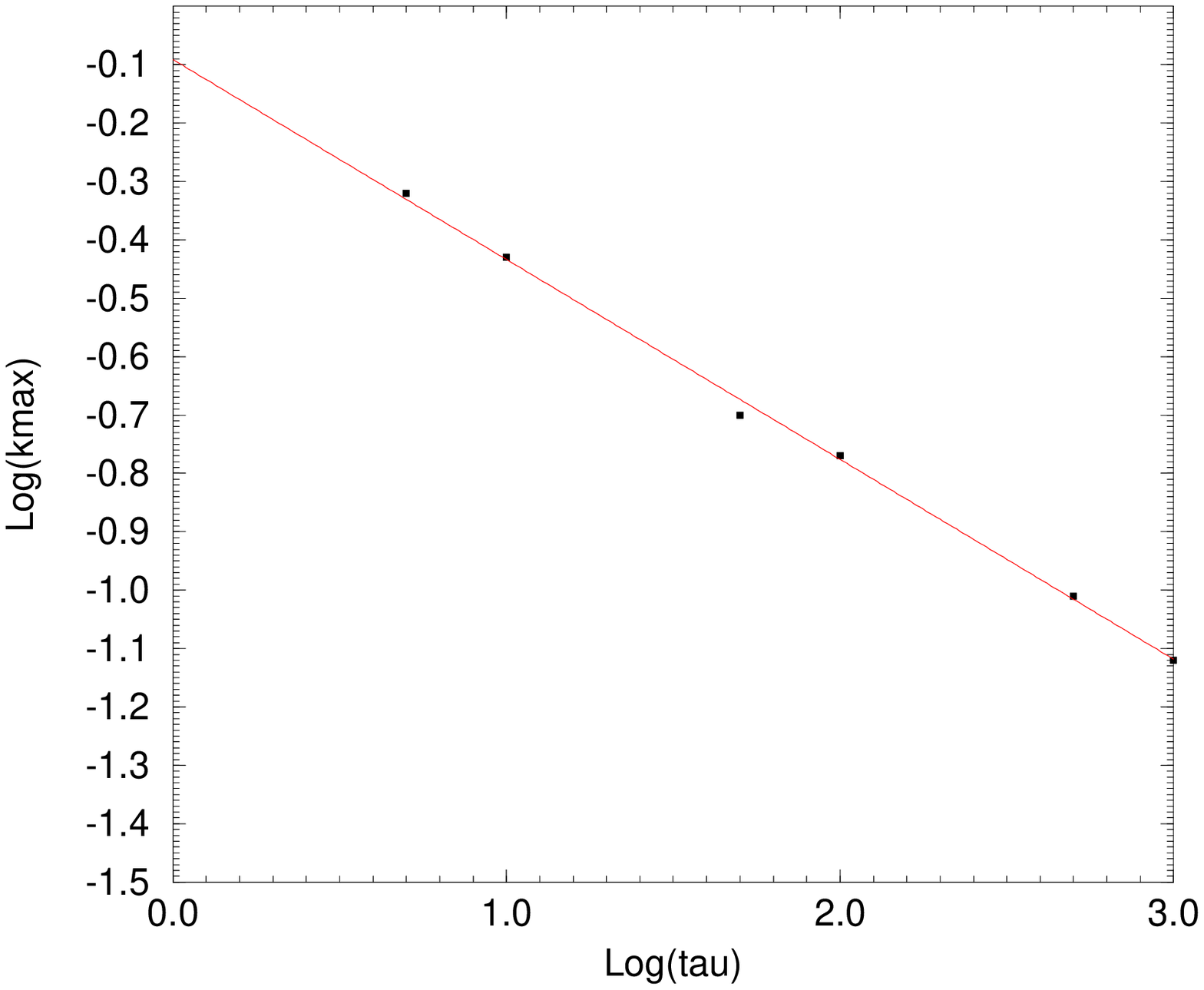,width=3.5in,angle=0}
\rmmcaption{Log-Log plot of $k_{max}$ vs.~$\tau$ in the analytic model, Eq.~(\ref{eq-amode}).  The maximum
$k_{max}$ was determined as soon as a peak was evident in the structure function $S(k,t)$. }
\label{fig:apowerlaw}
\end{center}
\end{figure}

We consider the linear equation
\begin {equation}
\label {eq-linear}
\left[ \frac{d^2}{dt^2}+3\frac{\dot a}a\frac d{dt}+\frac{K^2}{a^2} -m^2\right] F_k=0,
\end {equation}
\noindent where
\begin {equation}
K^2=k^2+k_0^2,
\end {equation}
and $k_0$ is related to the initial temperature,
\begin {equation}
k^2_0=\frac {\lambda} {24} T^2_0.
\end {equation}
In conformal time we have
\begin {equation}
\eta^2=4\tau \left( t+\tau \right),
\end {equation}
with conformal modes $f_k$ defined by
\begin {equation}
f_k\equiv F_k a.
\end {equation}
Equation (\ref{eq-linear}) may be rewritten in conformal time as
\begin {equation}
\left [ \frac {d^2} {d\eta^2} + K^2 -\frac {m^2\eta^2} {4\tau^2} \right ] f_k=0.
\end {equation}
This equation can be solved exactly in terms of parabolic 
cylinder functions.  However,
a simpler solution is obtained for times near the critical point.
We further approximate by expanding around the critical point,
\begin {eqnarray}
\left[ \frac{d^2}{d\eta ^2}-\frac{mK}\tau \left( \eta -\eta _k\right) \right] f_k=0, \\ 
\eta _k\equiv\frac{2K\tau } {m}.
\end {eqnarray}
The properly normalized solution with vacuum boundary conditions
before the instability is
\begin {equation}
\label {eq-amode}
f_k=i\sqrt{\frac{2x}{3\pi }}K_{1/3}\left[ -\frac {2} {3} \sqrt{\frac{mKx^3} {\tau} } \right], 
\end {equation}
where
\begin {equation}
x=\eta -\eta _k,
\end {equation}
and $K_{1/3}$ is a modified Bessel function.  
We now use the analytic solution Eq.~(\ref{eq-amode})
to examine the dependence of the position of the peak in the 
Fourier space structure factor $k^2G(k,t)$ on the quench parameter $\tau$.
The location of this peak $k_{max}(t)$ redshifts throughout 
the phase transition, moving towards lower momentum as the domains coarsen.
In order to compare domains formed in the linear model with those of the 
Hartree evolution we compare the position of the peaks at very early times when a peak
is first identifiable in $S(k,t)$.  Plots of $S(k,t)$ with the analytic modes were made with
{\it Mathematica}.  The results are 
shown in Fig.~\ref{fig:apowerlaw}.  Filled squares are 
measurements from plots of the analytic modes.  The solid line is a plot of the best fit linear function to the 
data, 
\begin{equation}
\log_{10}(k_{max})=-0.34\log_{10}(\tau)-0.09.
\end{equation}
The scaling of domains with the 
quench parameter $\tau$ with the same power law exponent as seen in the full 
numerical simulation for slow quenches is evident.

\section{Summary and future work}
\label{sec:ch2sum}
Using a two-loop truncation of the 2PI-CTP equations of motion for 
the two-point function of a quantum scalar field undergoing a phase 
transition in a 3+1 dimensional spatially flat, radiation-dominated FRW universe we have 
shown that the size of correlated domains, measured as the maximum of the
peak of the infrared part of the spectrum of $k^2 G(k,t)$, scales as a power 
of the quench rate $\tau$ as
\begin {equation} 
\xi_{domains} \sim \tau^{0.35},
\end {equation}
for slow, underdamped quenches ($\tau \geq 1.0$) and
\begin{equation}
\xi_{domains} \sim \tau^{0.28},
\end{equation}
for fast, overdamped quenches ($\tau<1.0$).  The observed power-law scaling of correlated domains is 
quantitatively consistent with the freeze-out hypothesis.  In both overdamped and underdamped cases, the value of 
the power-law exponent extracted from evolution of the two-point 
function is in good agreement with the value calculated in the freeze-out
scenario using the critical scaling exponents for a $\lambda \Phi^4$ 
theory in the Hartree-Fock approximation.

To further explore the behavior of the quantum system near the 
critical point we introduced an approximate linear model valid for slow quenches and short times 
after the onset of spinodal instability.  In this linear model, the size of 
correlated domains scales with the quench rate $\tau$ to the same 
power as observed in the underdamped numerical simulations and predicted by the freeze-out proposal.  
This provides analytical evidence for the freeze-out hypothesis. 

In order to place these results in physical context it is important to understand the validity of the 
two-loop truncation of the 2PI-CTP effective action.  
The truncation of the effective action to any finite loop order is an approximation.  
On some time scale we expect the contribution from higher loop terms to modify the two-loop dynamics.  
Unfortunately, the higher-loop equations are time-nonlocal and an exact determination of this time scale 
is not possible.  We provide instead an approximate analysis. 
The higher-loop terms in the 2PI-CTP effective action contain additional powers of the coupling constant $\lambda$.
In the perturbative regime and in the small-coupling limit these terms are suppressed.
This is the case early in the evolution, when the effective mass is positive.
However, the dynamics of the second-order phase transition is dominated by spinodal instabilities, 
the exponential growth of long-wavelength fluctuations visible as a peak in $k^2G(k,t)$. 
At times slightly beyond the onset of the phase transition, the effective mass is negative and 
long wavelength modes are growing.   At these early times the growing modes have yet 
to affect the dynamics and higher-loop terms remain negligible.  Eventually, the growth of 
fluctuations causes the effective mass to grow until they dominate at the spinodal point 
(when the effective mass reaches zero from below),
\begin{equation}
\lambda \int dk k^2G(k,t) \sim m^2.
\end{equation}
Since, for example, the three-loop piece of the effective action is
\begin{equation} 
\Gamma_{3-loop} \sim \lambda^2 G^4,
\end{equation}
higher loop terms are likely to be important beyond the spinodal point.  Whether these terms actually
alter the size of domains during the late stages of the phase transition is 
an open question. The two-loop truncation of the 2PI-CTP 
effective action, in the absence of a mean field, is also {\it exact} to leading order of a $1/N$ approximation.  
Although there are no defects in the leading-order Large-N model, domains can form in a symmetry-breaking
phase transition.  While the interpretation of our model as a large-N approximation provides 
mathematical rigor, it is not clear to what {\it physical} system the large-N approximation applies.  

In spite of these difficulties our approach remains, to date, the most comprehensive treatment of the dynamical formation
of domains in nonequilibrium quantum field theory.  In contrast to previous work we have allowed for the 
process of back reaction, which permits the quantum field to exit the region of spinodal instability.  
The power law exponent is the same whether it is measured at very early times with the analytic linear model 
(Fig.~\ref{fig:apowerlaw}) or, in the numerical simulations, at early times in the middle of 
the spinodal region at the minimum value of $m_{eff}^2$ 
(Fig.~\ref{fig:eslowpowerlaw}) or, at 
late times in the stable region at the maximum value of $m_{eff}^2$ (Fig.~\ref{fig:lslowpowerlaw}).  
This is consistent with the idea that the relevant processes for the growth 
of domains occur near the critical point.  However, back reaction is 
{\it essential} to the freezing in of the value of the power-law exponent and
to the freeze-out hypothesis. Although the power-law is 
accurately recorded in the early-time analytical 
model, if back reaction is ignored and the linear model is (incorrectly)
extrapolated to late times, the predicted scaling exponent is different
\begin {equation}
k_{max} \sim \tau^{-\frac {1} {2}}.
\end {equation}
The results derived from the numerical simulation of the mode function equation with back reaction 
and the analytical solution around the critical point are evidence supporting 
the first verification of the freeze-out scenario in 3 spatial dimensions 
and in a realistic system relevant to the early universe. It is at first 
surprising that arguments based on {\it classical equilibrium} critical 
scaling apply in the context of {\it dynamical} quantum field theory. 
A possible explanation rests with the high temperature initial conditions 
and the properties of classical $\lambda\Phi^4$ theory in FRW spacetime. 
Since the initial temperature is much higher than the initial effective mass
\begin {equation}
T_0 >> m_{eff}(0) 
\end {equation} 
the field is approximately conformally invariant.  
In an FRW universe, a conformally invariant field initially in thermal
equilibrium will remain in equilibrium at a redshifted temperature
\begin {equation}
\label {eq-temp}
T(t) = \frac {T_0} {a(t)}.
\end {equation}
A well-known example is the redshifting of the blackbody
spectrum of the cosmic microwave background radiation as the universe expands.
To illustrate the approximate conformal invariance in our model we 
compare the dynamical effective mass from the simulations 
with the effective mass of a theory with the equilibrium redshifted 
temperature of Eq.~(\ref{eq-temp})  
\begin{equation}
\label{eq-equilmass}
M^2_{eq}=-m^2+\frac {\lambda} {24} \frac { T^2_0} {a^2(t)}.
\end{equation}
\begin{figure} 
\begin{center}
\strut\psfig{figure=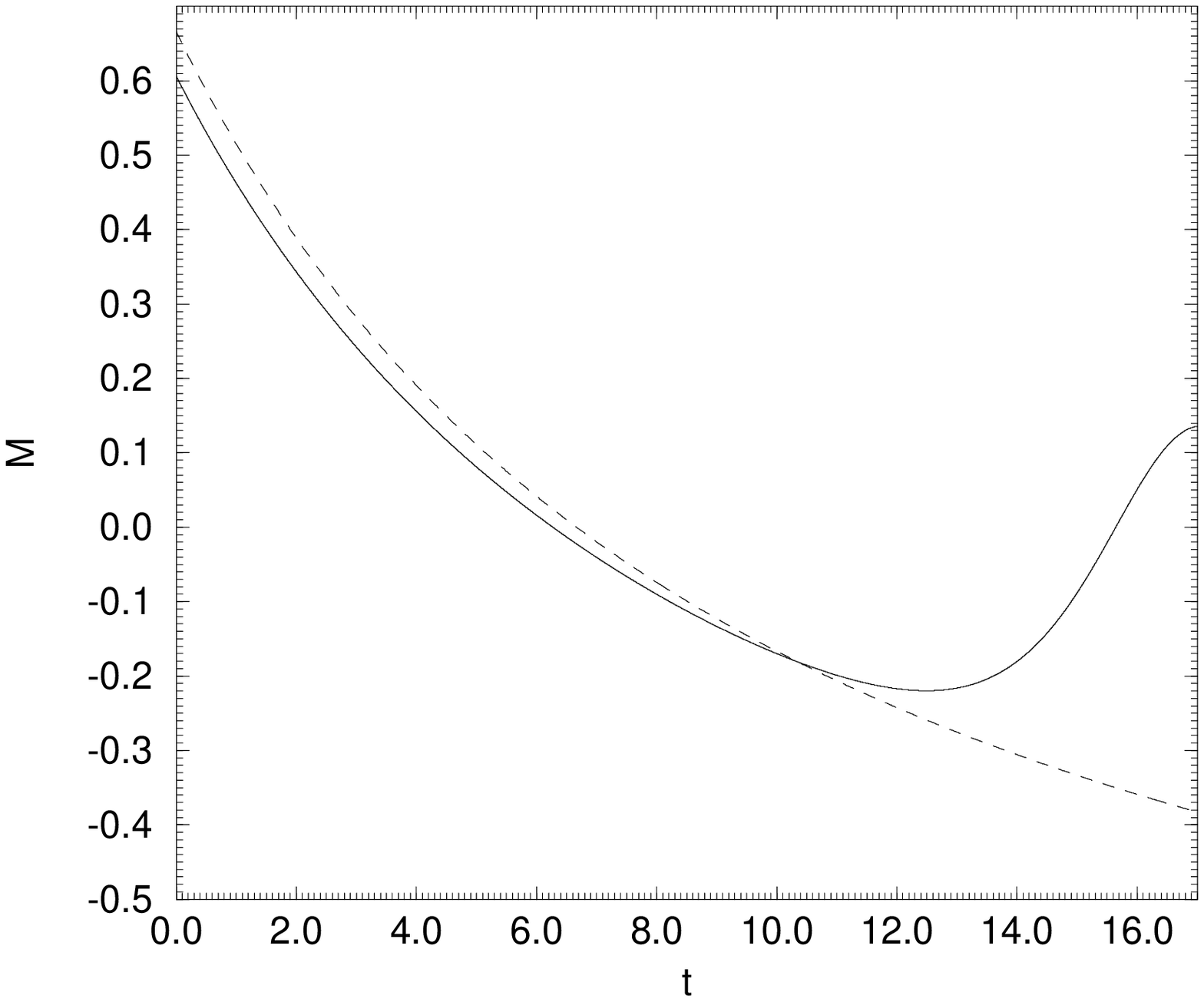,width=3.5in,angle=0}
\rmmcaption{Plot of the square of the effective mass $M=m^2_{eff}$ 
vs. cosmological time $t$ in the equilibrium 
case (dashed line) where
$m^2_{eff}=-m^2+\frac {\lambda} {24} \frac { T^2_0} {a^2(t)}$ and the in 
nonequilibrium case (solid line) where $m^2_{eff}$ is determined by
the full numerical simulations.}
\label{fig:equilibrium}
\end{center}
\end{figure}
The results are shown in Fig.~\ref{fig:equilibrium}.
The slight difference in the equilibrium and nonequilibrium curves at 
the beginning of the evolution is due to the difference between the general Hartree-Fock effective mass given by 
Eq.~(\ref{eq-minit}) and 
the high temperature, small $\lambda$ limit given by Eq.~(\ref{eq-highTm}). Only after the onset of the
spinodal instability do the equilibrium and dynamical effective 
mass differ significantly.  The approximate conformal invariance of 
the theory means that the finite-temperature effective potential and the 
critical behavior derived from it are approximately valid until 
times near the onset of the spinodal instability.  The high temperature 
initial conditions also offer a possible explanation for the observed 
classical behavior of the system.  At such high temperature, thermal fluctuations 
dominate over quantum vacuum fluctuations. 

The agreement in the value of the power law exponent between the 
microscopic evolution equations of the quantum field theory and
phenomenological critical scaling supports the contention that 
quantum critical systems in the early universe share, in certain circumstances, many of the 
properties of their classical counterparts.  It would be very interesting to further 
explore these connections.  To do so, however, it is necessary to go beyond the (relatively) strong 
coupling and high temperature conditions used in this research.

The results derived here represent a preliminary step in a 
first-principles approach to the calculation of the 
topological defect density immediately following the completion of a 
second-order phase transition in the early Universe.
Our work uses a microscopic quantum field theory and incorporates 
realistic initial conditions and a physical quench mechanism.  However, the 
domain wall defects formed in this model are inconsistent with 
cosmological observations. It is not difficult in principle to apply 
the methods used in this paper to more cosmologically realistic theories
such as cosmic string models.  The analytical and numerical evidence for the 
power-law scaling of the domains is expected to hold in a more realistic model.

The formalism of the 2PI-CTP effective action and related techniques can be used 
to probe more general questions related to the physical aspects of quantum 
critical dynamics \cite{calzetta:1995}. To observe and isolate quantum processes it is
necessary to relax the assumption of high temperature initial conditions, 
allowing quantum vacuum fluctuations to dominate over thermal fluctuations.  Also important are
the detailed properties of the system-bath interaction.  A system-bath interaction such as proposed in 
\cite{calzetta:1989} could address such issues as critical slowing down and the role of dissipation and noise.  
Work is in progress along these lines.

\chapter{Classical critical dynamics: Formation and evolution of topological textures}
\section{Introduction}
In Chapter 3 we used the techniques of nonequilibrium quantum field theory to examine
the dynamics of a phase transition in the early universe.  Unfortunately, the approximation methods
necessary to obtain even a numerically soluble quantum dynamical system preclude a detailed 
look at topological defects themselves. In this chapter, that in part represents work published in
\cite{stephens:2000}, we examine the phase transitions of a 
classical field theory in which topological defects emerge as localized, stable solutions.  
Since the {\it full} dynamics of a classical field can be simulated numerically, this approach
offers another window into the nonperturbative processes important for the formation and interaction
of topological defects. Specifically,
we study the formation of topological textures in a classical O(3) scalar field theory
in $2+1$ dimensions.  Systems containing topological textures are of particular interest as 
they provide an arena in which to explore myriad dynamical issues ranging from nonequilibrium 
phase transitions to the coarsening dynamics of quantum field theory. 

Topological textures form in $d$ spatial dimensions when the vacuum manifold $M$ is such that $\Pi_d(M)$ is nontrivial.
Unlike other topological defects, textured fields never leave the vacuum manifold but instead are ``knots'' of
gradient energy, given relative stability by their topological winding around the vacuum.  Like cosmic strings, 
global textures have been studied extensively as candidates for generators of large-scale 
structure formation \cite{turok:1989,spergel:1991}.  In condensed matter, topological textures are studied in a 
wide variety of (mostly two-dimensional) systems, from superconductors \cite{knighavko:1999} and quantum hall 
ferromagnets \cite{travesset:1998} to superfluid $^3He$ \cite{solomaa:1987}.

\subsubsection{Nonequilibrium phase transitions and topological textures}
The classical dynamics of textured systems provides an ideal study for nonequilibrium phase transitions.
The richness of texture interactions involves multiple dynamical length scales. 
Unlike other topological defects, textures do not have a fixed size and the evolution of a textured system 
is enriched both by the interactions between textures and by the changing 
scale of the texture itself.   In $2+1$ dimensions it is observed that the phase-ordering of a system containing 
topological textures is characterized at late times by at least 
{\it three} length scales: the (average) texture size, 
texture-texture separation and texture-antitexture separation \cite{zapotocky:1995,rutenberg:1995a}.  In 
distinction, the coarsening dynamics of systems containing
domain walls, vortices or monopoles are effectively single-scale
and described, at late-times, only by the average defect separation.  In a nonequilibrium phase transition
the multiple independent length scales of textured systems are determined by the phase transition dynamics.  
However, the arguments of the Kibble-Zurek mechanism are ambiguous when applied to textured systems.
Which, if any, of the multiple length scales observed in a texture distribution should we equate 
with the single frozen correlation length?   Although the dynamics of a textured system can be
qualitatively different, the details of the formation of topological textures in a 
nonequilibrium phase transition are mostly unexplored (see however \cite{dziarmaga:1998}).

\subsubsection{Topological textures and the large-N expansion}
A detailed understanding of the field dynamics of a classical
textured system can also be used to better understand the large-N expansion, a common tool
in the study of nonequilibrium quantum field theory (see e.g. \cite{boyanovsky:1999} and references therein). 
The large-N expansion truncates both the dynamic
and quantum content of the full quantum field theory by replacing the physical
system under study with one containing a very large number of scaler fields.  In doing so, the dynamics of
the quantum system are dominated by the dynamics of Goldstone modes and the density matrix is 
truncated to Gaussian form \cite{cooper:1997}.  However, real quantum systems contain only a small 
number of fields and are described by density matrices that may be far from Gaussian.  It is therefore important
to understand to what extent the large-N approximation reflects the physical processes of a real system.
  
Phase ordering in $d$ spatial dimensions of $N\geq d+1$ scalar fields provides 
a particular example of a useful theoretical laboratory in which to analyze the dynamical and quantum
content of the large-N approximation method.  If $N=d+1$, the dynamics of phase ordering is
the dynamics of interacting topological textures.  If $N>d+1$, no 
topological defects are formed.  In either case, phase ordering is driven by massless Goldstone dynamics.
However, the Goldstone dynamics of large-N can be radically different from a real system.  As we discuss
in the next section, the coarsening of a $2+1$ dimensional system containing topological textures is observed 
to depend on three separate length scales which grow with different powers of time.  A large-N analysis of the
same system misses the contribution of topological textures and 
predicts only one simple scaling length $L_{large-N}\sim t^{\frac {1}{2}}$. 

To understand how the large-N
approximation fails, the full classical equations of motion 
can be studied numerically.  For example, assume the dynamics of the 
coarsening system is described by an N-dimensional order parameter $\vec{\Phi}$ with fully 
dissipative time-dependent Landau-Ginzburg equation (TDGL),
\begin{equation}
\frac {\partial \Phi(x,t)} {\partial t}=-\Gamma \frac{\delta F[\Phi]} {\delta \Phi},
\end{equation}
with free energy
\begin{equation}
F[\vec{\Phi}]=\int d^dx[\frac{\lambda}{4}(|\vec{\Phi}|^2-m^2)^2].
\end{equation}
The dynamical content of the large-N approximation is revealed  
as this system is solved for larger and larger values of N.  In the formal limit of
infinite N, both the classical and quantum systems can be solved exactly (albeit numerically). 
In the infinite-N case, the comparison of classical and quantum ordering dynamics yields insight 
into the true quantum content of the large-N approximation.  The focus in this current 
work is on the formation of topological textures in a nonequilibrium phase transition of a classical field theory.  
We hope to return to the role of textures in the large-N approximation and nonequilibrium quantum field theory
in future reports.
\subsubsection{Organization}
This chapter is organized as follows. Section \ref{sec:tttsd} reviews the properties of topological textures in $2+1$ 
dimensions and the scaling violations observed in textured systems when quenched from a disordered phase.   
Section \ref{sec:ptncd} describes two models of a nonequilibrium phase transition, a dissipative quench and an external
quench. In addition we provide the numerical techniques used in the evolution and
the length scales used to characterize the texture distribution. 
Section \ref{sec:ucd} contains the main results of this chapter: an explanation of the quench rate 
dependence of the average texture separation and the average texture width observed near the end of 
the external quench phase transition.  We show that the Kibble-Zurek mechanism is recovered at early times but 
that by the end of the phase transition, formation dynamics and phase ordering dynamics are intrinsically linked.  
We use these results to argue that textured systems carry an imprint of the freeze-out 
correlation length to late times and suggest the possibility of novel, 
late-time measurements of dynamical critical phenomena.  Concluding remarks are made in Sec.~\ref{sec:ch3sum}.

\section{Topological textures in two spatial dimensions}
\label{sec:tttsd}

The properties of systems containing topological textures differ in many ways from those containing
other topological defects.  To illustrate these differences, consider an O(3) invariant model with classical action,
\begin{equation}
\label{eq-action}
S=\int d^3x [\partial_{\mu}\vec{\Phi}\cdot\partial^{\mu}\vec{\Phi}+\frac{\lambda}{4}(|\vec{\Phi}|^2-m^2)^2],
\end{equation}
and resulting equations of motion
\begin{equation}
\label{eq-fieldeq1}
\ddot{\Phi}_i-\bigtriangledown^2\Phi_i+\lambda\Phi_i(\vec{\Phi}^2-m^2)=0.
\end{equation}
\noindent For the classical dynamics that is of interest here, the value of the quartic self-coupling $\lambda$ and
the mass parameter $m^2$ may be scaled to unity with an appropriate scaling of the field and spacetime units,
\begin{equation}
\Phi_i \rightarrow \Phi_i m^2, \quad t \rightarrow t \sqrt{\lambda} m, \quad \vec{x} \rightarrow \vec{x} \sqrt{\lambda} m.
\end{equation}
\noindent The vacuum manifold is characterized by $\vec{\Phi}\cdot \vec{\Phi}=1$, topologically a
2-sphere. When the field is constrained to the vacuum, the action is that of a nonlinear sigma model (NLSM). 
For asymptotically uniform fields, a vacuum field configuration is a map 
\begin{equation}
\vec{\Phi}: S^2_{real \thinspace  space} \rightarrow S^2_{field \thinspace space}, 
\end{equation}
\noindent divided into distinct topological sectors by winding number $n$ \cite{belavin:1975}.  In two
 spatial dimensions these
windings are topological textures, topologically stable and static solutions with finite energy. For example, 
the following field configuration, a texture of negative unit winding, wraps around the vacuum 2-sphere exactly once, 
\begin{equation}
\label{eq-texture}
\Phi_1=\frac{4ar\sin(\phi)}{r^3+4a^2},\quad \Phi_2=\frac{4ar\cos(\phi)}{r^2+4a^2},\quad  \Phi_3=\frac{r^2-4a^2}{r^2+4a^2}.
\end{equation}
\noindent At the origin, $\vec{\Phi}$ points in the $-\Phi_3$ direction while for $r\rightarrow \infty$ it points in 
the opposite direction, $\Phi_3$.  In between, at $r=2a$, the field points radially outward like a hedgehog.  
The parameter $a$ characterizes the size of the texture and, in contrast to other defects such as strings
or monopoles, is independent of the parameters of the 
action Eq.~(\ref{eq-action}).  The NLSM also possesses an exactly conserved 
topological charge $Q$, which is expressed as the integral over a  
topological charge density $\rho$,
\begin{eqnarray}
\rho=\frac{1}{4\pi}\vec{\Phi}\cdot (\partial_x\vec{\Phi}\wedge \partial_y\vec{\Phi})\\
Q=\int d^2x \rho(x,y)
\end{eqnarray}
\noindent For the single winding texture, Eq.~(\ref{eq-texture}), the topological charge 
density has a particularly simple form
\begin{equation}
\rho=-\frac{1}{\pi}\frac{4a^2}{(r^2+4a^2)^2},
\end{equation}
 with total topological charge $Q=-1$. The energy $E=\frac{1}{2} 
\int d^2x (\vec{\partial}\Phi_i \cdot \vec{\partial}\Phi_i)$
of the single winding configuration Eq.~(\ref{eq-texture}) is simply $E=8\pi$, independent of the size $a$ of the
texture.  In accordance with Derricks theorem \cite{derrick:1964}, the energy of a texture configuration in higher dimensions 
scales as a positive power of its size, demonstrating their instability to collapse.  In lower dimensions textures are 
unstable to expansion.  The stability of a texture configuration in any dimension can be altered by 
adding higher derivative terms to action, as was done in the Skyrme model of
nucleons \cite{skyrme:1961}.  

Although the energy of a single isolated topological texture is independent of its size, systems with multiple
textures are not static.  Textured systems {\it order} under the constraint of conserved topological charge Q.
However, the details of texture interactions are not well understood.  
It is observed that textures and antitextures can annihilate with each other and that more isolated textures can
decay by unwinding \cite{zapotocky:1995,rutenberg:1995a}.  The details of the unwinding process 
depend on the model under consideration.  For the action Eq.~(\ref{eq-action}) 
textures can decay by pulling the field off the
vacuum manifold and unwinding.  This is analogous to isolated textures in higher dimensions which are unstable 
to shrinking and unwind when their gradient energy is large enough to move the fields off the vacuum manifold.  
In a NLSM where the fields must remain on the vacuum manifold, textures can unwind through
higher-derivative terms present in the lattice discretization.    

\subsubsection{Topological textures and the violation of dynamical scaling}

When a disordered system is quenched into the ordered phase (e.g. by the removal of thermal fluctuations),
the approach to a new equilibrium and associated long-range order is through a sequence of nonequilibrium states.
A simple example is the Ising ferromagnet in 3 spatial dimension.  Above the Curie temperature, spins are
randomly oriented and there is no net magnetization.  A snapshot immeadiately following the quench of this system 
to zero temperature shows a system containing many domains in which the spins are coherently aligned.  
The net magnetization remains zero because the domains are randomly oriented with respect to each other. 
However, the true $T=0$ ground state of this system consists of only one
domain and no domain walls.  The system approaches this equilibrium through the dynamic process of 
coarsening, during which small domains shrink and large domains grow.  The process of coarsening has been studied 
extensively in both condensed matter systems and in the early universe (for reviews see \cite{bray:1994} and 
\cite{hindmarsh:1995}).  
The scaling hypothesis arose from these studies and states that, at late times, and in a quench from a 
disordered state, the domain distribution is statistically characterized by a {\it single} dynamical scale $L(t)$ 
which grows in time.  For simple dissipative systems, the approach to ($T=0$) equilibrium  is 
governed by model-A time-dependent Landau-Ginzburg (TDGL) dynamics for a nonconserved order parameter 
\begin{equation}
\label{eq-tdgl}
\frac {\partial \Phi(x,t)} {\partial t}=-\Gamma \frac{\delta F[\Phi]} {\delta \Phi}.
\end{equation}
It has been shown both theoretically and experimentally that the average domain size $L(t)$
grows with time as
\begin{equation}
\label{eq-scaling}
L(t)=(t\Gamma)^{\frac {1}{2}},
\end{equation}
which is also the power-law predicted by a dimensional analysis of Eq.~(\ref{eq-tdgl}).
Not all systems coarsen in such a simple way and there is currently no general theoretical framework 
in which to explain why some systems scale while others do not.  In particular, systems containing 
topological textures in one and two spatial dimensions are strong exceptions to the 
scaling hypothesis  \cite{zapotocky:1995,rutenberg:1995a,rutenberg:1995b}.  
In a $2+1$ dimensional O(3) model with dissipative
dynamics and an instantaneous quench from the disordered phase, late-time coarsening of 
topological textures is described by at least {\it three} different length scales. 
These three length scales are the average defect-defect separation $L_{sep}$, the average defect width $L_w$, and the
average texture-antitexture separation $L_{tat}$.  The scaling hypothesis is violated since
at late times these three length scales are observed to grow with different powers of time
\begin{eqnarray}
L_{sep}(t)&=&\xi_0^{\frac{1}{3}}t^{\frac {1}{3}}, \\
L_w(t)&=&\xi_0^{\frac{2}{3}} t^{\frac {1}{6}}, \\
L_{tat}(t)&=& t^{\frac {1}{2}},
\end{eqnarray}
\noindent where $\xi_0$ is the correlation length in the disordered phase.  It is remarkable that length
scales describing the late-time ordering of textures retain a dependence on the
initial correlation length.  In distinction, systems that obey the 
scaling hypothesis erase any memory of their initial state and at late times
retain only a {\it dynamical} scale (for example Eq.~(\ref{eq-scaling}) in the case of an Ising ferromagnet). 
The scaling of systems that contain topological defects such as vortices implies that 
the vortex distribution at late times is independent of the dynamical details of the phase transition.  
For textured systems, it is {\it not} possible to have such a clean separation between the dynamics of 
the phase transition and the late-time texture distribution.
In the context of a nonequilibrium phase transition we argue that $\xi_0$ is determined {\it not} from
initial conditions but from the Kibble-Zurek mechanism, thus connecting early-time critical dynamics and 
late-time coarsening.  In Sec.~\ref{sec:ucd} we discus how this connection may provide late-time 
probes of nonequilibrium critical dynamics.

\section {Phase transitions and nonequilibrium classical dynamics}
\label{sec:ptncd}
Classical interacting field theories appear as useful physical descriptions in diverse systems ranging from
hydrodynamics and condensed matter to general relativity and the early universe.  
Since the dynamics of a classical field can 
often be simulated numerically, classical approaches offer a window into the nonperturbative dynamics of the
decohered sector of quantum field theory.  In this section we
describe two models of the classical dynamics of a nonequilibrium phase transition, a dissipative quench and an external
quench. It is useful, however, to first discuss the nature of a classical field in thermal equilibrium.

A field contains an infinite number of degrees of freedom.  For a field
confined in a cavity, these excitations are the (countably) infinite number of standing waves, satisfying
the cavity boundary conditions.  If the field is maintained in equilibrium (for example by keeping the walls at
constant temperature), each degree of freedom holds a finite average energy ($E\sim T$ for a free field) 
given by the equipartition theorem. A classical field in a finite cavity in thermal equilibrium thus contains an 
infinite amount of energy!   The energy density is also divergent. For black body electromagnetic radiation 
in a 3-dimensional cavity the spectral energy density $\rho_T(\nu)$ is given by the Rayleigh-Jeans formula,
\begin{equation}
\rho_T(\nu)d\nu \sim T \nu^2  d\nu.
\end{equation}
The divergence of the spectral energy density at high frequency is known as the 
ultraviolet catastrophe and its resolution led Planck to the foundations of quantum theory.

In light of the ultraviolet catastrophe, it what sense does equilibrium exist for a classical field theory?
If the classical theory is only a long-wavelength approximation to a quantum field theory, then a correct {\it quantum} 
accounting of the short-wavelength modes will suffice to render equilibrium meaningful. 
However, the classical Landau-Ginzburg field theories of interest here are a only 
continuum approximation to a fundamentally discrete condensed matter system.  This discrete nature provides a natural 
cutoff in the frequencies of field oscillations, rendering
the spectral energy density and total energy finite.  In principle, the observables computed from a thermal classical field
theory with a cutoff depend upon the value of cutoff.  In practice, for certain situations and {\it especially in the
vicinity of a second-order phase transition}, low frequency modes are remarkably independent of the behavior of the high 
frequency modes, and therefore insensitive to the cutoff.  In this case, observables computed from long-wavelength modes
are universal.    

\subsection{Dissipative quench}
\label{subsec:dquench}
As in Chapter 3, to study the dynamics of a phase transition, we are faced with the challenge of implementing a 
realistic quench.  One possibility is to allow energy dissipation of the field to occur
solely through a phenomenological dissipative term $\eta \dot{\phi}$ in the equations of motion.  The field 
evolves with equations of motion  
\begin{equation}
\label{eq-dqfieldeq}
\ddot{\Phi}_i-\bigtriangledown^2\Phi_i+\Phi_i(\vec{\Phi}^2-1)+\eta \dot{\phi}_i=0
\end{equation}
and initial conditions are chosen so as to correspond to a high-temperature symmetry restored phase.  
The initial conditions are implemented either through a Langevin equation or simply by assigning 
random values to $\vec{\phi}$ and $\frac{\partial \vec{\phi}}{\partial t}$.  The advantage of this approach is that the 
quench proceeds naturally.  Energy is lost through dissipation but the dynamics of the fields are otherwise
unaffected.  In particular, and in distinction to other approaches in the literature \cite{laguna:1997,yates:1998}, 
no {\it a priori} form of the quench rate or the effective mass is assumed.  

\begin{figure}
\begin{center}
\strut\psfig{figure=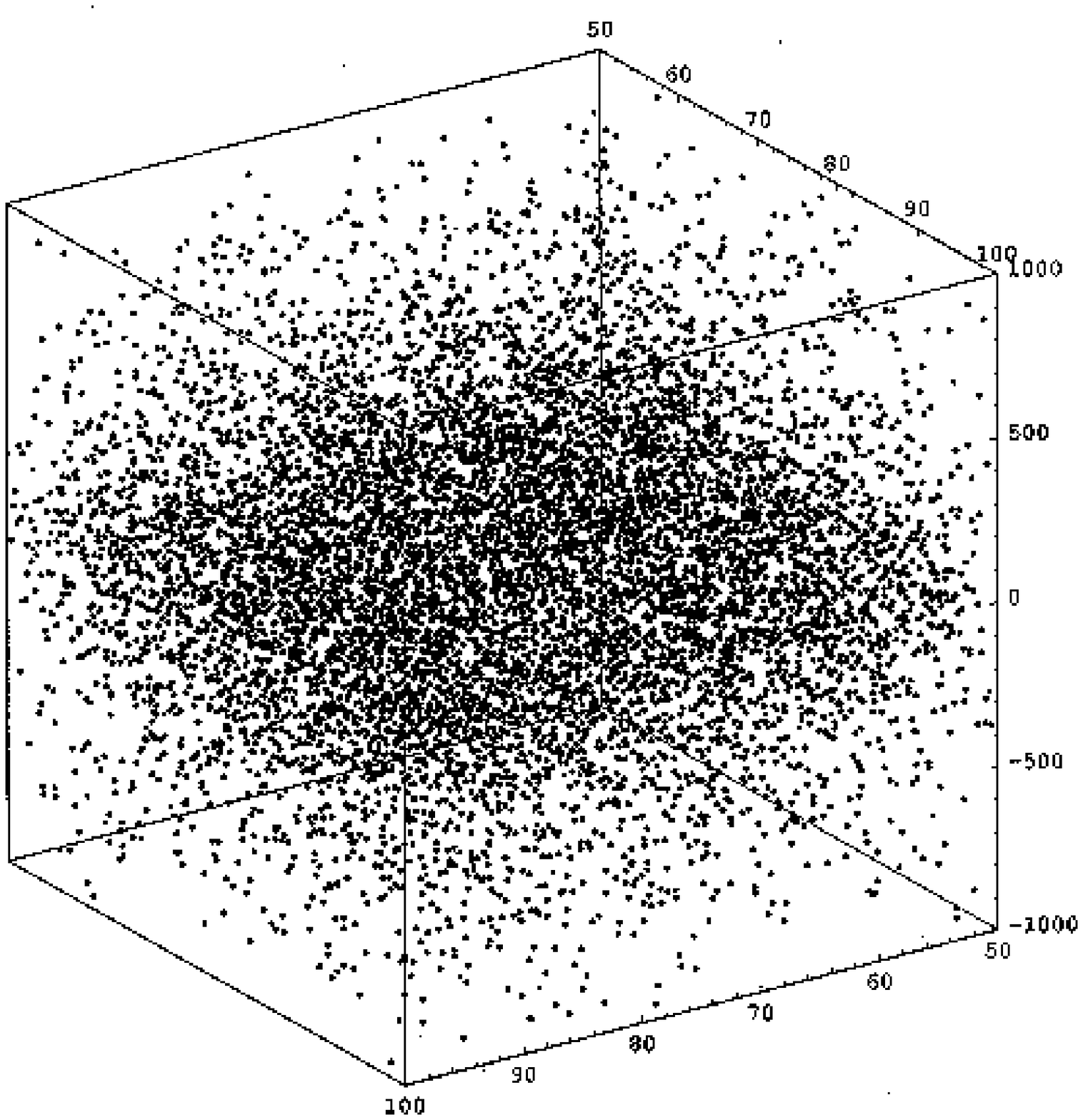,width=3.0in,angle=0}
\rmmcaption{Plot of the topological charge density for a small section of the lattice
at a time near the onset of the quench, $t=0.0$.} 
\label{fig:dq0.0}
\end{center}
\end{figure}

\begin{figure}
\begin{center}
\strut\psfig{figure=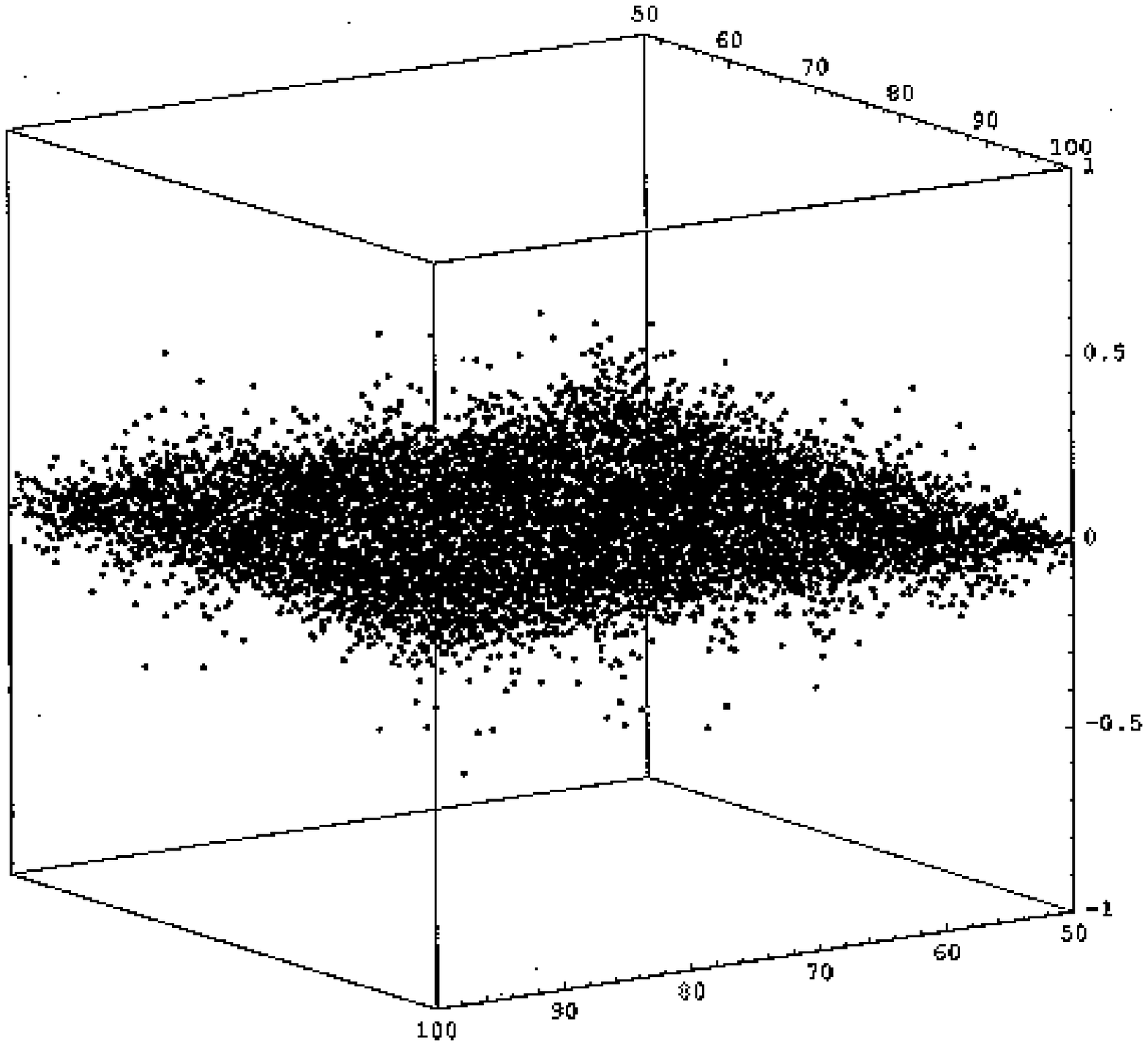,width=3.0in,angle=0}
\rmmcaption{Plot of the topological charge density at time $t=9.0$.}
\end{center}
\label{fig:dq9.0}
\end{figure}

\begin{figure}
\begin{center}
\strut\psfig{figure=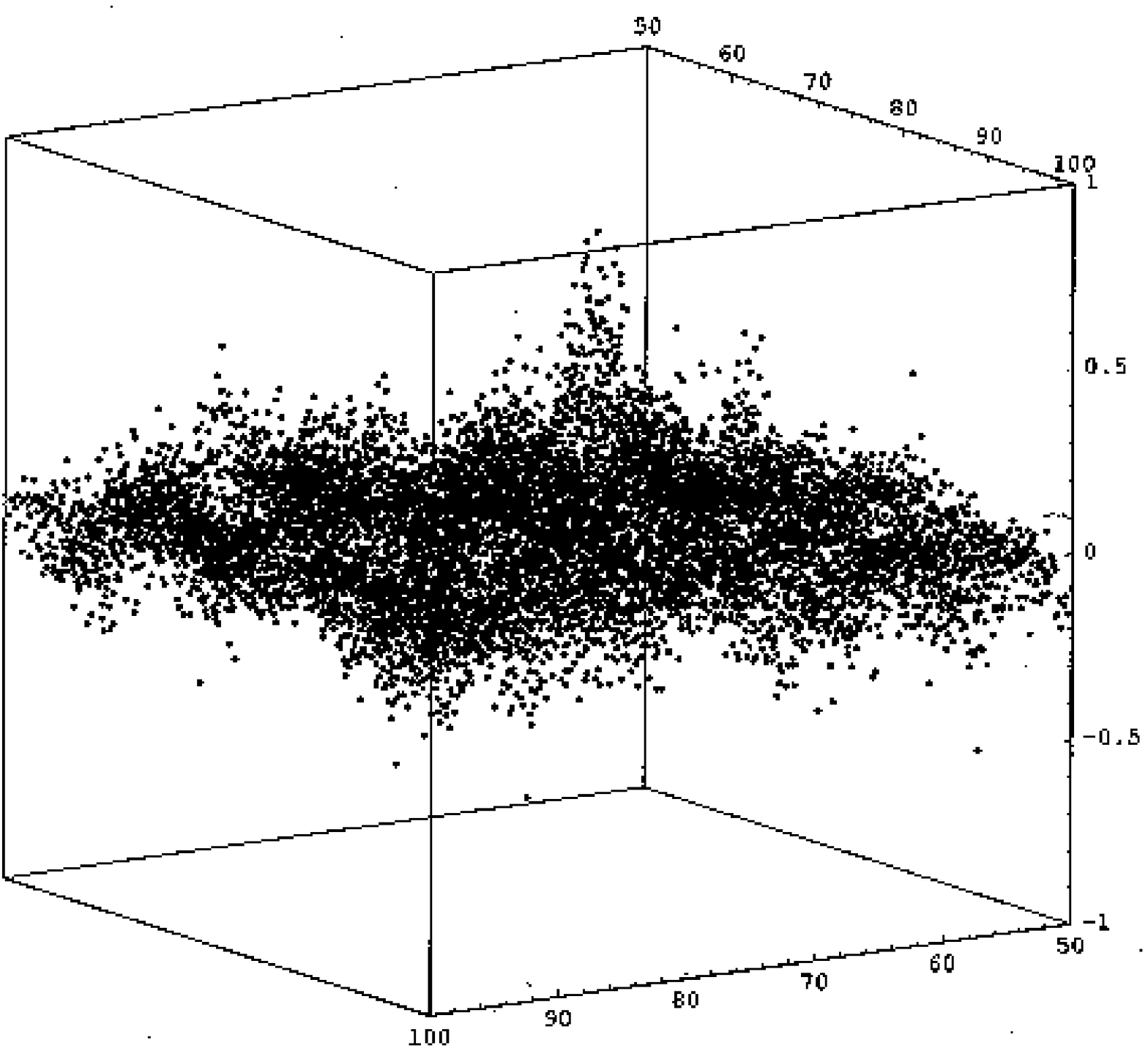,width=3.0in,angle=0}
\rmmcaption{Plot of the topological charge density at time $t=10.5$.} 
\label{fig:dq10.5}
\end{center}
\end{figure}

\begin{figure}
\begin{center}
\strut\psfig{figure=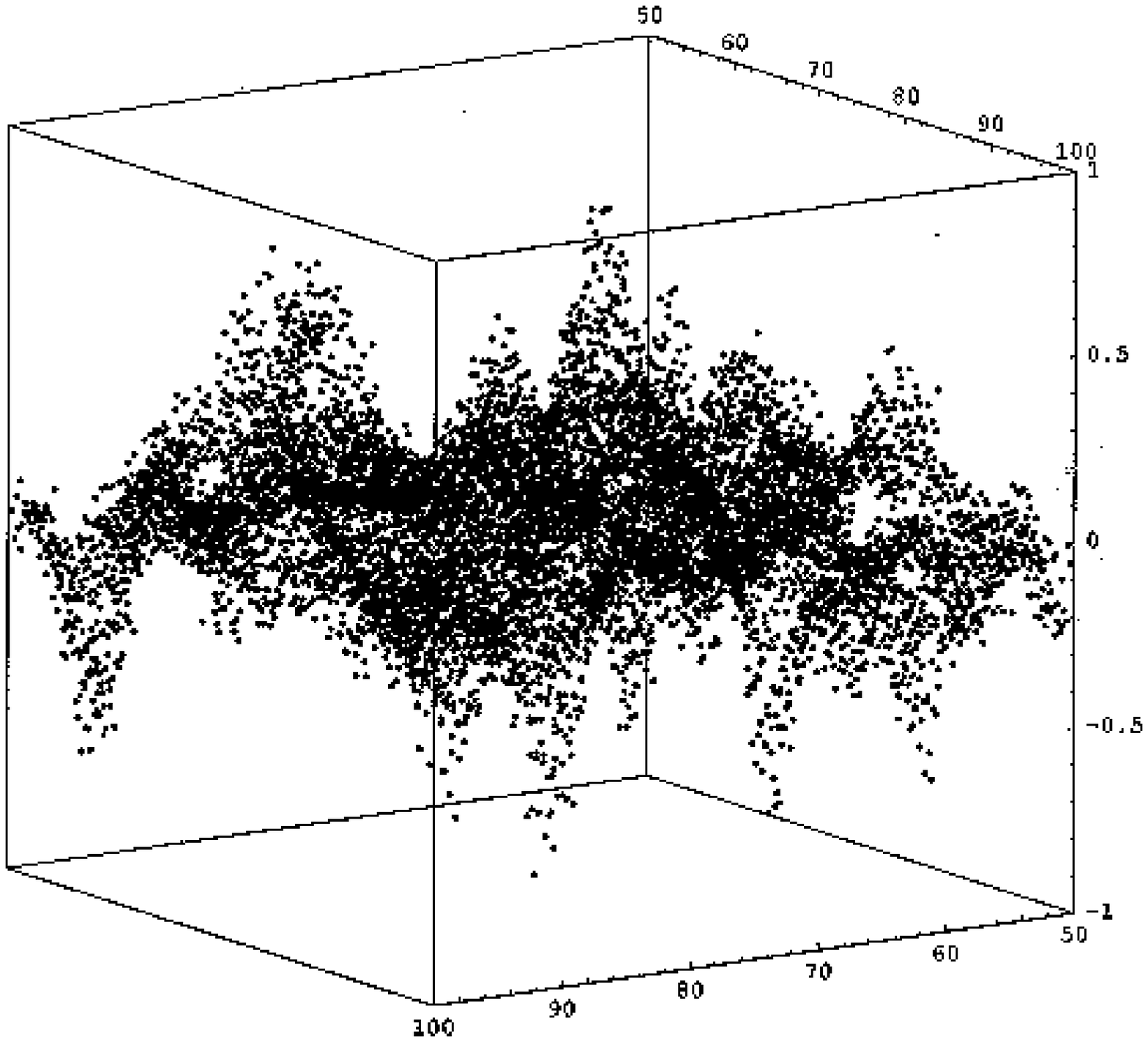,width=3.0in,angle=0}
\rmmcaption{Plot of the topological charge density at time $t=12.0$.} 
\end{center}
\label{fig:12.0}
\end{figure}

\begin{figure}
\begin{center}
\strut\psfig{figure=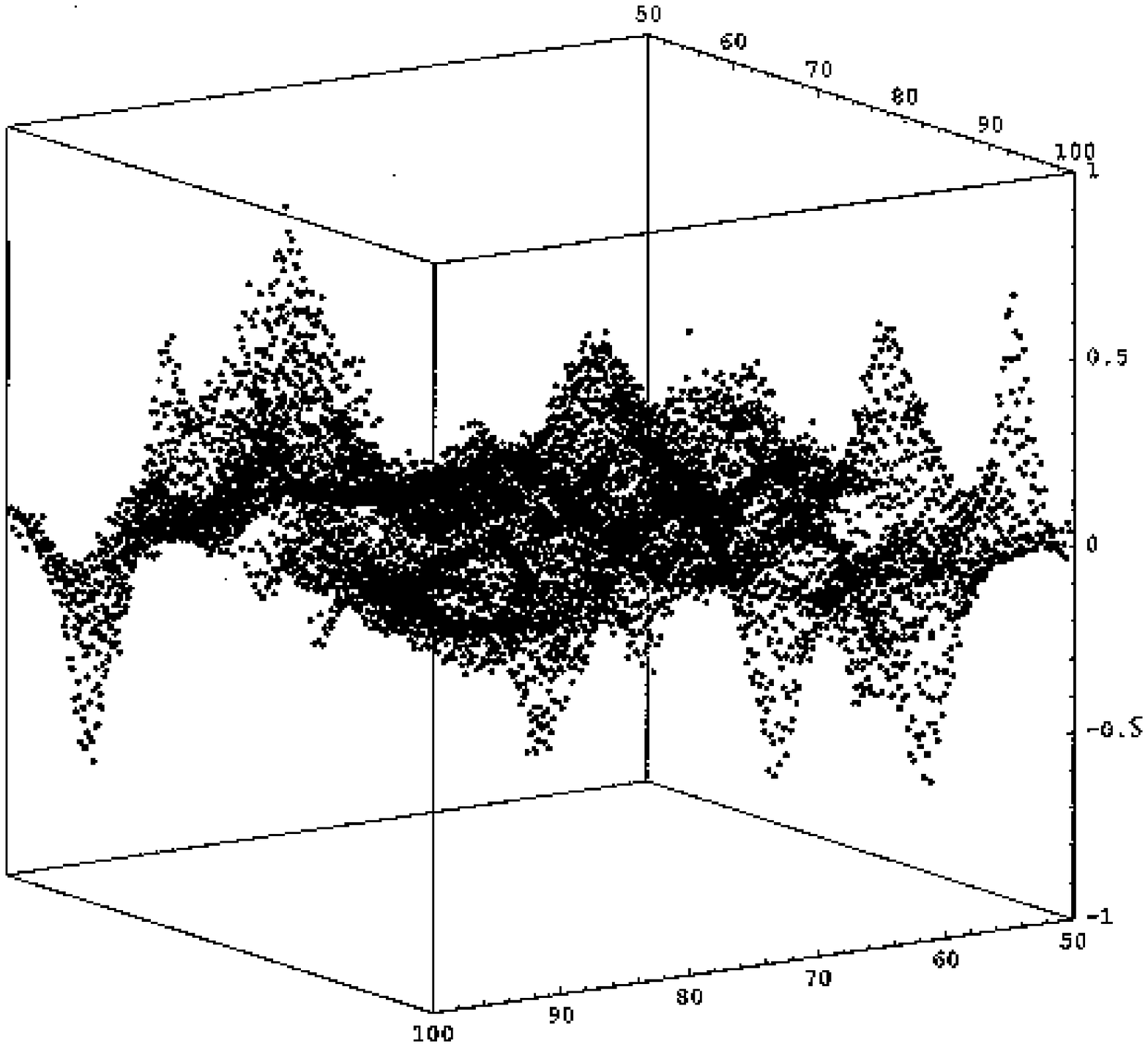,width=3.0in,angle=0}
\rmmcaption{Plot of the topological charge density at time $t=13.5.$} 
\label{fig:dq13.5}
\end{center}
\end{figure}

\begin{figure}
\begin{center}
\strut\psfig{figure=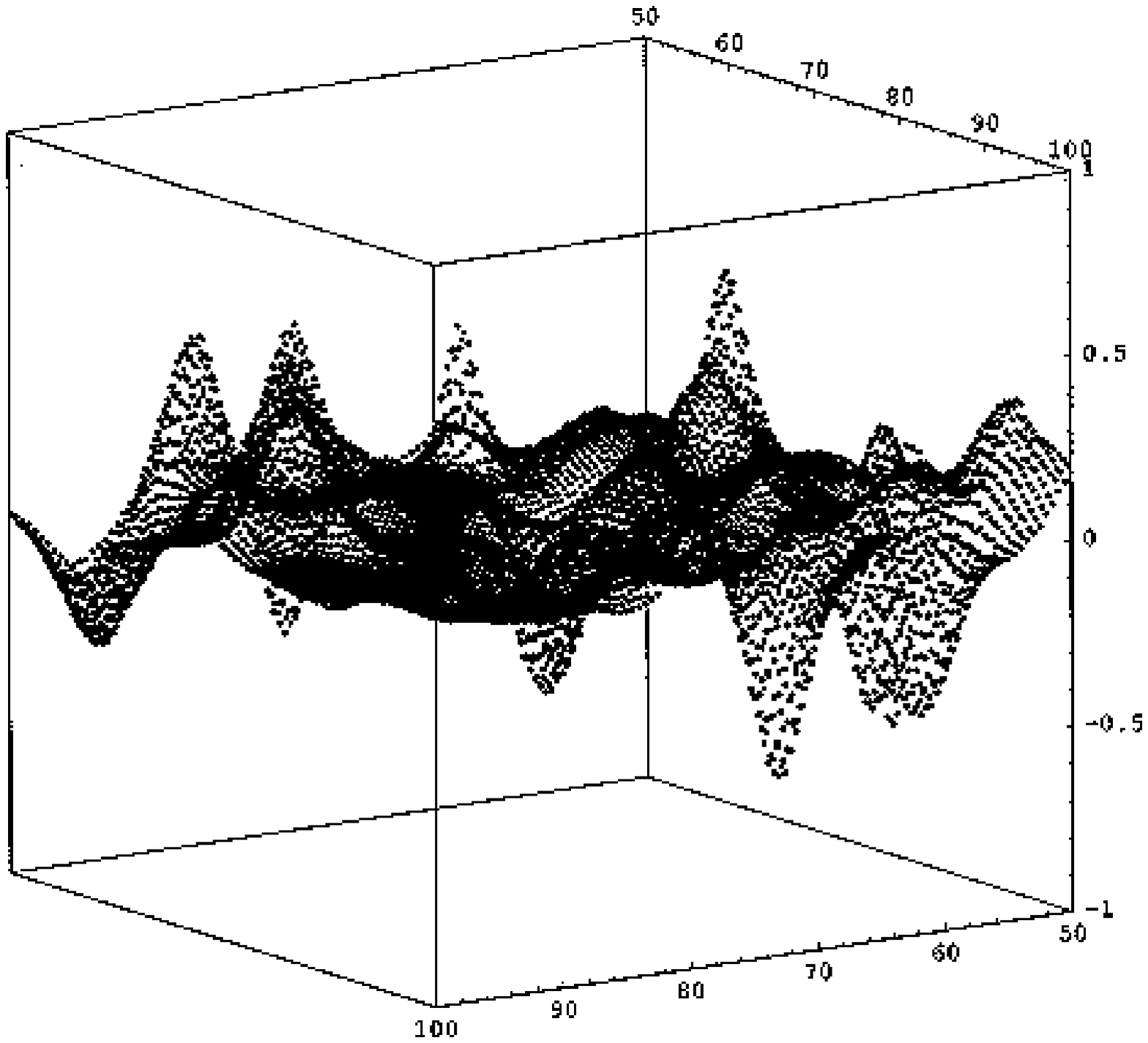,width=3.0in,angle=0}
\rmmcaption{Plot of the topological charge density at time $t=17.5$.} 
\label{fig:dq17.5}
\end{center}
\end{figure}

Using the numerical techniques explained in Sec.~\ref{subsec:equench}, details of the dynamics of the dissipative quench 
phase transition and
the distribution of topological textures were extracted from a numerical solution of Eq.~(\ref{eq-dqfieldeq}).
A series of snapshots of the topological charge density for the above dissipative quench model with $\eta=0.8$ is shown in
Figs.~\ref{fig:dq0.0}-\ref{fig:dq17.5}.  Horizontal axes label location
on the spatial lattice while the vertical axis labels the value of the topological charge density.  

\begin{figure}
\begin{center}
\strut\psfig{figure=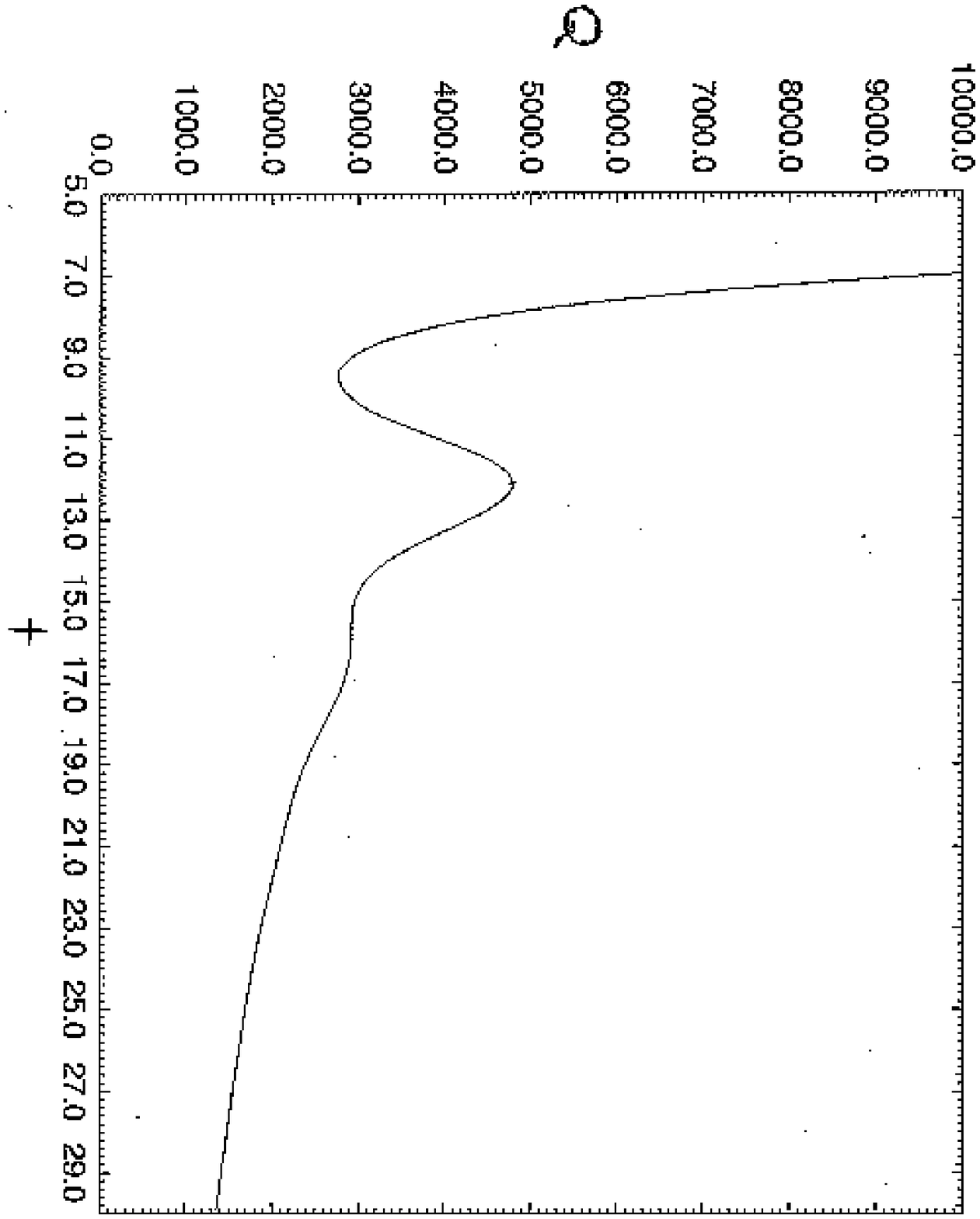,width=4.0in,angle=90.3}
\rmmcaption{Plot of the total number of textures $Q$ vs. time $t$ for a quench with $\eta=0.7$} 
\label{fig:dqQt}
\end{center}

\end{figure}

\begin{figure}
\begin{center}
\strut\psfig{figure=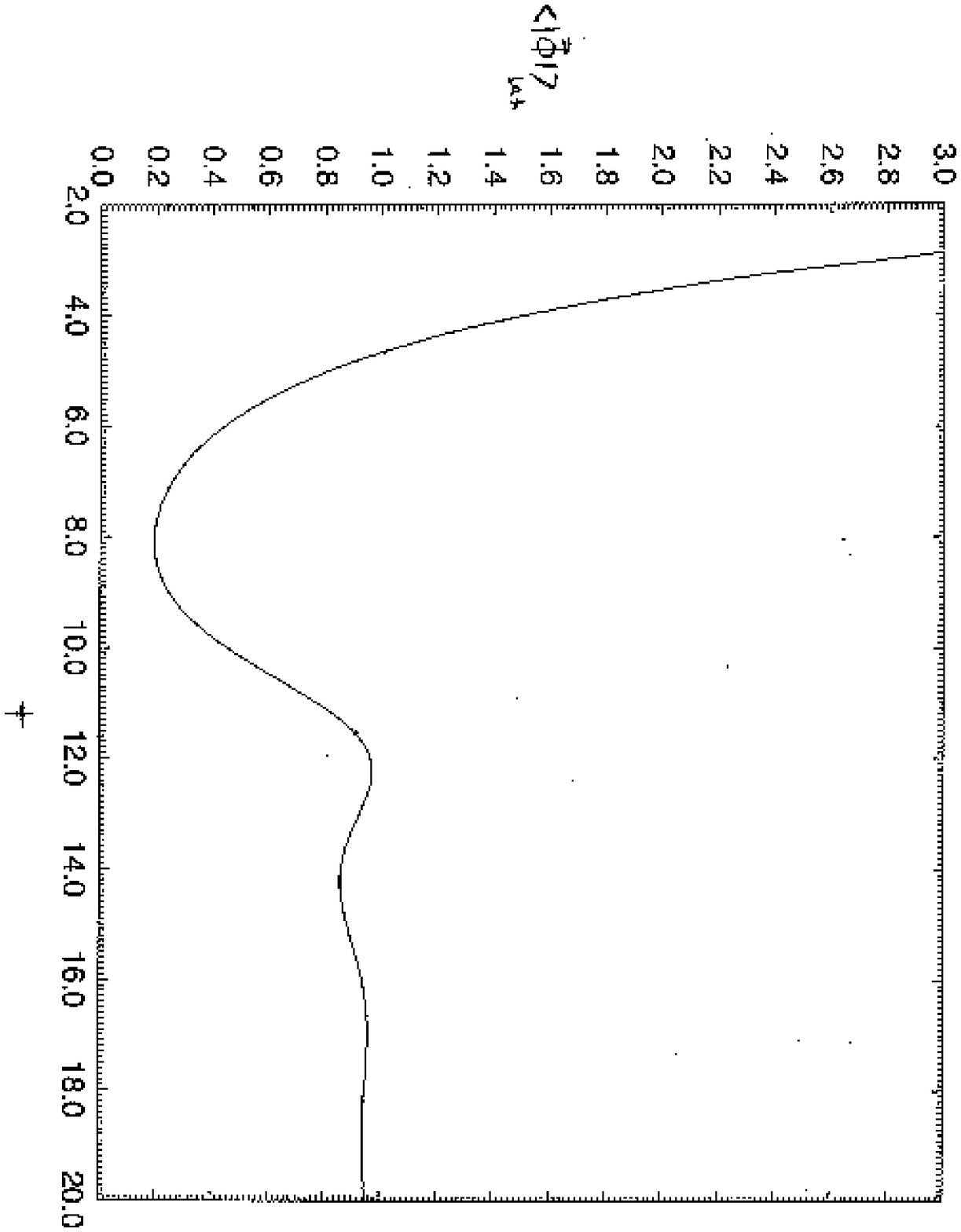,width=4.0in,angle=90.4}
\rmmcaption{Plot of the order parameter $\langle |\vec{\Phi} | \rangle_{lat}$ vs. time $t$ 
for a quench with $\eta=0.7$} 
\label{fig:dqPhit}
\end{center}
\end{figure}

Though the $2+1$ dimensional theory defined by Eq.~(\ref{eq-dqfieldeq}) has no finite-temperature equilibrium phase 
transition, \cite{mermin:1966} there can still be an {\it effective} dynamical
transition.  As energy is dissipated, the high temperature initial state dominated by short wavelength thermal 
fluctuations eventually evolves into a state dominated by long wavelength topological textures (well-defined peaks
in the topological charge density).  Textures become 
quasi-stable excitations when the effective temperature drops substantially below the texture energy of $8\pi$. 
The dynamics of the phase transition is also evident in the time evolution of the total number of textures $Q$ and
of the order parameter, $\phi=\langle |\vec{\Phi}| \rangle$, where brackets denote an average over the lattice.
In Figs.~\ref{fig:dqQt} and \ref{fig:dqPhit} we exhibit the total number of textures and the order parameter
for a quench with $\eta=0.7$.  In the early stages of the phase transition both $Q$ and $\phi$ decrease exponentially
as dissipation dampens the fields.  However, when enough energy is removed, the potential term in Eq.~(\ref{eq-dqfieldeq})
becomes important and the system evolves to the stable ground state, $\phi=1.0$.  Texture formation occurs primarily
in the interval from $t=9.0$ to $t=12.0$.


\begin{figure}
\begin{center}
\strut\psfig{figure=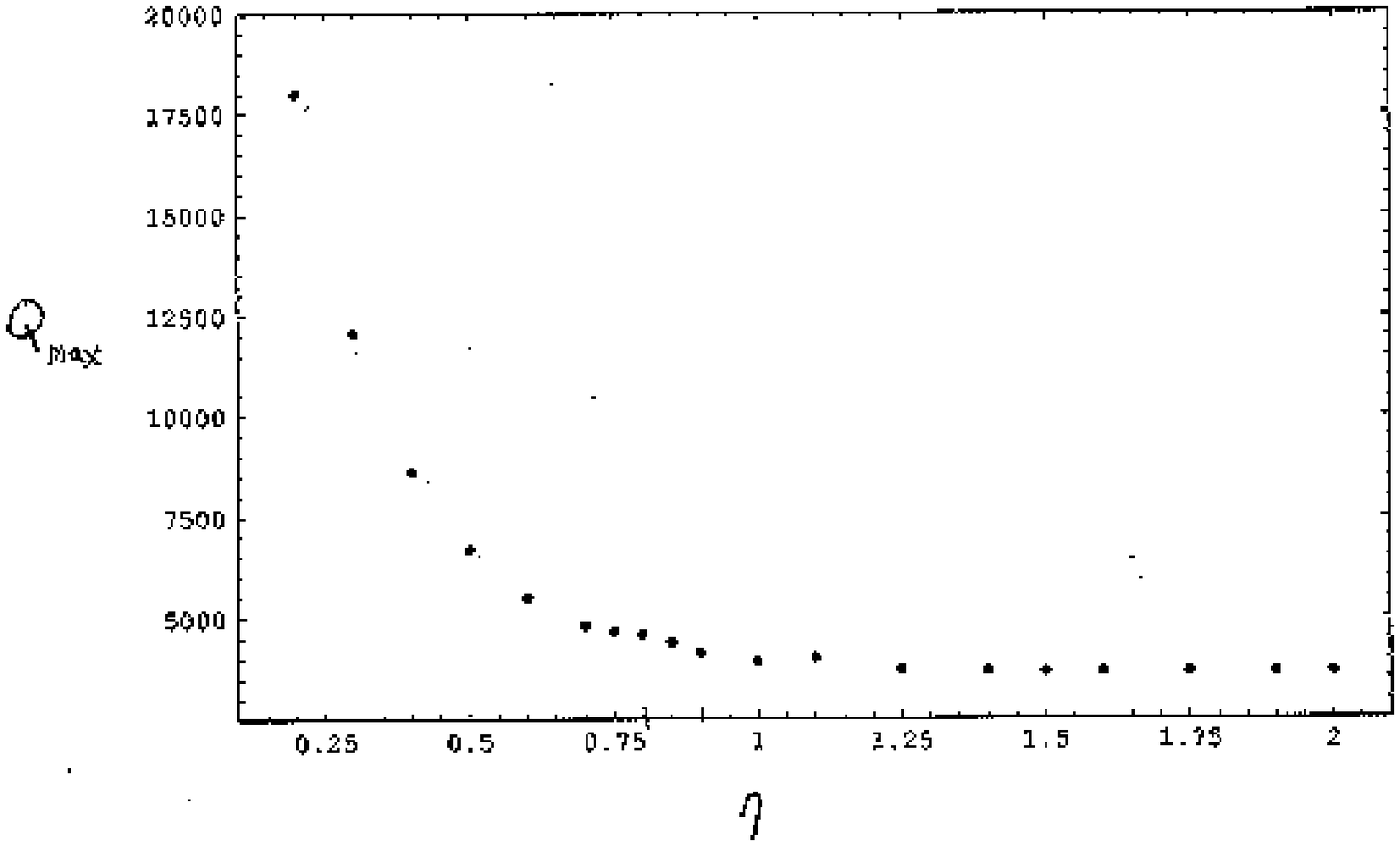,width=5.0in,angle=-0.45}
\rmmcaption{Plot of the total number of textures $Q_{max}$ vs. dissipation constant $\eta$} 
\label{fig:dqQeta}
\end{center}
\end{figure}

The dissipative quench model has one free parameter, the dissipation constant $\eta$.  In Fig. \ref{fig:dqQeta}
we show the number of textures measured at the local maximum $Q_{max}$ for different values of $\eta$.  Larger $\eta$
results in a faster quench.  Surprisingly, it appears that, contrary to the Kibble-Zurek mechanism, faster 
quenches lead to a smaller number of topological defects.  Unfortunately, we have yet to develop a convincing
explanation for this behavior.  It is possible that since this system has no equilibrium phase structure
the arguments of the Kibble-Zurek mechanism simply do not apply.  It is also possible that the identification
of textures through the topological charge density is inadequate: we may mistake thermal fluctuations for stable
defects.  It was our original intent to study the Kibble-Zurek mechanism in the context of the multiple length scales
of a textured system.  In the following section, motivated by similar models in the literature \cite{laguna:1997,yates:1998},
we introduce and analyze a $2+1$ dimensional texture model with a phase transition implemented through an
{\it imposed} equilibrium phase structure.   The dissipative quench provides a useful 
physical model to which we hope to return our attention in future work.

\subsection{External quench}
\label{subsec:equench}
In order to study the dynamical formation of topological textures in the absence of strong thermal fluctuations,
the equations of motion Eq.~(\ref{eq-fieldeq1}) are
modified by the addition of an externally controlled time-dependent effective mass and a stochastic environment.
The effective mass is
\begin{equation}
m^2_{eff}(t)=\tanh \left (\frac{t_c-t}{\tau} \right ),
\end{equation}
\noindent where $\tau$ is the quench rate and $t_c$ defines the critical point where 
spinodal instabilities, characteristic of a second order phase transition, begin to grow.  Both
in the early universe and in the laboratory, texture formation occurs in an environment containing many
other degrees of freedom.  The effect of the environment is assumed to be that
of a stochastic (Gaussian white noise) driving term $\xi(x,t)$ and simple ohmic dissipation, obeying a fluctuation-dissipation
relation.  The external quench equations of motion are    
\begin{equation}
\label{eq-pqfieldeq}
\ddot{\Phi}_i-\bigtriangledown^2\Phi_i-\eta\dot{\Phi}_i+\Phi_i(\vec{\Phi}^2-m^2_{eff})=\xi(\vec{x},t),
\end{equation}
where the fluctuation-dissipation relation is
\begin{equation} 
\label{eq-fd}
\langle \xi(\vec{x},t) \xi(\vec{x}',t') \rangle = 2T\eta\delta^2(\vec{x}-\vec{x}')\delta(t-t').
\end{equation}
The fluctuations, controlled by the parameter $T$ are weak and serve only to give a random kick as the fields roll
to the true vacuum.  The dissipative term $\eta \dot{\Phi}_i$ is strong for short wavelengths and 
helps desensitize the classical field model to modes near the lattice cutoff (used below in the numerical simulations).  
Independence from the cutoff was also verified operationally by halving the lattice spacing and observing
no change in the output. 
The use of a time-dependent mass in the equations of motion of this $2+1$ dimensional model 
deserves comment.
At $t=0$ the effective mass is positive and the system fluctuates around the false vacuum $|\vec{\Phi}|=0$. When
t reaches $t_c$ the system is destabilized and the fields begin to roll to the true vacuum $|\vec{\Phi}|=1$. 
This behavior mimics that of a phase transition in $3+1$ spacetime dimensions.  It is well-known 
that there is no continuous symmetry breaking and accompanying long range order in spatial dimensions
of two or less \cite{mermin:1966}.  Therefore the time dependence of the effective mass does {\it not} come
from changing a thermodynamic variable, such as temperature.  Instead, we use the external 
time-dependent effective mass to allow for a controlled passage between the fluctuation-dominated and 
texture-dominated phases.  Within this controlled
setting the role of the quench rate in the formation of texture can be isolated and studied effectively.  It is our
hope that the lessons learned are applicable to more realistic phase transitions, both in the early 
universe and in the laboratory.

\subsubsection{Numerical parameters and techniques}
The dynamics of the non-linear stochastic system were solved numerically.  Equation (\ref{eq-pqfieldeq}) was discretized on
a 2-dimensional lattice using a standard second-order leapfrog method for the time evolution 
\cite{press:1989,spergel:1991,antunes:1997} 
and a second-order spatial discretization for the laplacian $\bigtriangledown^2$,
\begin{eqnarray}
\vec{\Pi}_{n+1}[j][k]&= &\frac {1-\chi}{1+\chi}\vec{\Pi}_n[j][k]+ \frac{dt}{(1+\chi)} (\frac{1}{dx^2}(\vec{\Phi}_n[j+1][k] -4\vec{\Phi}_n[j][k]\nonumber \\
& & +\vec{\Phi}_n[j-1][k]+\vec{\Phi}_n[j][k+1] +\vec{\Phi}_n[j][k-1])-\vec{\Phi}_n[j][k](\vec{\Phi}_n^2[j][k]   \nonumber \\
& &  +m_{eff}^2[ndt])+\xi_n[j][k]) \nonumber \\
\vec{\Phi}_{n+1}[j][k]&= &\vec{\Phi}_n[j][k]+dt\vec{\Pi}_{n+1}[j][k] 
\end{eqnarray}
\noindent where $\vec{\Pi}=\frac{\partial\vec{\Phi}}{\partial t}$ and $\chi=\eta dt$.  The labels $[j][k]$ identify 
the spatial lattice point and $n$ labels the time step. At each time step and each lattice point the noise 
is generated through a sum of $M$ randomly distributed numbers \cite{alexander:1993}
\begin{equation}
   \xi[j][k]=\sum\limits_{i=1}^M{\frac{\theta[j][k]}{M} \left (\frac {24M\eta T} {dx^2dt}\right )^{\frac{1}{2}}},
\end{equation}      
\noindent where $-0.5<\theta[j][k]<0.5$ are a set of $N^2$ random numbers.  In the limit when
$M\rightarrow\infty$ the $\xi[j][k]$ distribution approaches that of a Gaussian with the variance required for
the fluctuation-dissipation relation Eq.~(\ref{eq-fd}).  In practice $M=24$ was used and no change in the dynamics was
observed for larger $M$.  The timestep was $dt=0.02$ and the lattice spacing was fixed at $dx=0.5$ on a 
square lattice with $N=1000$ sites per side.
There are three physical parameters in the system, the quench rate $\tau$, the dissipation constant $\eta$,  
and the strength of the fluctuations $T$.  The quench rate was varied between $\tau=10$ and $\tau=150$.  
Faster quenches (smaller $\tau$) deviate from the observed power-law behavior and asymptote 
to the behavior characteristic of an instantaneous quench as has been previously observed \cite{yates:1998}.  
Longer quenches are in principle possible but take substantially more computer time.  
The dissipation constant was chosen at
$\eta=1.0$ to produce relatively overdamped behavior.  The evolution equations were supplemented with initial conditions
generated by solving the equilibrium 
dynamics of Eq.~(\ref{eq-pqfieldeq}) with a value of
the effective mass {\it fixed} at one quench timescale from the critical point,
\begin{equation}
 m_{eff}^2(t<0)= \tanh{(1.0)}.
\end{equation}
\noindent Run times for the system to reach an order parameter of $\phi=0.95$ on a DEC 500 Mhz workstation varied 
from hours for $\tau=10$ to days for $\tau=150$.

\subsubsection{Properties of the texture distribution}
Field configurations obtained from numerical simulations provide an embarrassment of riches.  In order to 
fully explore the 
participation of topological textures in the dynamics of the phase transition, it is necessary to cull, from the 
abundance of information contained in the snapshots of the topological field configurations, a few 
well-chosen functions, whose time evolution provides a clear 
indication of the dynamics.  The late-time coarsening of a textured system is described by at least 
{\it three} different length scales, the average defect-defect separation $L_{sep}$, the average 
defect width $L_w$, and the average texture-antitexture separation $L_{tat}$.  In the formation dynamics studied here
we were able to measure reliably only $L_{sep}$ and $L_w$.  These length scales are defined through the 
the topological charge density and the topological defect two-point correlation function.   
We consider the two-point defect density correlation function
\begin{equation}
C_1(r,t)=\langle |\rho(\vec{x})||\rho(\vec{x}+\vec{r})| \rangle -\langle|\rho(\vec{x})|\rangle ^2,
\end{equation}
\noindent where $\langle \rangle$ denotes average over the lattice.  We define the normalized correlation function
\begin{equation}
G(r,t)\equiv \frac {C_1(r,t)} {C_1(0,t)}.
\end{equation}
\noindent $G(r,t)$ is a statistical measure of the ``lumps'' in the defect density function $|\rho(\vec{x})|$. 
The average texture width $L_w$ is proportional to the half-height scale $r_{\frac{1}{2}}$ of $G(r,t)$ defined by
\begin{equation}
G(r_{\frac{1}{2}}, t)=1/2.
\end{equation}
\noindent To measure the average texture separation we consider
\begin{equation}
Q_{total}=\int d^2x |\rho|.
\end{equation}
\noindent $Q_{total}$ counts the total number of topological defects in the system without distinguishing 
between textures and antitextures.  On dimensional grounds,
\begin{equation}
Q_{total}\sim \frac{1}{L^2_{sep}}L_{sys}^2,
\end {equation}
where $L_{sep}$  measures the average separation of the topological defects and $L_{sys}$ is
the physical size of the lattice which did not vary between runs.   These two length scales, $L_{w}$ and $L_{sep}$, 
provide a detailed characterization of the topological texture network at the end of the transition. In principle, each scale 
offers a different window into the nonequilibrium dynamics of the phase transition.

\section{Unraveling critical dynamics}
\label{sec:ucd}
In this section we explore the details of the external quench contained within the dynamics of  
Eq.~({\ref{eq-pqfieldeq}).
Plots of the time evolution of the order parameter $\phi\equiv \langle\vec{\Phi}^2 \rangle$
and the average number of topological defects $Q=\int d^2x \mid \rho(x,t) \mid$ are shown in Figs.~\ref{fig:order} 
and \ref{fig:charge} respectively. 
\begin{figure}
\begin{center}
\strut\psfig{figure=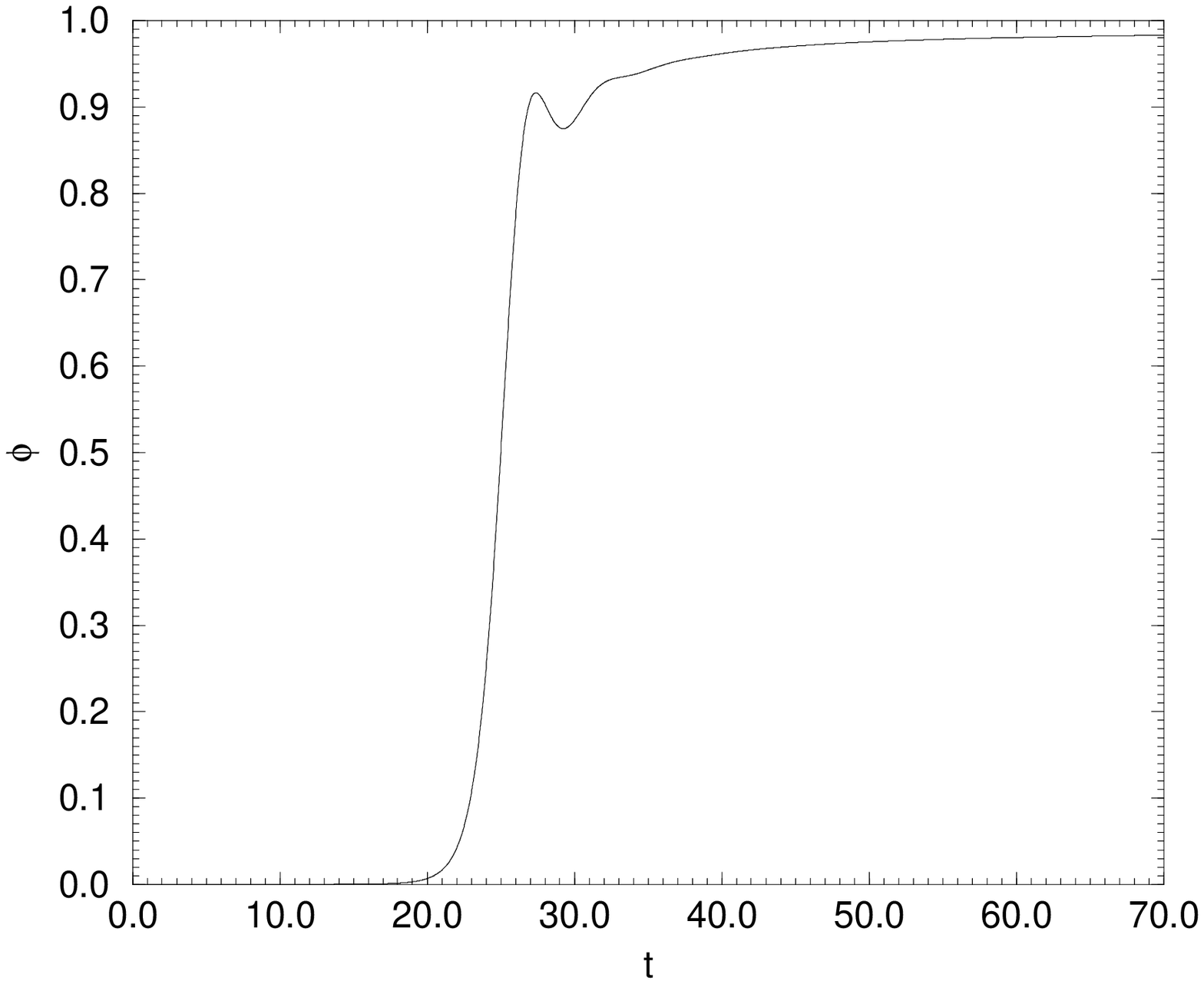,width=4.0in,angle=0}
\rmmcaption{Plot of the order parameter $\phi(t)=\langle |\vec{\Phi}|^2 \rangle$ 
versus time $t$ for quench parameter $\tau=10$.} 
\label{fig:order}
\end{center}
\end{figure}
\begin{figure}
\begin{center}
\strut\psfig{figure=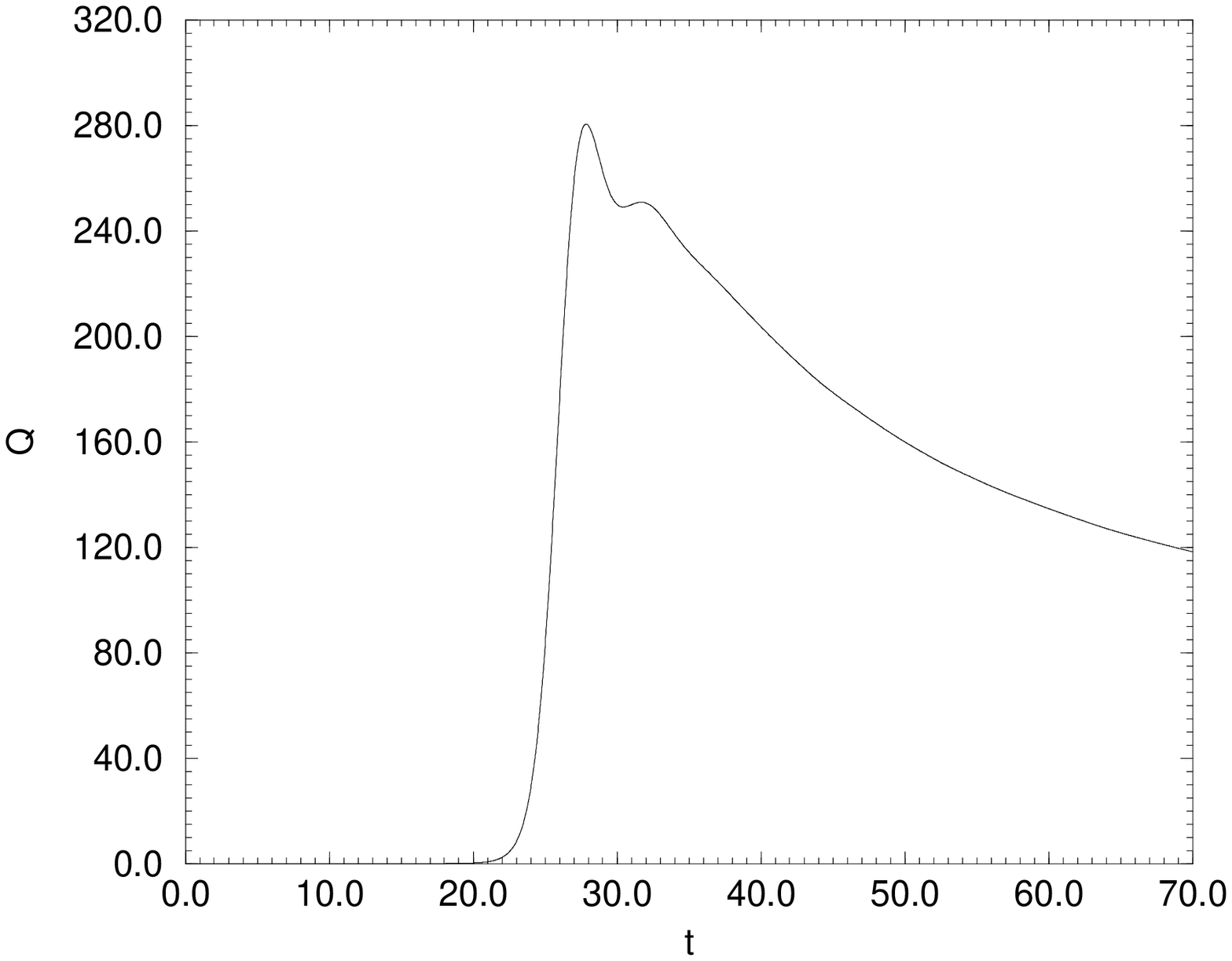,width=4.0in,angle=0}
\rmmcaption{Plot of the average number of textures  $Q(t)=\int d^2x |\rho(\vec{x},t)|$ 
 versus time $t$ for quench parameter $\tau=10$.} 
\label{fig:charge}
\end{center}
\end{figure}
The plots correspond to an evolution with
quench parameter $\tau=10$.  In the symmetric state
$(t<10.0)$ both $\phi$ and $Q_{total}$ are close to zero.  When $t>10.0$ the external effective mass is negative
and the fields, kicked by the small stochastic force $\xi(\vec{x},t)$, begin to roll to the true vacuum $\phi=1$.
The order parameter $\phi$ is a measure of the number of lattice points where the fields have fallen to the 
true vacuum.  As $\phi$ increases both the gradient energy and the number of defects increase.  However 
as $\phi\rightarrow 1$ most of the lattice is near the ground state vacuum and the defect 
formation process is over. Although the system is overdamped near freeze-out, the small oscillations in 
$\phi$ and $Q$ near the end of the transition indicate that 
oscillations around the true vacuum are not completely overdamped.
After formation, the defect density does not remain constant but decreases slowly as 
texture-texture and texture-antitexture interactions dominate.

\begin{figure}
\begin{center}
\strut\psfig{figure=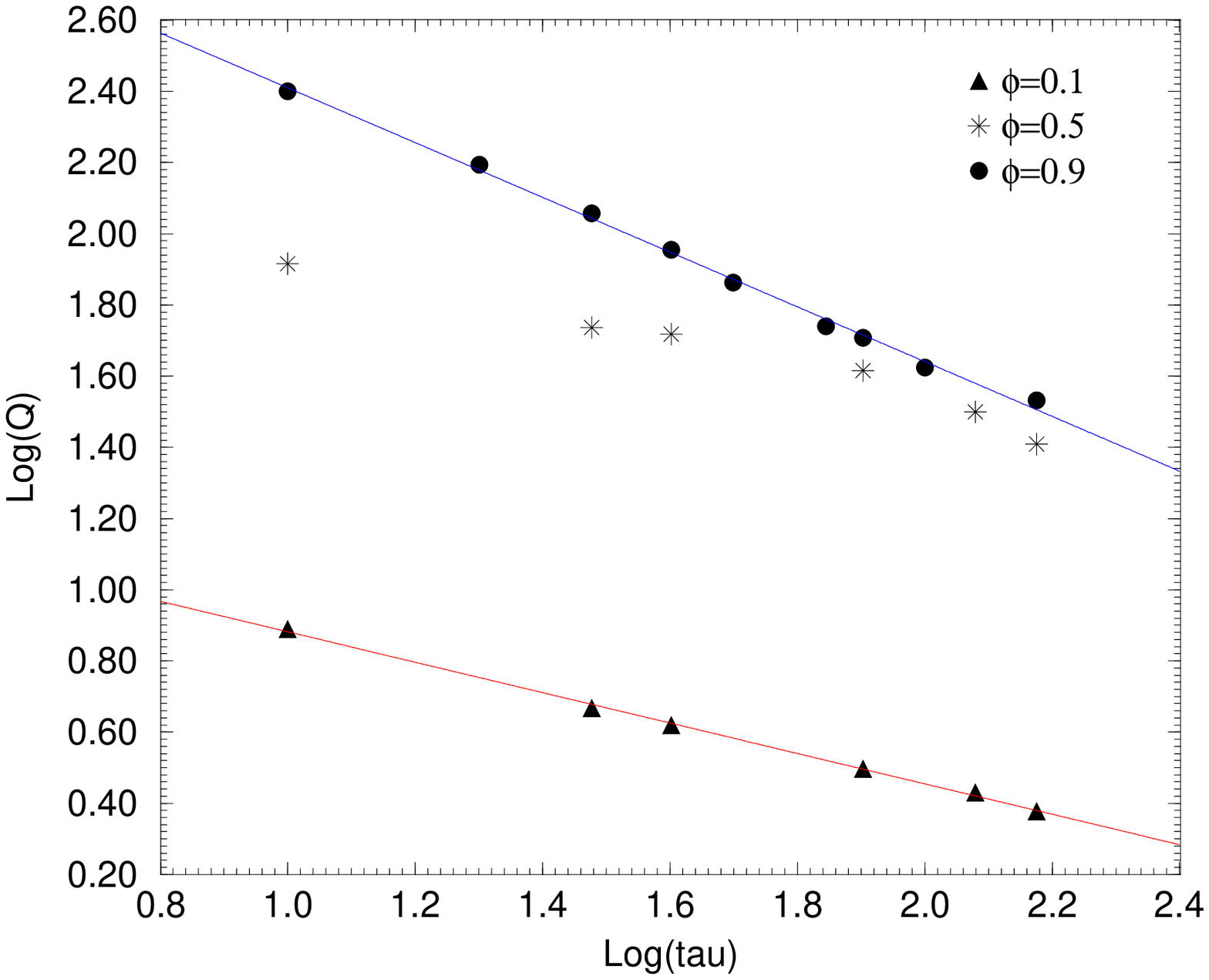,width=4.0in,angle=0}
\rmmcaption{Log-Log plot of the average number of textures
$Q=\int d^2x |\rho(\vec{x},t)|$ versus quench parameter $\tau$ for
different values of the order parameter $\phi=\langle \vec{\Phi}^2 \rangle$.}
\label{fig:Qevolution}
\end{center}
\end{figure}

\begin{figure}
\begin{center}
\strut\psfig{figure=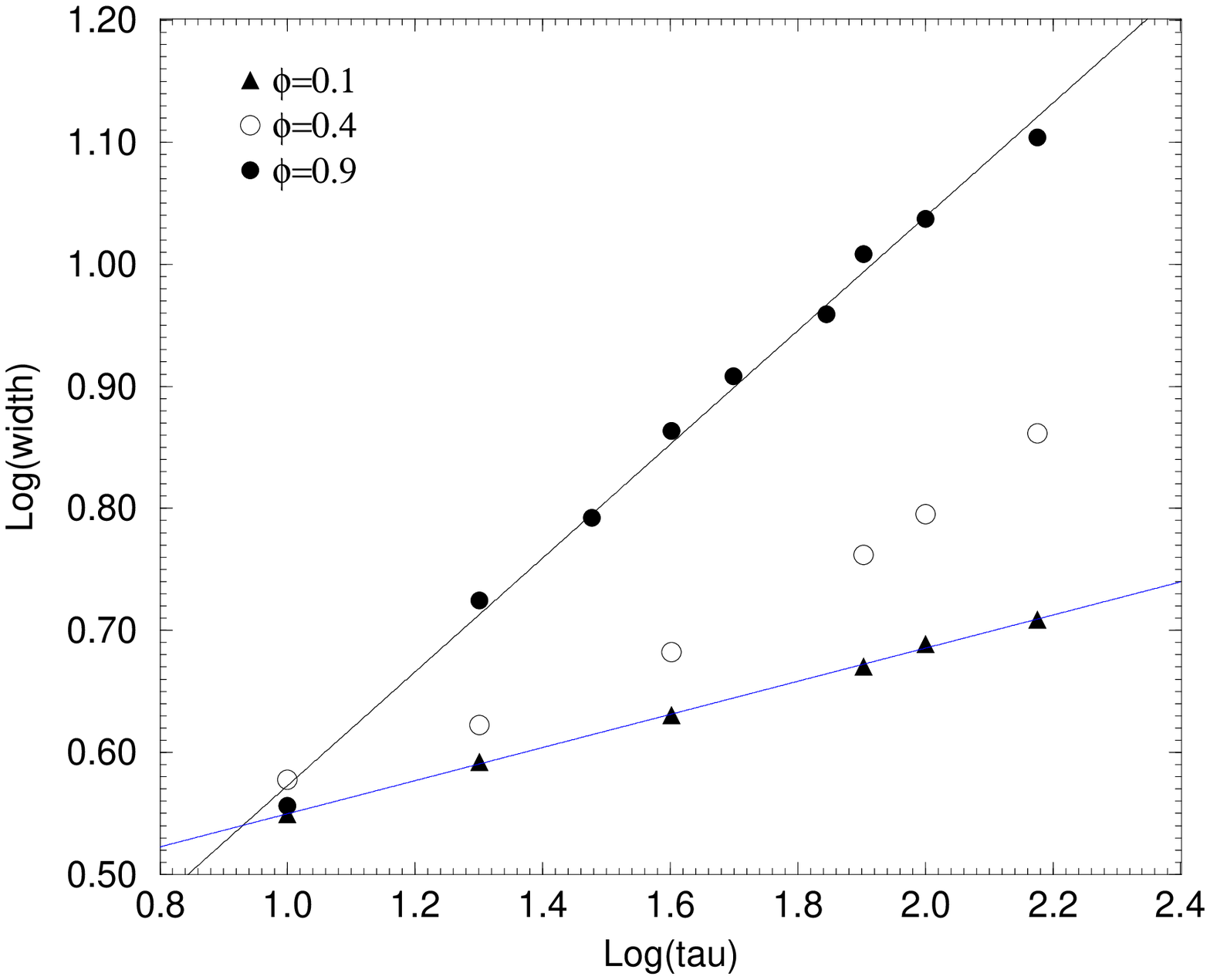,width=4.0in,angle=0}
\rmmcaption{Log-Log plot of the average texture width versus quench parameter $\tau$ for different values of the
order parameter $\phi=\langle \vec{\Phi}^2 \rangle$.}
\label{fig:Qwidthevolution}
\end{center}
\end{figure}

Figs.~\ref{fig:Qevolution} and \ref{fig:Qwidthevolution} plot the 
average number of textures and the average texture width versus the quench rate.  In each figure,
symbols denote lattice measurements while solid lines denote best-fit power-laws. In Fig.~\ref{fig:Qevolution}
lattice measurements were made at early times $(\phi=0.1)$ denoted by solid triangles on the lower graph,
at intermediate times $(\phi=0.5)$ denoted by stars on the middle graph and
at late times $(\phi=0.9)$ denoted by solid squares on the upper graph.
The best-fit power-law for times early in the phase transition is
\begin{equation}
\log(Q)_{early}=(1.31\pm 0.02)- (0.43 \pm 0.01 )\log(\tau).
\end{equation}
At late times the best-fit power law is
\begin{equation}
\log(Q)_{late}=(3.17 \pm 0.03) - (0.77\pm 0.02)\log(\tau).
\end{equation}  
A crossover is apparent at intermediate times when the number of textures $Q(\tau)$ is not
given by a single power-law of the quench rate.
In Fig.~\ref{fig:Qwidthevolution}
lattice measurements were made at early times $(\phi=0.1)$ denoted by solid triangles on the lower
graph,
at intermediate times $(\phi=0.4)$ denoted by open circles on the middle graph and
at late times $(\phi=0.9)$ denoted by solid squares on the upper graph.
The best-fit power-law for times early in the phase transition is
\begin{equation}
\log(L_w)_{early}=(0.42\pm 0.02) +(0.14\pm 0.01)\log(\tau).
\end{equation}
At late times the best-fit power-law is
\begin{equation}
\log(L_w)_{late}=(0.11\pm 0.02) + (0.47\pm 0.01)\log(\tau).
\end{equation}  
At intermediate times $L_w(\tau)$ is growing at a faster rate for longer quenches and is not
given by a single power-law of the quench rate.  

At late times the texture separation and texture width are well-described by the power-law scalings 
\begin{eqnarray}
\label{eq-latescale}
L_{sep} &\sim& \tau^{0.38 \pm 0.01}, \\
L_{w}   & \sim& \tau^{0.47 \pm 0.01}. \nonumber
\end{eqnarray} 
The best-fit exponents were derived using equal weighting for all lattice measurements.
As we found no reliable means to estimate the error in individual lattice measurements, 
the errors in the scaling exponents arise from statistical errors in the fit.  
Certainly, power-law scaling is expected from the freeze-out picture.  However, the Kibble-Zurek mechanism 
also provides a precise prediction for the {\it value} of the power-law exponent as derived from equilibrium 
critical exponents.  The model under consideration is a {\it mean field} model and the only role of the
small fluctuations is to seed texture formation.  Since the effective mass is controlled 
externally and $m_{eff}^2 \sim t_c-t$ near the critical point, it is expected that $\nu=\frac{1}{2}$,
the mean field value of equilibrium correlation length exponent.  This expectation was verified
by equilibrium lattice simulations.  With $\eta=1.0$ and $\tau \geq 10$ the dynamics is overdamped
at freeze-out.  Therefore $\mu=2\nu=1$ and the Kibble-Zurek mechanism predicts the scaling
\begin{equation}
\xi_{KZ}(\tau) \sim \tau^{\frac{1}{4}},
\end{equation}
in obvious disagreement with Eq.~(\ref{eq-latescale}).         

The larger observed exponent means that for longer quenches topological textures are both less numerous and bigger
than the Kibble-Zurek mechanism predicts.  However, slower quenches take longer for the phase transition to complete, 
since as $\tau$ increases, the order parameter rolls more slowly. In fact, observations of the dynamics show that the 
time $t_{0.9}$ (measured from the critical point) it takes for the order parameter to reach 
$\langle \vec{\Phi}^2 \rangle =0.9$ increases with $\tau$ in a simple way $t_{0.9} \sim 1.5 \tau$.
With more time to evolve between freeze-out and the end of the transition, slower quenches allow   
for more interactions among the texture distribution.  This suggests that the apparent power-laws observed at
the end of the transition are {\it not} indicative of a single scale, but are a combination of the Kibble-Zurek 
mechanism and the dynamical length scales of the evolving texture network.

Evolution of the power law exponents of the texture length scales is clearly indicated in the two lower graphs of
Figs.~\ref{fig:Qevolution} and \ref{fig:Qwidthevolution}.  At these early times the effect of coarsening
dynamics is insignificant and power-law scaling indicative of the Kibble-Zurek mechanism is evident,
\begin{eqnarray}
\label{eq-earlyscale}
L_{sep} &\sim& \tau^{0.22 \pm 0.01}, \\
L_{w}    &\sim& \tau^{0.14 \pm 0.01}. \nonumber
\end{eqnarray}
\noindent Although the power-law exponent of $L_w$ is slightly smaller than predicted by the Kibble-Zurek
 mechanism, the difference is not likely to be significant.  Textures are not well-formed at these early times 
and the method for measuring $L_w$ is sensitive to the details of the two-point function.    
In the middle graphs of each figure, the number of textures
and the texture width were measured at intermediate times.  Both show the crossover from
formation dynamics to coarsening dynamics as longer quenches develop a steeper power-law indicative
of texture interactions.  A similar picture of the evolution of $L_{sep}$ is given by the average number of
textures as measured by the total gradient energy, Fig.~\ref{fig:Gevolution}.
As before, symbols denote lattice measurements and the two solid lines denote 
best-fit power-laws.  Early in the phase transition $\phi=0.1$
the best fit power-law is 
\begin{equation}
\log(G)_{early}=(2.34 \pm 0.01)- (0.46 \pm 0.01)\log(\tau), 
\end{equation}
in agreement  with the Kibble-Zurek mechanism. At late times $\phi=0.9$, 
\begin{equation}
\log(G)_{late}=(3.65\pm 0.10) - (0.94\pm 0.06)\log(\tau).
\end{equation} 
As in Fig.~\ref{fig:Qevolution} a crossover is
apparent at intermediate times $\phi=0.45$.  The late-time power law is slightly 
steeper than that determined from the topological charge because the gradient energy
also includes topologically trivial configurations which
decay more rapidly than textures.
\begin{figure}
\begin{center}
\strut\psfig{figure=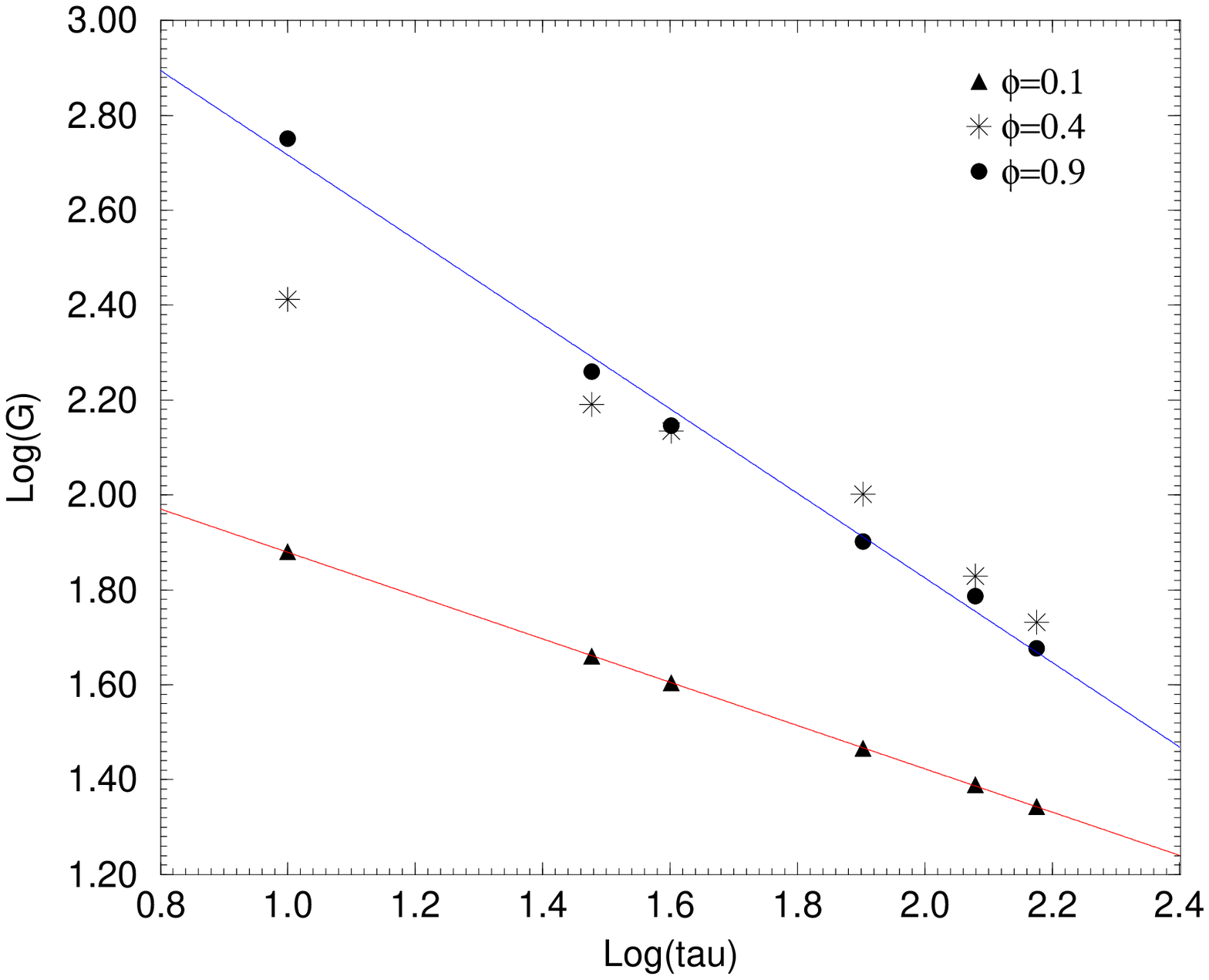,width=4.0in,angle=0}
\rmmcaption{Log-Log plot of the average number of textures determined from the gradient energy
$G=\frac{1}{16\pi}\int d^2x (\vec{\partial}\Phi_i \cdot \vec{\partial}\Phi_i)$
versus quench parameter $\tau$ for different values of the order parameter $\phi=\langle \vec{\Phi}^2 \rangle$.} 
\label{fig:Gevolution}
\end{center}
\end{figure}
\subsubsection{Interactive texture formation}
Figures \ref{fig:Qevolution}, \ref{fig:Qwidthevolution} and \ref{fig:Gevolution} provide clear evidence of the evolution of
the power-laws characterizing the length scales of the texture distribution.  We can use 
the knowledge of
the coarsening dynamics of texture interactions to provide a {\it quantitative} understanding of the power-laws observed
at the end of the transition.  As discussed in Sec.~\ref{sec:tttsd}, the phase ordering of a textured system results in a 
dynamical length scale characterizing the separation of defects, $L_1(t)=\xi_0^{\frac{1}{3}}t^{\frac {1}{3}}$.  
Figure ~\ref{fig:longcharge} shows the dynamics of this length scale for a fast quench 
(which enters the coarsening regime earlier). The solid line denotes data from lattice measurements.  
The dashed line is the best-fit power law of the late-time coarsening dynamics, 
\begin{equation}
Q(t) \sim t^{-0.72}.
\end{equation} 
\begin{figure}
\begin{center}
\strut\psfig{figure=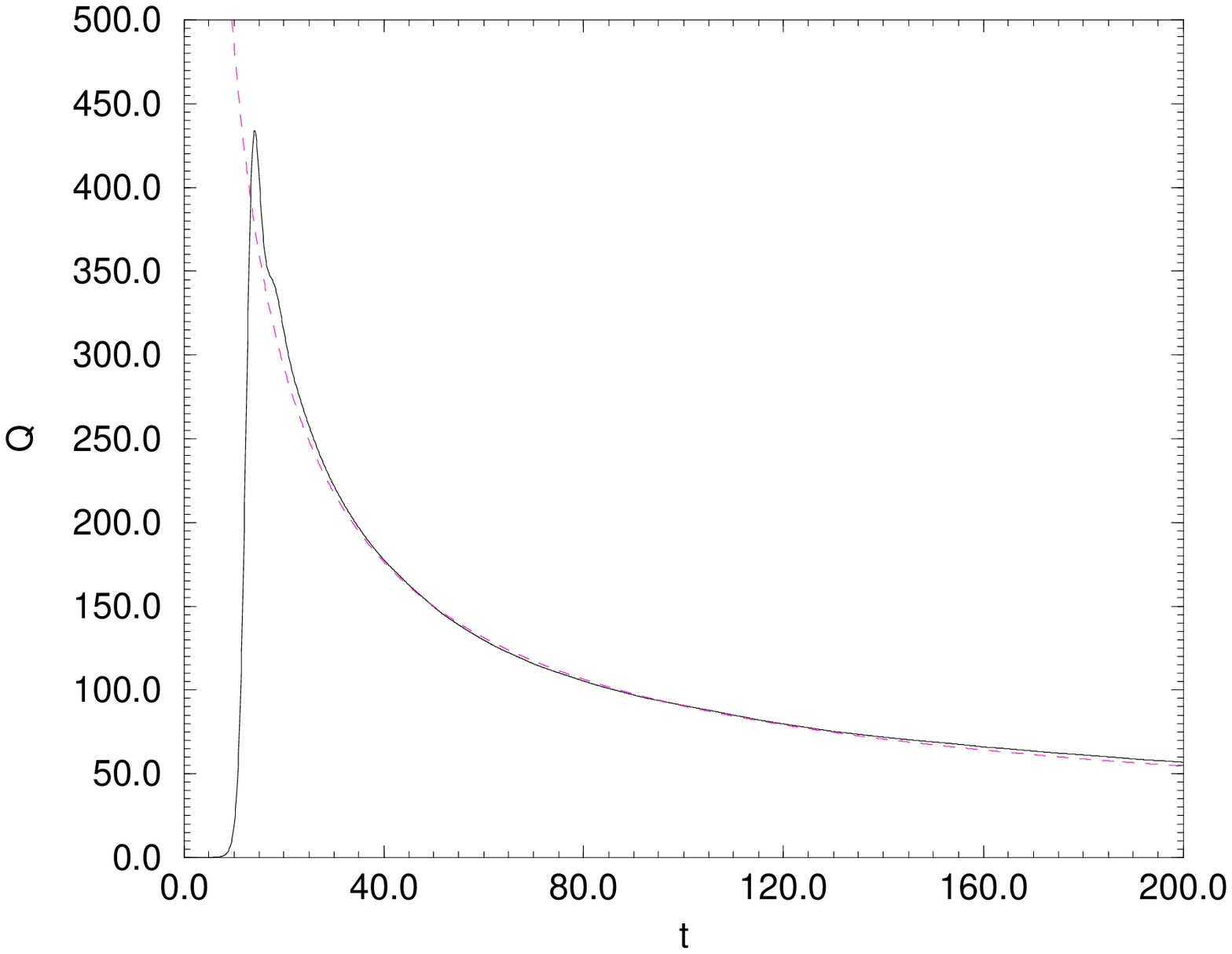,width=4.0in,angle=0}
\rmmcaption{Plot of the average number of textures $Q(t)=\int d^2x |\rho(\vec{x},t)|$ versus time 
for a fast quench $\tau=1$.}
\label{fig:longcharge}
\end{center}
\end{figure}
The coarsening exponent $\alpha=-0.72$ agrees favorably with previous studies which found
$\alpha=-\frac{2}{3}$.
In the instantaneous quenches of \cite{rutenberg:1995a} $\xi_0$ was identified 
with the initial correlation length.  In a dynamical
quench it is more natural to identify $\xi_0$ with the freeze-out correlation length.  Combining the Kibble-Zurek
mechanism with phase ordering dynamics we suggest that the evolution of the scale characterizing the separation of 
topological textures is given by
\begin{equation}
\label{eq-Lsep}
L_{sep}(t)=\xi_{freeze}+ (\xi_{freeze})^{\frac {1}{3}}(t-t_{freeze})^{\frac {1}{3}},
\end{equation}
\noindent where $\xi_{freeze} \sim \tau^\alpha$ is the length scale determined by the Kibble-Zurek
mechanism, $t_{freeze} \sim \tau^{2\alpha}$ is the time from freeze-out
and t is the time from the critical point. Although Eq.~(\ref{eq-Lsep}) appears slightly complicated, its
form is tightly constrained and the only free parameter is $\alpha$, the exponent of the freeze-out
correlation length.  To test Eq.~(\ref{eq-Lsep}) we match the expected number of defects 
$Q \sim \frac{1}{L^2_{sep}}$ at time $t=t_{0.9}\sim 1.5\tau$ against the number of defects determined from the 
numerical simulations.  A plot of the data and the best-fit length scale is shown in Fig.~\ref{fig:Lsepfit}.  
\begin{figure}
\begin{center}
\strut\psfig{figure=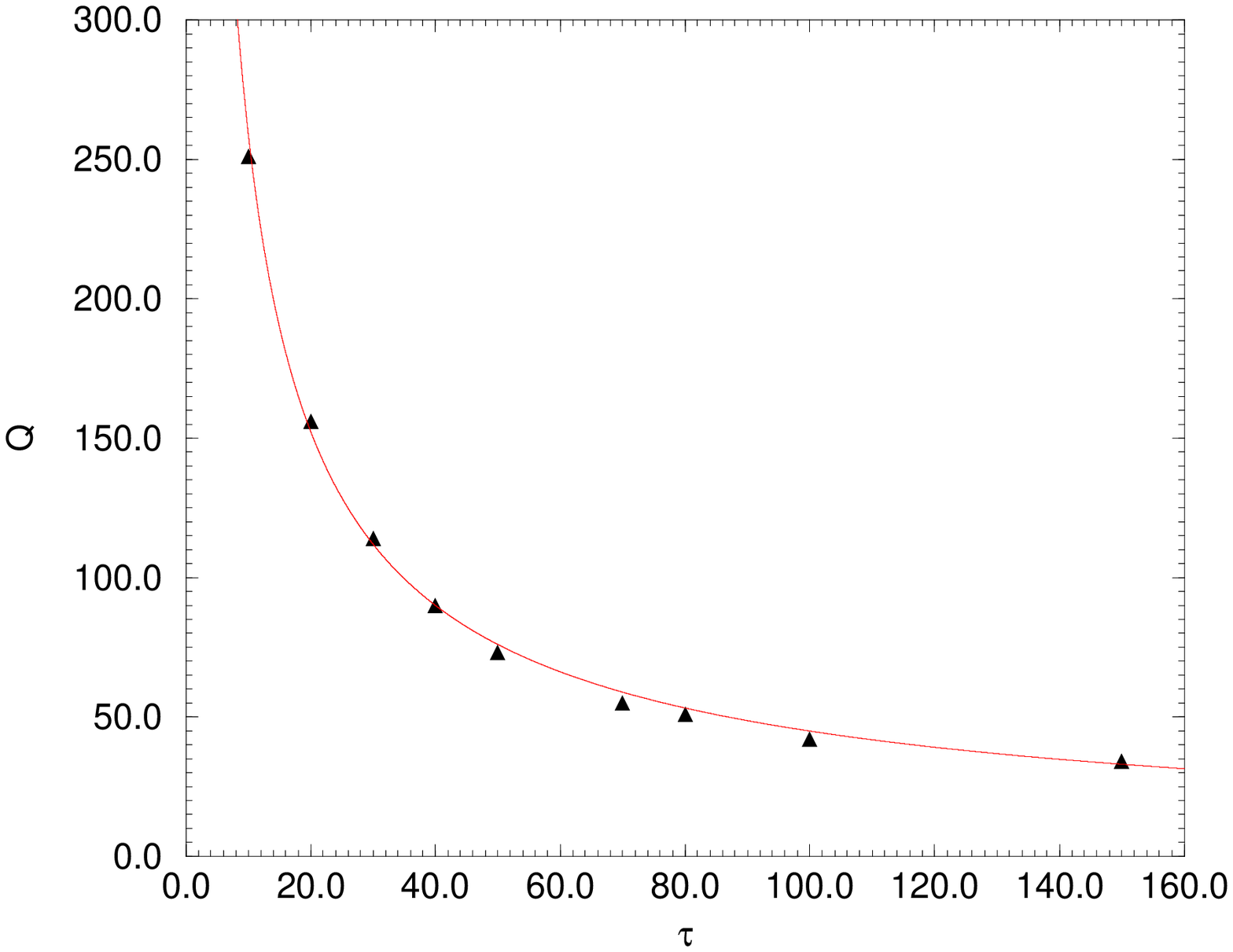,width=4.0in,angle=0}
\rmmcaption{Plot of the average number of textures $Q=\int d^2x |\rho(\vec{x})$ versus
quench parameter $\tau$.} 
\label{fig:Lsepfit}
\end{center}
\end{figure}
Filled triangles denote lattice measurements while
the solid line denotes the theoretically expected number of textures computed from Eq.~(\ref{eq-Lsep}).
To obtain the fit one data point is used to normalize the scale and 
a value of $\alpha$ is determined by minimizing $\chi^2$.  By using, in turn, each data point as a normalization
a spread in the value of $\alpha$ is obtained.  The best-fit value of $\alpha$ is taken as the average over all
normalizations and the error is the standard deviation.  As determined from these fits, $\alpha=0.24 \pm 0.02$ 
in excellent agreement with $\alpha=0.25$ expected from the Kibble-Zurek mechanism.  

The analysis of the dynamics of $L_w$, the length scale characterizing the size of topological textures, is more
complicated.  Fig.~\ref{fig:longlogwidth} provides an example of texture width dynamics for a
single quench with $\tau=80$. 
\begin{figure}
\begin{center}
\strut\psfig{figure=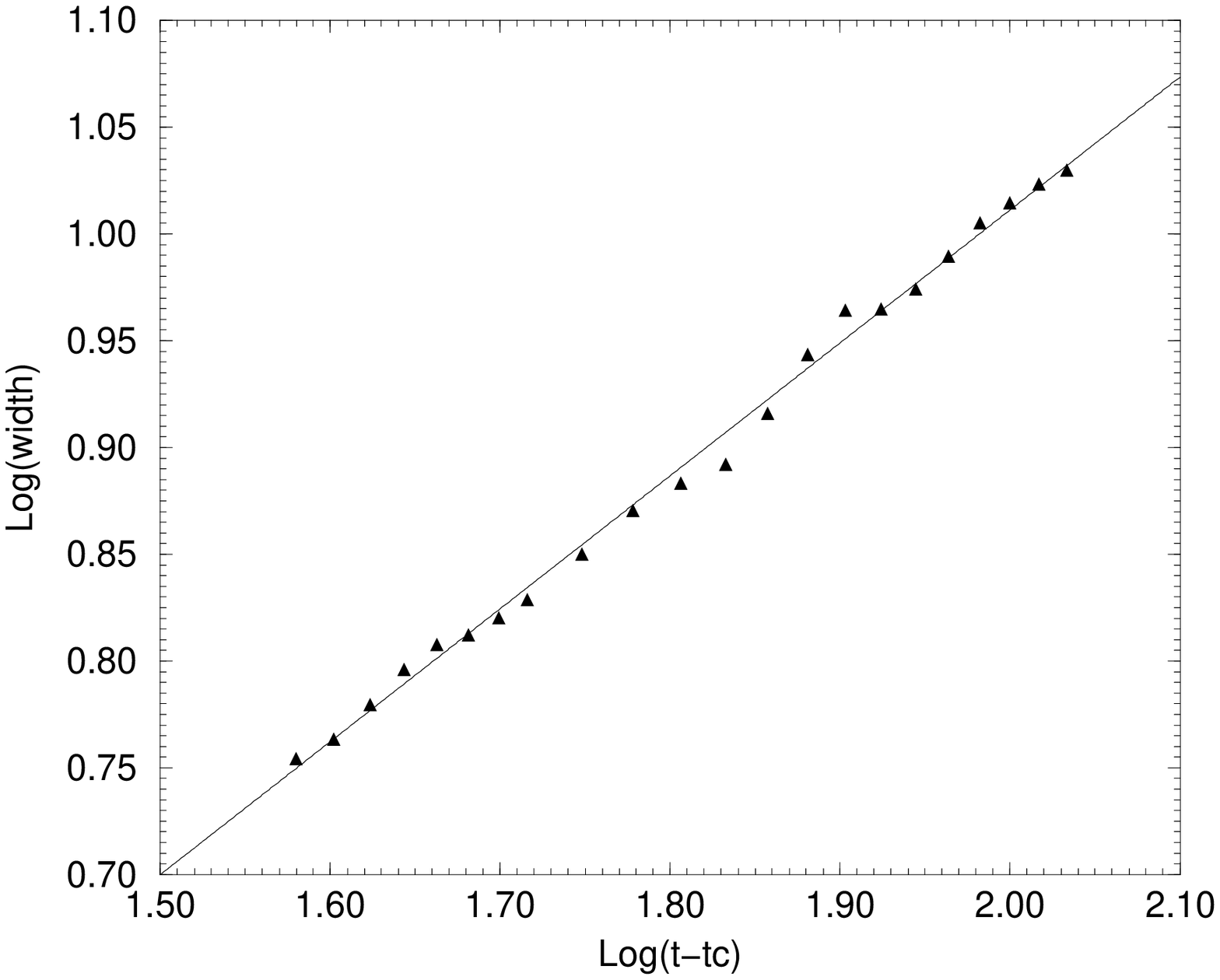,width=4.0in,angle=0}
\rmmcaption{Log-Log plot of the texture width versus time for quench parameter $\tau=80$.  Time $t$ is 
measured from the critical point, $t_c=80$.}
\label{fig:longlogwidth}
\end{center}
\end{figure}
Solid triangles denote lattice measurements and the straight  
line is the best-fit power law of the coarsening dynamics, 
\begin{equation}
L_w(t-t_c) \sim (t-t_c)^{-0.62}. 
\end{equation}
The coarsening exponent $\alpha=-0.62$ is close to the simple scaling exponent $\alpha=-\frac{1}{2}$.
In the simulations the texture width was observed to grow in time with a power law
near the simple coarsening dynamics predicted for an overdamped model $L \sim t^{\frac{1}{2}}$. 
However, this growth rate is different from previously reported studies of the late-time dynamics 
of the texture width.  Consistent with previous studies
\cite{rutenberg:1995a}, it is possible that $L_w(t)$ reaches its asymptotic dynamics only at later times.  It is also possible 
that $L_w(t)$ is contaminated at early times by topologically trivial spin configurations whose decay then dominates 
the dynamics.  The
resolution of this issue requires a more accurate method of counting textures e.g. by looking explicitly at
the winding around the vacuum manifold.  Whatever the mechanism, Fig.~\ref{fig:longlogwidth} shows 
that $L_w$ can evolve significantly between the early and late periods of the phase transition.  
We suggest that the evolution is given by
\begin{equation}
\label{eq-Lwidth}
L_{w}(t)=\xi_{freeze}+ (t-t_{freeze})^{\frac{1}{2}}.
\end{equation} 
As before, $\xi_{freeze} \sim \tau^\beta$ is the length scale determined by the Kibble-Zurek
mechanism, $t_{freeze} \sim \tau^{2\beta}$ is the time from ``freeze-out'' 
and t is the time from the critical point.  To test Eq.~(\ref{eq-Lwidth}) we match the expected width of the textures 
at time $t=1.5\tau$ against the width determined from the numerical simulations.  A plot of the data 
and the best-fit length scale is shown in Fig.~\ref{fig:Lwidthfit}.  
Filled triangles denote lattice measurements.  The solid line denotes the theoretically expected width computed
from Eq.~(\ref{eq-Lwidth}).
The fits were obtained in the same way as for $L_{sep}$.  As determined from these fits, $\beta=0.22 \pm 0.07$ 
also in agreement with $\beta=0.25$ expected from the Kibble-Zurek mechanism.
\begin{figure}
\begin{center}
\strut\psfig{figure=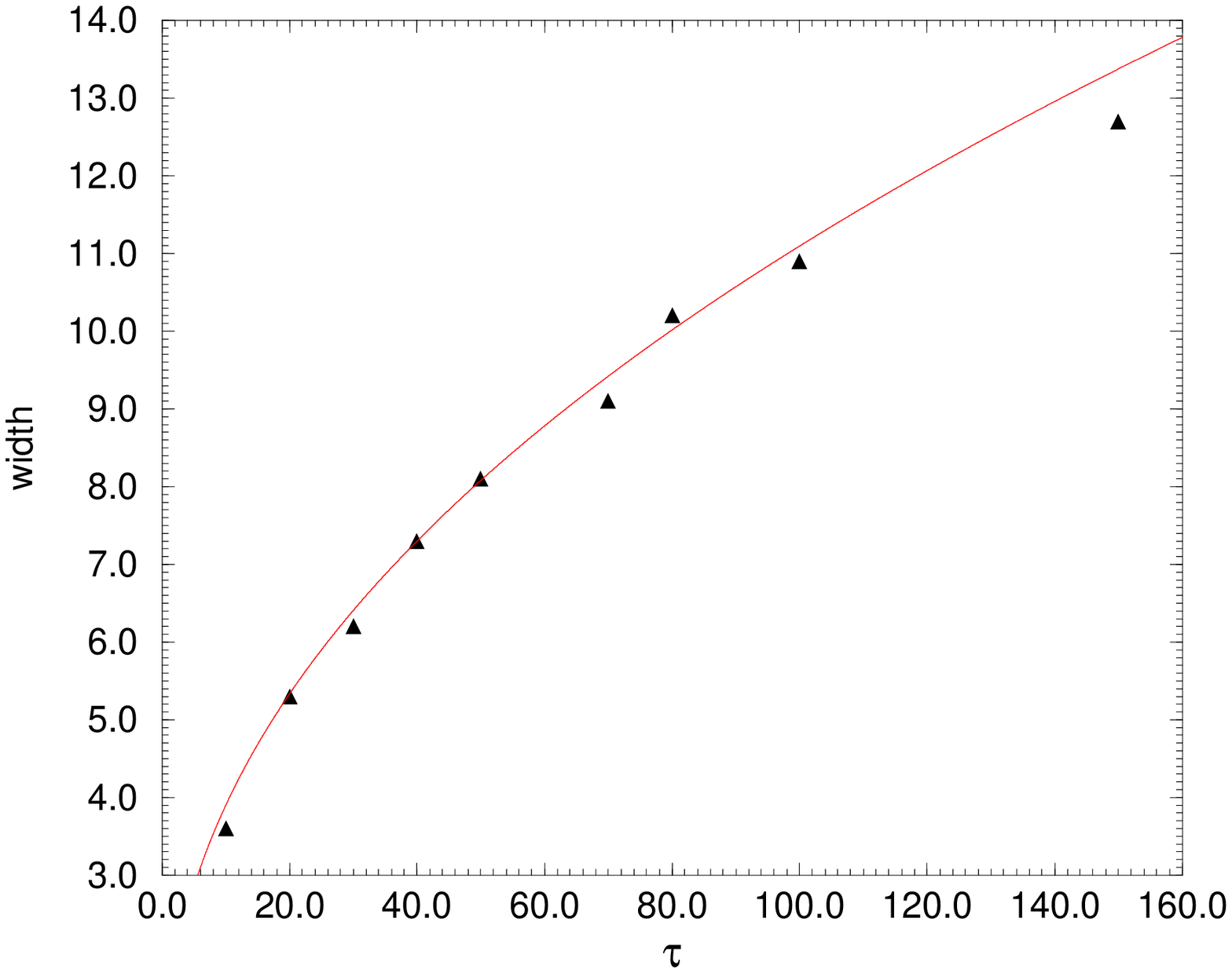,width=4.0in,angle=0}
\rmmcaption{Plot of the average texture width versus
quench parameter $\tau$.} 
\label{fig:Lwidthfit}
\end{center}
\end{figure}
It is clear from Figs.~\ref{fig:Lsepfit} and \ref{fig:Lwidthfit} that the power-laws
of Eq.~(\ref{eq-latescale}) can be explained by incorporating texture dynamics.  Eqns.~(\ref{eq-Lsep})
and (\ref{eq-Lwidth}) offer a conservative, simple guess to the proper combination of formation and coarsening dynamics.  
A more complete analysis, with better techniques for identifying textures, would fit $L_{sep}(t)$ and $L_w(t)$ 
to functions with both general formation {\it and} coarsening exponents.  Nonetheless it is remarkable that these
simple equations are in good agreement with the Kibble-Zurek mechanism. Even if the exact form of 
Eqns.~(\ref{eq-Lsep}) and (\ref{eq-Lwidth}) changes,  the observation remains that the length scales characterizing
the topological defect distribution at the end of the phase transition are a combination of defect interaction
and formation dynamics. 

It is not surprising 
that defect interactions can influence the properties of the defect network
before the end of the transition.  Although the long-wavelength dynamics is very slow until after the 
freeze-out time $t_{freeze} \sim {\tau}^{\frac{1}{2}}$ (in the overdamped
case), the time to complete the phase transition is $t_{0.9} \sim \tau$ allowing plenty of time for 
even partially formed defects to influence each other.  In previous numerical simulations in 1 and 2 
dimensions \cite{laguna:1997,yates:1998} this effect was not noticeable (in overdamped evolutions)
because the interactions between defects were extremely weak. 
In a recent simulation of global vortex formation in 3 dimensions \cite{antunes:1999} one might expect 
an observable change in the scaling exponents since the defect interactions are much stronger.  However,
the transition was at relatively high temperature and the defects were at least partially 
screened by thermal fluctuations.  

\subsubsection{Texture coarsening and the Kibble-Zurek mechanism}
The close agreement between Eqns.~(\ref{eq-Lsep}) and (\ref{eq-Lwidth}) and the length scales observed at the end of the 
transition provides evidence that the {\it a priori} independent length scales, $L_{sep}$ and $L_{w}$, are formed
by the same nonequilibrium dynamics of the Kibble-Zurek mechanism.  That this should be so is not 
obvious.  In fact, the Kibble-Zurek mechanism can be questioned even in models which at late times depend 
only upon a single scale.  In \cite{karra:1997} it was argued
argued that the length scales defined by the field correlation length and vortex 
density, in principle, derive from different attributes of the power spectrum.  
Only under restricted circumstances does the defect separation follow the field correlation length
and scale with the quench rate.  The initial analysis of the dissipative quench in Sec.~\ref{subsec:dquench} may
support the contention that strong fluctuations can alter the Kibble-Zurek scenario.
The external quench, however, occurred in the absence of strong fluctuations.
In this case, as was the case in Chapter 3, the early-time power spectrum is dominated by
a single peak with momentum $k=k_{max}$ in the low-momentum modes.  When the power spectrum
is strongly peaked, the freeze-out correlation length $\xi_{freeze} \sim \frac{1} {k_{max}}$
determines a unique scale at early times to which all other length scales are connected.    

\subsubsection{Late-time dynamics and experiments}
The strong violation of the scaling hypothesis observed in the coarsening of $2+1$ dimensional
textured systems suggests the possibility of novel, late-time measurements of early-time
critical dynamics.  As discussed in Sec.~\ref{sec:tttsd}, the coarsening of a textured system
with an instantaneous quench retains ``memory'' of a length scale $\xi_0$, the correlation
length in the disordered phase.  In a nonequilibrium phase transition it is the freeze-out 
correlation length, determined by the Kibble-Zurek mechanism, and {\it not} the correlation 
length of the initial state, that scars the texture distribution at late times.  For overdamped coarsening 
dynamics at late times,
\begin{equation}
\frac{\xi^2_w(t)}{\xi_{sep}(t)} = constant=\xi_{freeze}.
\end{equation}  
\noindent The freeze-out correlation length can be measured by a late-time measurement of the (average) 
texture width and texture separation.  It is the particular (and not well-understood) dynamics of 
texture-texture interactions that produces strong scaling violations and unusual memory effects.  
In contrast, experimental probes of vortex formation in nonequilibrium phase transitions of $^3He$ and $^4He$ 
are hampered by the details of the interacting vortex distribution.  Because these systems approach a scaling solution
at late times, vortex interactions work to erase any memory of the initial state.  This is a 
particular problem in the $^4He$ experiments where the vortex network is experimentally observable only at later 
times and 
the number of initial defects must be inferred using detailed assumptions about the decay of the vortex tangle.

Textured systems in $2+1$ dimensions can be created in the laboratory.  If their dynamics are
approximately described by Eq.~(\ref{eq-pqfieldeq}) then late-time experimental measurements of
the Kibble-Zurek mechanism are possible.  Topological textures can also be formed during phase
transitions in $3+1$ dimensions in superfluid $^3He$ and the early universe.  In $3+1$
dimensions without additional higher derivative terms in the action, topological textures are unstable.
In this case the dynamics of coarsening is complicated by texture collapse.  
However, as long as textures exist in the system scaling violations and associated dynamical memory
are possible.    

\section{Summary}
\label{sec:ch3sum}

In this chapter, we exploited the multiple length scales of a textured system to study a
nonequilibrium phase transition of an overdamped classical 0(3) model in $2+1$ dimensions.
We introduced a dissipative quench model in which the dynamics of the phase transition occur
through the dissipative cooling of the scalar fields, initially at high temperature.  Our analysis
of the topological textures formed in the dissipative quench seems to indicate a departure from the Kibble-Zurek mechanism.
Unfortunately an understanding of these results is complicated by the presence of large thermal fluctuations and is
not yet complete.  We then introduced an external quench model in which the dynamics of the
phase transition proceed through the controlled change of a time-dependent effective mass.

In the external quench at very early times, we identified power-law scaling characteristic of the Kibble-Zurek mechanism
and a single freeze-out scale.  This suggests that the multiple length scales characteristic of the late-time ordering of a 
textured system derive from the critical dynamics of a {\it single} nonequilibrium correlation length. 
When observed near the end of the phase transition, we found the scaling of the texture separation $L_{sep}(\tau)$ and 
the texture width $L_w(\tau)$ result instead from a competition between the length scale determined at freeze-out 
and the ordering dynamics of a textured system. We expect that this
observation is not restricted to systems of 
topological textures or to low dimensions but is relevant anytime defect
interactions are significant.   It is not surprising that defect interactions can significantly
modify the defect distribution even before the end of the phase transition. Well before defects are fully formed,
they frustrate the system from its true minimum energy ground state and interact with each other.

The Kibble-Zurek mechanism provides a useful connection between critical dynamics 
and the length scales of the topological defect distribution observed at the end of the phase transition.
However, power-law scaling provides a rough and rather opaque window into
the complicated nonequilibrium processes of the phase transition. As demonstrated explicitly here, power-law scaling 
can also arise from completely different mechanisms such as defect interactions.  To understand the characteristics
of the topological defect distribution at the end of the transition it is therefore important to look closely 
both at the nonequilibrium dynamics of the phase transition {\it and} the evolution of the defect network.

\chapter{Black hole phase transitions}
\section{Introduction}
As discussed extensively in previous chapters, thermal fluctuations can induce  
phase transitions in which the zero-temperature degrees of freedom are reorganized into a 
qualitatively different form.  While there are many familiar examples such as the boiling of water,
phase transitions also occur in more exotic systems like spacetime geometry.
In this chapter, which represents work in progress
\cite{stephensA:2000,stephensB:2000,stephensC:2000}, we study a spacetime phase transition evident 
through the spontaneous formation of
a black hole in equilibrium with a thermal environment.  We seek the answer to two important questions:
\begin{quote}
1. Why does a black hole phase transition occur?

2. By what dynamical scenario does the phase transition take place?
\end{quote}
To answer these questions we will necessarily cast a wide net, drawing insight from diverse subjects
ranging from Euclidean quantum gravity and string theory to vortices in condensed matter.

The abstract nature of the black hole phase transition is a departure from the more physical
systems considered in earlier chapters.  No laboratory is yet equipped with the tools necessary to
probe the formation of a black hole in a phase transition of thermal spacetime, although such situations
may have existed in the very early universe.  Here, our motivation is {\it not} the modeling of an existing 
physical system.  Rather we intend to use our knowledge gained from the previous study of quantum and classical 
phase transitions to peer into the complicated workings of general relativity and semiclassical gravity.
General relativity is a highly nonlinear theory and comparatively little is known about its detailed behavior away
from exact solutions.  For example, in the extremely energetic environment of the early universe,  
general relativity may exhibit a disordered phase dominated by black holes, wormholes, geons and other 
nonperturbative gravitational excitations.  At even higher energies near the Planck scale it has
long been speculated that spacetime appears as a foam or froth \cite{wheeler:1957}.  While the picture of spacetime
foam is vivid, it is very hard to realize in quantitative detail: in part because a theory
of quantum gravity does not yet exist. However, it is our hope that understanding the phase structure of 
semiclassical gravity provides at least a preliminary step on the road to quantum gravity.

Black holes bear some similarity to topological defects.  They are both stable, nonperturbative 
solutions of the classical theory.  Both black holes and topological defects carry with them a 
remnant of the high temperature phase; symmetry is restored in the core of topological defects and 
we are likely to find a quantum phase of spacetime in the high-curvature region near the black hole singularity. 
In addition, in results reported in this chapter, we show that the black hole phase transition is qualitatively
similar to the defect-mediated Kosterlitz-Thouless phase transition in condensed matter and the Hagedorn transition
in string systems.

The analogy between black holes and topological defects is only part of a larger relationship
existing between condensed matter physics and ``fundamental'' physics such as general relativity, particle 
physics and cosmology \cite{hu:1995a, hu:1995b, volovik:2000}.  Such apparently disparate subjects are
unified by the complex organizing behavior of their constituent elements, whether atoms and molecules in condensed matter,
or spacetime itself in quantum gravity.  Condensed matter systems are usually more completely understood than
their high energy counterparts and we can exploit these analogies to illuminate the behavior of systems which are 
otherwise experimentally and theoretically intractable. For example, Bose-Einstein condensates may provide
a testable model of black hole Hawking radiation \cite{garay:2000}. The experimental tests of physics at high 
energy scales are (and are likely to remain) relatively few.  In this environment, the analogies between
condensed matter and other physical systems are increasingly important in our understanding 
of traditional fundamental physics such as semiclassical gravity.

\subsubsection{Organization}
The organization of this chapter is as follows.  In Sec.~\ref{sec:thermo} we review the thermodynamics
of the black hole phase transition, focusing on the calculation of the semiclassical free energy of a black hole
in thermal equilibrium.  In Sec.~\ref{sec:atom} we review an atomic model of the black hole, first
introduced as a quantum model of black hole microstates.  In original work we use this model to 
provide a statistical mechanics of the phase transition and highlight the important role of black hole
entropy.  In Sec.~\ref{sec:dynamics} we discuss the nonequilibrium dynamics of the black hole phase transition. 
We highlight the inadequacy of the homogeneous nucleation theory of first-order phase transitions and explore  
examples from condensed matter and string theory which appear qualitatively similar to the black hole system.  
A summary is provided in Sec.~\ref{sec:ch4sum}.  In this chapter, unless otherwise noted, 
we use Planck units for which $\hbar=G=c=1$.

\section{Black hole free energy}
\label{sec:thermo}
For normal matter, gravitational interactions are universally attractive and
a self-gravitating system is fundamentally unstable to collapse.   
For example, a nonrelativistic ideal homogeneous fluid with density $\rho$ and 
sound speed $v_s$ is unstable to long wavelength density perturbations.
Perturbations with wave vector 
\begin{equation}
k_J<\sqrt{\frac{4\pi \rho}{v_s^2}}
\end{equation}
will grow exponentially.  This is the Jeans instability.  Since gravity cannot be screened, gravitational
instabilities remain for a system in thermal equilibrium.  Compress an ideal, isothermal, self-gravitating gas 
below a critical volume and the gas will collapse. 

The combination of quantum theory and general relativity adds additional instabilities to a thermal gravitational 
system. In particular, hot, flat space is unstable to the (quantum) nucleation of black holes \cite{gross:1982}. 
Using the techniques of Euclidean quantum gravity, the nucleation of black holes was identified through the discovery of
a Schwarzschild instanton contributing an imaginary piece to the free energy of the system.  The 
nucleation rate is maximum for a black hole with mass $M=\frac{1}{8\pi T}$ and is (approximately) given by 
\begin{equation}
\Gamma \sim T^5 \exp {\left (-\frac{1}{16\pi T^2}\right )}.
\end{equation}
A black hole nucleated with 
temperature $T$ is in unstable equilibrium with a thermal environment
of the same temperature.  If, in a fluctuation,  
the black hole absorbs a small amount of radiation, its Hawking temperature decreases (black holes generally
have negative specific heat).  As the black hole grows it becomes colder still, 
absorbing more radiation and eventually engulfing the system.  
Thus, although $\Gamma$  is small except near the Planck scale, the negative specific heat of nucleated black holes 
renders the canonical ensemble of hot, flat space ill-defined.

Black hole systems become thermodynamically stable with the addition of
special boundary conditions or in spacetimes with
a negative cosmological constant \cite{hawking:1983}.  In the following we fix the temperature $T$ on an 
isothermal boundary of radius $r$ containing
a black hole of mass $M$.  In equilibrium,
the Hawking temperature measured on the boundary must equal the
boundary temperature,
\begin{equation}
\label{eq-bhT}
T(r)=\frac{1}{8\pi M} \frac{1}{\sqrt{1-\frac{2M}{r}}}.
\end{equation}
Eq.~(\ref{eq-bhT}) admits two real, nonzero solutions for the mass: a smaller, 
unstable black hole with mass $M_1$ and a larger, 
stable black hole with mass $M_2$,
\begin{eqnarray}
\label{eq-bhm}
M_1 &\simeq& \frac{1}{8\pi T} \left[1+\frac{1}{8\pi rT} \right],  \\  
M_2 &\simeq& \frac{r}{2}\left[1-\frac{1}{(4\pi rT)^2} \right ].
\end{eqnarray}
The isothermal boundary renders $M_2$ thermodynamically stable because of the temperature redshift.  A fluctuation
that increases $M_2$ also {\it increases} the temperature of the black hole as measured on the boundary,
giving $M_2$ a positive specific heat.  Surprisingly, for 
\begin{equation}
\label{eq-bhTc}
T<T_c=\frac{\sqrt{27}}{8\pi r}
\end{equation}
no real value $M$ can solve Eq.~(\ref{eq-bhT}) and no black hole can exist in the box.  It therefore
appears that as the temperature on the boundary is increased from
$T=0$, a phase transition to a black hole spacetime occurs at $T=T_c$.

To further elucidate the thermodynamics of the black hole system and the nature of any potential phase transitions
we consider the canonical partition function
defined through an Euclidean path integral \cite{hawking:1979},
\begin{equation}
Z[\beta]=\int D[g]e^{-\frac{I_E[g]}{\hbar}},
\end{equation}
where we have temporarily restored the $\hbar$ dependence in anticipation of the semiclassical limit.
The Euclidean action is obtained from the Lorentzian Einstein-Hilbert action (with boundary
terms) through the Wick rotation, $t\rightarrow -i\tau$.  The functional integration is taken over 
real Euclidean metrics, periodic in imaginary time coordinate $\tau$ with period equal to the inverse
temperature $\beta$.  Apart from artificial toy models, the full functional integral is intractable.
However, in the semiclassical limit $(\hbar \rightarrow 0)$, the integrand is highly peaked around
metrics that minimize the classical Euclidean action.  In the semiclassical limit we evaluate the
partition function for the system consisting of a single black hole in a finite cavity with boundary
topology $S^1 \times S^2$ (the partition function beyond the semiclassical approximation
was considered in \cite{whiting:1988}).  The action is
\begin{equation}
\label{eq-ebha}
I_E=\frac{1}{16\pi} \int R\sqrt{g}d^4x + \frac{1}{8\pi} \oint K\sqrt{\gamma}d^3x,
\end{equation}  
where $K$ is the trace of the extrinsic curvature of the boundary and $\gamma$ is
the determinant of the induced three-metric.  The free energy of a single Schwarzschild black hole of mass $M$
within the cavity is given by $F=\beta^{-1}I_E$ where the Euclidean action Eq.~(\ref{eq-ebha}) is evaluated for the metric
\begin{equation}
ds^2=\left(1-\frac{2M}{r} \right)d\tau^2 + \left ( 1-\frac{2M}{r} \right )^{-1} dr^2 + r^2d\Omega^2.
\end{equation}
and  $\tau$ is a Euclidean time coordinate with period $\beta$.
Normalized so that $F=0$ when $M=0$, the semiclassical free energy is 
\begin{equation}
\label{eq-bhf}
F(M,r,T)=r-r\sqrt{1-\frac{2M}{r}}-4\pi M^2 T.
\end{equation}
\begin{figure}
\begin{center} 
\strut\psfig{figure=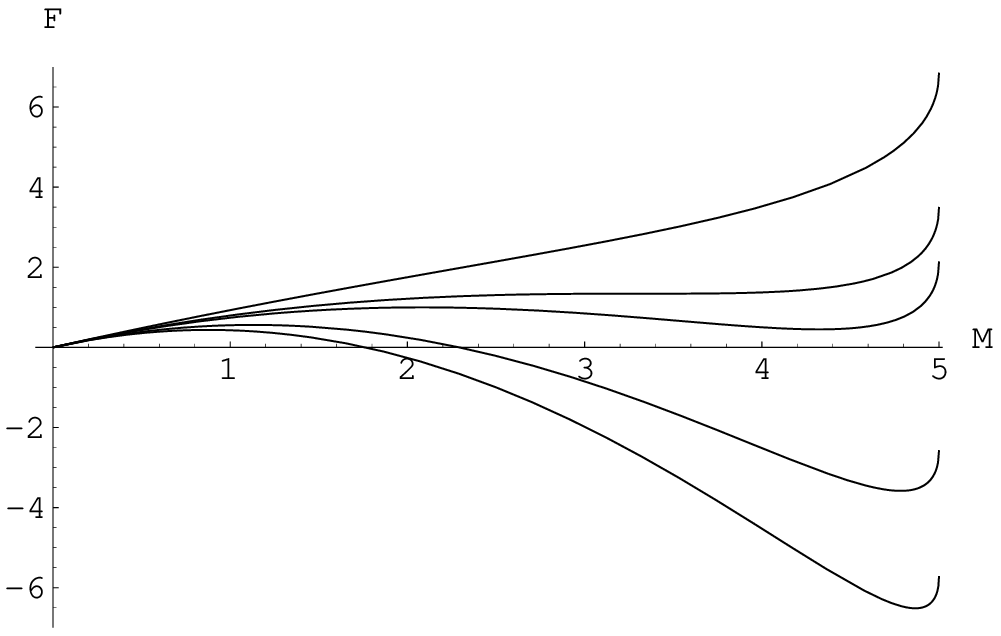,width=3.0in,angle=0}
\rmmcaption{Plot of the free energy vs. mass of the black hole system.} 
\label{fig:bhF}
\end{center}
\end{figure}
In Fig.~\ref{fig:bhF} we plot the free energy $F(M)$ at various temperatures for a system with box size $r=10$.  
From the top down, the first curve is for 
temperature $T=0.01$, below the critical temperature $T_c=0.021$.  The next
curve is for $T=T_c$.  The lower three curves are for $T=0.25$, $T=0.4$ and $T=0.5$ respectively.
For temperatures below $T_c$ no black hole is present.  At $T=T_c$ an extremum
appears at $M=M_1=M_2$ (where the free energy is flat). For temperatures above $T_c$ there are two extrema, 
the unstable black hole with mass $M_1$ and the stable black hole with mass $M_2$, in agreement with the simple 
arguments at the beginning of the section.  Figure \ref{fig:bhF} also suggests that the transition to
a black hole spacetime above $T_c$ occurs discontinuously.  Above $T_c$ a finite mass black hole can be nucleated
in equilibrium with the walls of the box.

The nucleation of a stable black hole at $T=T_c$ does {\it not} necessarily signal a phase transition.   
A system in thermodynamic equilibrium always resides in a state of lowest free energy. A phase transition occurs
only if the free energy of the system with the black hole is lower than the free energy without the black hole.  
At temperatures below $T_c$ no black hole is present
and the free energy is approximately that of a box filled with thermal gravitons (hot, flat space),
\begin{equation}
F_{hfs}\sim -T^4 r^3.
\end{equation}
The free energy of hot flat space is  negative.  Therefore a phase transition from hot flat spacetime to a black hole
spacetime can only occur at temperatures $T>T_c$ for which $F(M_2) < F_{hfs}$.  In Fig.~\ref{fig:bhF} 
the temperature is low enough that $F_{hfs} \approx 0$ and a black hole phase transition occurs when
$F(M_2)<0$.  

The nucleation of a black hole in equilibrium with the walls of the box does not appear to be
the result of classical gravitational collapse. If the box is uniformly filled with 
massless thermal radiation, we approximate the collapse temperature
as the temperature for which the Schwarzschild radius for the thermal energy of the radiation 
is equal to the box size,
\begin{equation}
\label{eq-ctemp}
T_{collapse}^4  r^3\sim r.
\end{equation}
Thus $T_{collapse} \sim \frac{1}{\sqrt{r}}$, qualitatively distinct from the nucleation 
temperature.  

The phase transition does not occur at a high temperature characteristic of quantum gravity.
In fact, as the size of the box is increased the critical temperature decreases and is arbitrarily low
for arbitrarily large boxes.  The stable black hole formed at temperatures above $T_c$ is large, 
with a mass $M_2$ on the order of the size of the box.  These 
considerations seem to be paradoxical.  How can such a large energy ($E\sim M$) fluctuation actually 
{\it lower} the free energy of the system at such low temperatures $T\ll M$?  
The answer lies in the enormous entropy black holes hold within their horizon.  The free energy
results from a competition between the internal energy and 
the entropy, $F=E-TS$. For a black hole, $E=M$ and the entropy is proportional to the 
area of the horizon, $S=4\pi M^2$.
Because the entropy is growing with mass faster than the energy, there is always a 
critical temperature above which the entropy completely
compensates for the energy cost of making a black hole and the black hole spacetime is the lowest 
free energy state.  
\section{Black hole atoms}
\label{sec:atom}
The black hole phase transition is {\it entropically} driven.  Therefore,  an understanding of the nature of black hole
entropy offers potential insight into the details of the transition. In this 
section we study a toy model for black hole microstates, focusing on their implications for the thermal
black hole system.   

The origin of black hole entropy has been an outstanding problem since 
Bekenstein first introduced the concept \cite{bekenstein:1973}.  A complete resolution likely requires
a consistent theory of quantum gravity, which has so far proved elusive.  However,
just as semiclassical reasoning such as the Bohr model was important in the early 
development of quantum theory and the interpretation of atomic spectra, a similar approach 
may be fruitful in understanding some of the microscopic features of black hole entropy \cite{bekenstein:1997}.    
It is not our intention to survey the large number of semiclassical black hole models. However common to many 
is the quantization of the horizon area into equally spaced levels (see \cite{kastrup:1997} and references therein),
\begin{equation}
A=\alpha n; \;\;\;\;\;\; n=1,2\ldots 
\end{equation}
where $\alpha$ is dimensionless but as yet unspecified and the area is given in units
of Planck area.  For a black hole of mass $M$ the entropy,
\begin{equation}
\label{eq-entropy}
S=\frac{A}{4}=4\pi M^2,
\end{equation}
and the mass of the black hole is also quantized, 
\begin{equation}
\label{eq-masslevel}
M =\gamma \sqrt{n},
\end{equation}
where $\gamma= \sqrt{\frac{\alpha}{16\pi}}$.  If the energy levels have degeneracy
$g(n)$, the black hole entropy may simply count the number of available
microstates for a fixed energy level $n$, $S=\ln{g(n)}$.  In this case,
\begin{equation}
\label{eq-degeneracy}
g(n)=e^{\frac{\alpha n}{4}}.
\end{equation}  
Since $g(n)$ must also be an integer, $\alpha$ is restricted,
\begin{equation}
\label{eq-alpha}
\alpha=4\ln{k}; \;\;\;\;\;\;\; k=2,4\ldots
\end{equation}
Equations (\ref{eq-masslevel}), (\ref{eq-degeneracy}) and (\ref{eq-alpha}) define a semiclassical black hole 
model in analogy with
the Bohr model of an atomic system. If $k=2$ the large degeneracy $g(n)$ can be thought of as the number of ways to
make a black hole in the nth level by starting in the ground state \cite{mukhanov:1986}, though there are
other interpretations \cite{danielsson:1993,kastrup:1999}. Our interest in the atomic black hole model is its behavior in 
thermal equilibrium.    

The partition function for the quantum 
black hole atom in the canonical ensemble is
\begin{equation}
\label{eq-abhZ}
Z[T]=\sum_{n=0}^{\infty} k^n e^{-\frac{ \gamma \sqrt{n}}{T}}.
\end{equation}
Without modification, the sum in Eq.~(\ref{eq-abhZ}) does not converge.  Convergence is obtained upon 
analytic continuation but the partition function acquires a imaginary piece \cite{kastrup:1999}.  In light of 
the discussion
of the previous section this is not at all surprising.  Without either special boundary conditions or special
spacetimes, the thermodynamics of black holes is not well-defined.  In fact, Im(Z) is due precisely to
the nucleation of black holes.  To obtain a well-defined and real partition function we take a new approach
and place the quantum black hole atom into a box of radius $r$.  The box is realized as a sharp cutoff 
in the energy levels accessible to the black hole,
\begin{equation}
n_{max}=\frac{r^2}{4\gamma^2}.
\end{equation}
This is reasonable as the box functions to remove energy levels with Schwarzschild radius larger than $r$.  
The canonical partition function for the
quantum black hole
atom in a box is
\begin{equation}
\label{eq-boxabhZ}
Z[T]=\sum_{n=0}^{\frac{r^2}{4\gamma^2}} k^n e^{-\frac{\gamma \sqrt{n}}{T}}.
\end{equation}
In Figs.~\ref{fig:abhfig1}, \ref{fig:abhfig2} and \ref{fig:abhfig3} we plot the average energy, specific heat and 
entropy as a function of temperature for a quantum black hole atom in a box with radius $r=10$.
The plots were made with rescaled variables $T\rightarrow \gamma T$, $r\rightarrow \gamma r$
and for the particular choice $k=2$. All thermodynamic quantities were calculated 
using the partition function Eq.~(\ref{eq-boxabhZ}).
\begin{figure}
\begin{center}
\strut\psfig{figure=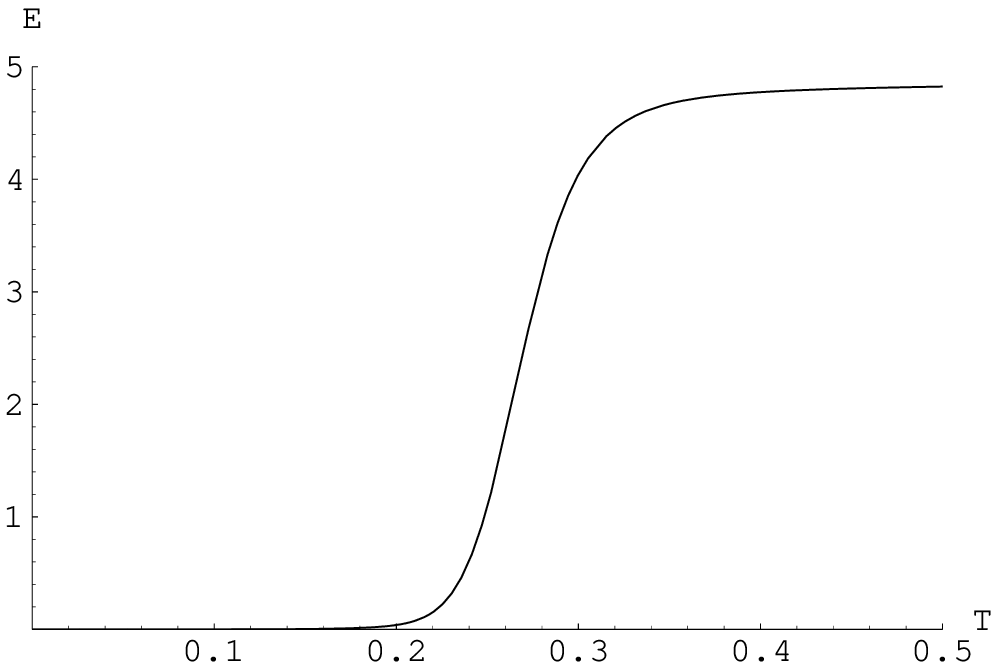,width=3.0in,angle=0}
\rmmcaption{Plot of the average energy $E$ vs. temperature $T$ for the black hole atom.} 
\label{fig:abhfig1}
\end{center}
\end{figure}
\begin{figure}
\begin{center}
\strut\psfig{figure=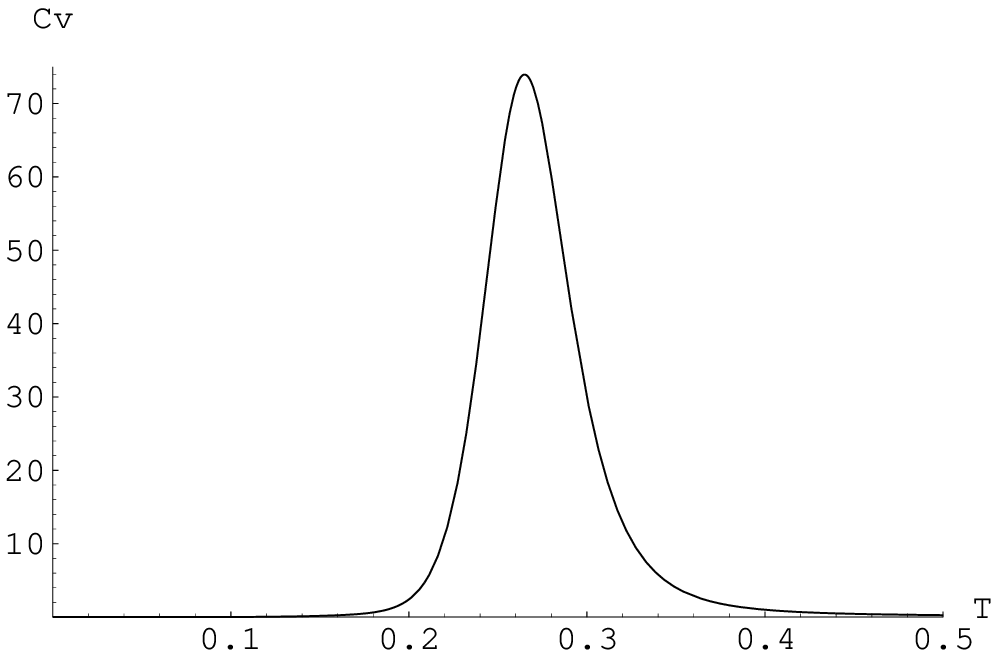,width=3.0in,angle=0}
\rmmcaption{Plot of the specific heat $C_v$  vs. temperature $T$ for the black hole atom.} 
\label{fig:abhfig2}
\end{center}
\end{figure}
\begin{figure}
\begin{center}
\strut\psfig{figure=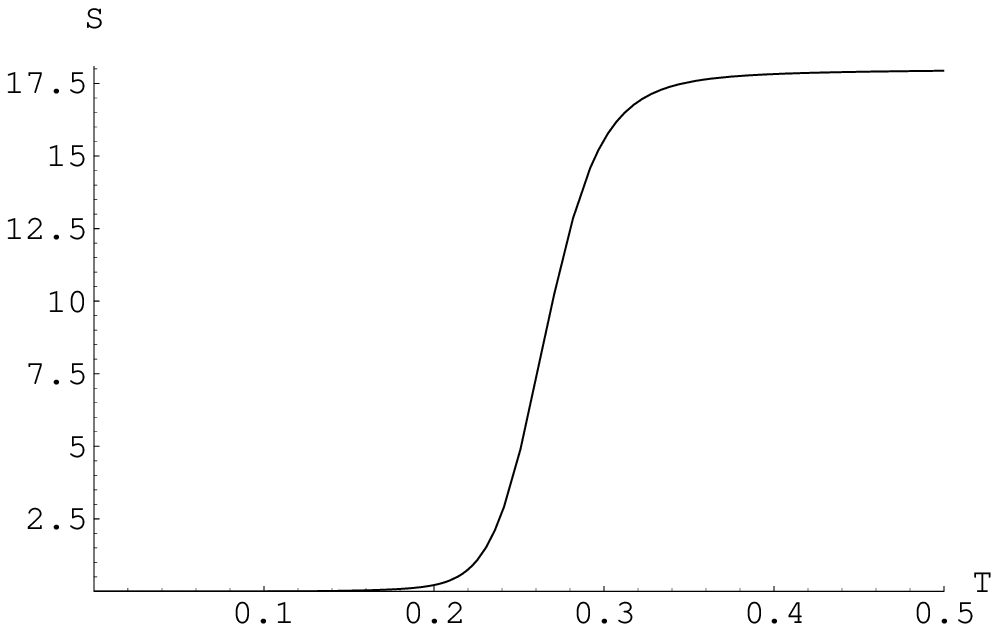,width=3.0in,angle=0}
\rmmcaption{Plot of the entropy $S$ vs. temperature $T$ for the black hole atom.} 
\label{fig:abhfig3}
\end{center}
\end{figure}
The inspection of Figs.~\ref{fig:abhfig1}-\ref{fig:abhfig3} reveals a sharp transition from the ground state to
a highly excited state in the black hole atomic system.   This is reminiscent of the phase transition behavior that we saw in
the previous section.  To quantify this behavior consider the effective Boltzmann factor of level $n$,
\begin{equation}
f(n)=-\frac{\gamma \sqrt{n}}{T}+n\ln{k}.
\end{equation}
For small $n$, $f(n)$ is negative and higher energy levels are suppressed, as is usually the case.
However, since the degeneracy grows as $n$, there is always an energy level $n_c(T)$ defined by $f(n_c)=0$
beyond which degeneracy compensates for the higher energy.  Levels beyond $n_c$ are {\it enhanced}, not suppressed.  
At low $T$, $n_c(T) > n_{max}$ and these enhanced levels are not part of the allowed spectrum.  
As T increases $n_c(T)$ decreases and a transition to the higher levels
occurs when $n_c(T) = n_{max}$.  If we adopt this picture of the black hole phase transition the critical temperature is
\begin{equation}
\label{eq-bhaTc}
T_c=\frac{1}{2\pi r}.
\end{equation}
The average energy at the critical point is
\begin{equation}
\label{eq-bhaM}
E(T_c) \sim e^{f(n_c(T_c))}\gamma \sqrt{n_{max}}=\frac{r}{2}.
\end{equation}
These equations reveal two important details.  First, although the numerical factors are slightly
different they have the same qualitative structure as their thermodynamic counterparts,
Eqns.~(\ref{eq-bhTc}) and (\ref{eq-bhm}).  Second, neither equation contains the one free parameter $k$  
of the black hole atom.  In fact Eqns.~(\ref{eq-bhaTc}) and (\ref{eq-bhaM}) are largely independent of the 
details of the spectrum of the black hole atom and apply even when the horizon area is quantized into nonuniform levels.

At first glance, our study of the quantum black hole atom in thermal equilibrium appears to only slightly 
advance our understanding of the black hole phase transition.  We have simply reaffirmed the fact that a
system with finite energy (limited by the boundary conditions) and obeying a thermodynamic relation $S=4\pi E^2$ will
always have a transition to the highest allowed energy state, the details of which are largely model independent.  
But in principle the quantum black hole atom allows us to go farther. We can now imagine treating dynamical
processes such as emission and absorption much as we do with more common atomic systems.  In fact only
two levels, the ground state and the highest allowed excited state, are significantly populated below
and above the transition and we may further approximate our system as a quantum two-level atom, a 
popular system of study. Since the quantum black hole transition occurs as the excitation of a high energy state,
the dynamics of this atomic excitation approximate
the dynamics of the black hole phase transition.  We hope to report on further work on this topic in
the near future \cite{stephensA:2000}.

Our analysis of the quantum black hole atom in thermal equilibrium underscores the entropic nature of
the black hole phase transition.  This result is useful in itself.
Following the dynamics of the black hole phase transition through a fully nonequilibrium 
formulation of semiclassical gravity is a very hard problem to which, at the moment, there is no direct method of attack.  
However, the study of simpler systems with a similar entropic transition is likely to yield 
insight into the dynamics of the black hole phase transition.  In the next section we turn our attention to these
dynamical issues.

\section{Nonequilibrium dynamics}
\label{sec:dynamics}
Figure \ref{fig:bhF} appears qualitatively similar to the free energy of a 
system undergoing a (strongly) first-order phase transition.  The field theoretic treatment of the dynamics of
first-order phase transitions is based upon the homogeneous nucleation theory developed by Langer 
(see for example \cite{langer:1992}).  In
homogeneous nucleation, widely separated spherical bubbles of the stable phase nucleate in a background of the
unstable phase.  If the bubbles are larger than a critical radius, the volume
energy of the stable phase inside the bubble is less than the surface energy and the droplet will grow.  
The phase transition completes as droplets of the stable phase expand to fill the volume of the system.  In 
this picture,
the black hole phase transition occurs when an unstable black hole with mass $M_1$ nucleates 
(with probability $P \sim e^{-\beta F(M_1)}$) and grows to form the stable mass $M_2$ through the 
absorption of thermal radiation.  There are reasons to believe, however, that homogeneous nucleation is {\it not} the 
correct description, as we indicate below.

The free energy of the black hole system is calculated under the assumption of spherical symmetry, adequate 
only for a single 
black hole.  In equilibrium, a single black hole is preferred because it maximizes the entropy.   For example, 
in a state with
two black holes,
\begin{equation}
S_{m_a}+S_{m_b} \sim m_a^2+m_b^2 < S_{m_a+m_b} \sim (m_a+m_b)^2.
\end{equation}
However the dynamics of the phase transition may involve multiple black holes.  Black holes are strongly 
interacting gravitational systems, unscreened by thermal fluctuations.  It is possible that the dynamics of the
phase transition proceeds by the (exponentially more probable) nucleation of small black holes with mass $m \ll M_1$
which then merge to form larger holes. Phase transitions in which there are strong 
interactions between bubbles of the new phase lie beyond the scope of homogeneous nucleation.

\subsection{Black hole gas}
To study the possibility that the dynamics of the phase transition proceeds through the
nucleation and merger of small black holes we propose a model in which black holes are treated
as a gas of gravitationally interacting particles. 
The number of particles is {\it not} fixed, but changes 
through the stochastic nucleation of small black holes, mergers and Hawking evaporation.  To incorporate the interaction of
black holes with the thermal environment the mass of each particle is not constant but changes in proportion to the free energy
of a single black hole,
\begin{equation}
\frac{dm(t)}{dt} \sim - \frac{\delta F}{\delta m}.
\end{equation}
From $F(m)$ given by Eq.~(\ref{eq-bhf}) we obtain the phenomenological relation
\begin{equation}
\label{eq-massloss}
\frac{dm(t)}{dt}=-\frac{1}{\sqrt{1-\frac{2m}{r}}} +8\pi m T,
\end{equation}
where $T$ is the temperature and $r$ is the size of an effective boundary.  
With a time-dependent mass given by Eq.~(\ref{eq-massloss}) black holes are stable
only above a critical temperature, as we expect from previous discussions.  The difference and advantage 
of this model is that above the critical temperature, a stable black hole can form by small mergers,
in addition to a single large fluctuation.   A conceptually similar approach but with
very different emphasis was considered in \cite{ellis:1999}. We hope to report on the analysis of this 
black hole gas model in the near future \cite{stephensB:2000}.  

\subsection{Entropic transitions}
Another approach to the dynamics of the black hole phase transition is through the study of similar
physical systems.  The black hole phase transition is driven by the large amount of black hole
entropy $S \sim E^2$.  A thermodynamic relation where the entropy grows rapidly with energy is very unusual.
For example, a classical ideal gas has entropy $S\sim \ln{(E)}$.   However there are examples of condensed matter
systems that exhibit similar entropic transitions. In particular we examine the Kosterlitz-Thouless (KT) transition 
in a global $O(2)$ model in $2+1$ dimensions \cite{kosterlitz:1973} and the Hagedorn transition 
\cite{hagedorn:1965} in string systems.  

\subsubsection{Kosterlitz-Thouless transition}
We consider a global $O(2)$ scalar field model with Hamiltonian,
\begin{equation}
H=\int d^2x \left ( \frac{1}{2}|\nabla \vec{\phi}|^2 + \frac{\lambda}{8} (\vec{\phi}^2-\eta^2)^2 \right ).
\end{equation}
This model admits vortex topological defects.  The energy of a vortex of single winding is
\begin{equation}
\label{eq-Ev}
E \approx \pi \eta^2 \left (\ln{\frac{R}{a}}+\lambda \eta^2 a^2 \right ),
\end{equation}
where $a \sim \frac{1} {\sqrt{\lambda} \eta}$ is the vortex size and $R$ is the size of the system.  If
$R \gg a$ the energy of the vortex is dominated by gradient energy, the first term in Eq. (\ref{eq-Ev}).
If a vortex can be nucleated anywhere in the system then the entropy 
\begin{equation}
S=\ln{\left (\frac{R}{a}\right )^2},
\end{equation}
and the free energy of the system with a single vortex is
\begin{equation}
F = (\pi \eta^2 -2T)\ln{\left (\frac{R}{a} \right )}.
\end{equation}
Inspection of the free energy reveals that above the critical temperature 
\begin{equation}
T_c=\frac{\pi \eta^2}{2},
\end{equation}
it is thermodynamically favorable to nucleate vortices.  This is the KT transition, a transition
from (algebraic) order to disorder. 
The dynamics of the KT transition proceeds through the interaction of vortices.
At low temperatures vortices exist as a low density of tightly bound
vortex-antivortex pairs with zero net topological charge.  As the temperature increases, both the density of
pairs and the average vortex-antivortex separation also increase.  
In addition, as the separation increases, thermal fluctuations are more effective in screening the 
interaction between defects in a vortex-antivortex pair. 
When the critical temperature is reached, screening is complete
and the vortex-antivortex pairs unbind leaving a gas of essentially free topological defects.

The KT transition exhibits both similarities and differences when compared with the black hole system.
Analogous to the black hole transition, the nucleation of a vortex with energy $E$ produces a 
large amount of entropy $S \sim E$ which drives the phase transition.
In distinction however, the vortex entropy results from the 
possible locations of the vortex and not, as in the black hole, from internal states.
The natural analogy of an antivortex in the black hole system is a {\it white} hole, a spacetime
region that expels all worldlines.  A white hole is the time-reverse of a black hole, just as 
(e.g. in superfluid helium) an antivortex is the time-reverse of a vortex.
The analogy of the KT transition suggests that the black hole phase transition might be 
understood as the unbinding of black hole-white hole pairs.  It is far from clear that this viewpoint is 
correct.  However, at the very least, the analogy of the KT transition suggests a new angle towards
the dynamics of the black hole phase transition.

\subsubsection{Hagedorn transition}
A system of fundamental closed bosonic strings in thermodynamic equilibrium possesses an 
exponential density of states \cite{hagedorn:1965}
\begin{equation}
\nu(E) \sim \exp{(\beta_H E)},
\end{equation}
where $\beta_H$ is the model and dimension dependent Hagedorn temperature.
Above the Hagedorn temperature the exponential density of states renders the
partition function ill-defined and a physical description of the system is unclear.  

Insight into the interpretation of the Hagedorn transition is obtained by examining the relationship between 
string theory and Quantum Chromodynamics (QCD).  QCD is the (nonabelian) SU(3) gauge theory of strong interactions.
The fundamental degrees of freedom are three colored fermionic quarks interacting via bosonic gluons.  
QCD displays asymptotic freedom, at high energies the QCD coupling is weak and quarks are effectively free. 
However, at low energies QCD is strongly coupled and quarks exist only in tightly bound color singlet states, 
hadrons and mesons.  Between the two regimes, QCD is expected to undergo a deconfinement
phase transition at which hadrons and mesons melt into their constituent quarks.
The large-N \cite{thooft:1974} and strong coupling \cite{wilson:1974} limits of QCD may be described by a fundamental string
theory and the link between QCD and string theory provides a physical picture of the Hagedorn transition:
the proliferation of string states at the Hagedorn temperature corresponds to the 
emergence of quark color degrees of freedom at deconfinement \cite{pisarski:1982,olesen:1985,salomonson:1986}.

Surprisingly, string theory also offers the possibility that the Hagedorn transition in large-N 
SU(N) gauge theory is linked through a hypothesized AdS/CFT duality 
to the formation of a black hole in anti-deSitter (AdS) spacetime (see \cite{aharony:2000} for a comprehensive review).
The thermodynamics of a semiclassical black hole system in a spacetime with negative 
cosmological constant is qualitatively similar to the black hole in a box discussed previously in Sec.~\ref{sec:thermo}.
In particular, a black hole phase transition, the Hawking-Page transition \cite{hawking:1983} occurs 
at a critical temperature 
\begin{equation}
T_c \sim \sqrt{|\Lambda |}
\end{equation}
where $\Lambda $ is the (negative) cosmological constant.
Exploiting the AdS/CFT duality, Witten argued \cite{witten:1998} that the enormous entropy released in 
the deconfinement transition of a supersymmetric gauge theory defined on the boundary of AdS 
spacetime has a dual bulk interpretation as a black hole forming through the Hawking-Page phase transition.

A Hagedorn transition also occurs in systems of 
cosmic strings and condensed matter vortices.  In this case, the exponentially growing density of states simply
reflects the ``wiggles'' present in each string and the Hagedorn transition signals the abrupt formation
of infinite string.  The thermodynamic behavior of strings in a classical 
U(1) scalar field theory with symmetry breaking was examined in \cite{antunes:1998a,antunes:1998b}.  
The formation of infinite string was observed at the field critical point, suggesting that, as in the KT
transition, strings play a fundamental role in the phase transition.  

Driven by the exponential density of states, the Hagedorn transition in a classical U(1) field theory
provides a tractable analogy to the black hole phase transition.  Visible as the
formation of infinite string, we propose to study the detailed dynamics of this transition \cite{stephensC:2000}. 
A U(1) scalar theory provides a simple realization of a system exhibiting vortex excitations 
and also belongs to the same important universality class as $^4He$ and type-II superconductors.
The use of a classical field theory allows for detailed numerical observation
of the formation and interaction of the nonperturbative string structures important for the transition.
As discussed above, a Hagedorn transition also describes the deconfinement
phase transition in large-N or strongly-coupled QCD.  Through the AdS/CFT correspondence the deconfinement phase transition
corresponds to the Hawking-Page black hole phase transition.  It is therefore possible that 
the dynamics of the formation of infinite string is similar to the dynamics 
of black hole formation in the Hawking-Page transition.  Of course, it is impossible to proceed rigorously 
in such a long chain of analogies.  However, the dynamics of the formation of infinite string in a classical system
is an interesting physical problem in its own right which, to our knowledge, has not yet been addressed.  

\section{Summary}
\label{sec:ch4sum}
In this chapter we have built the foundation for a thorough analysis of black hole phase transitions
in semiclassical gravity.  We reviewed the semiclassical thermodynamics of the black hole system and
determined that the black hole phase transition is entropically driven: it
occurs because the large entropy of a black hole compensates for the energy cost of formation and
lowers the total free energy of the system.  This view was reinforced through the analysis of an 
atomic model of a quantum black hole in thermal equilibrium.  In this model, the phase
transition occurs as the abrupt excitation of a high energy state above the critical temperature.  Thus, the
study of atomic emmision and absorption may provide intuition about the black hole system, a direction we are
currently pursuing.  The detailed {\it dynamics} of the black hole 
phase transition remain an open problem.  We argued that homogeneous nucleation does not apply to phase transitions
in which there are strong interactions among bubbles of new phase, as is the case for the
black hole system.  We then appealed to the study of similar entropic transitions 
and observed that the black hole phase transition contains elements of both the Kosterlitz-Thouless and Hagedorn phase 
transitions.  Reasoning by analogy, we argued that the dynamics of the
black hole phase transition might be similiar to the dynamical formation of long string
in the nonequilibrium quench of a classical U(1) scalar field theory.  Work in these directions is
in progress.

\chapter{Conclusion}
In this dissertation we have analyzed the nonequilibrium formation of topological defects during a
second-order phase transition of both a quantum field in the early universe and a classical field appropriate
for the description of a condensed matter system.  We also discussed the existence and 
dynamics of a black hole phase transition in semiclassical gravity.  

We first studied the symmetry-breaking phase transition of a quantum scalar field 
in the early universe.  Using equations of motion derived from the two-loop 2PI-CTP effective
action we identified domains in the infrared portion of the 
momentum-space power spectrum.  We found that the domain size scales as a power-law with the 
expansion rate of the universe. The observed power-law scaling, for both overdamped and underdamped evolution,  
is in good agreement with the predictions of the Kibble-Zurek mechanism.

We then studied the formation and interaction of topological textures in a nonequilibrium phase transition
of a classical O(3) scalar field theory in $2+1$ dimensions. 
We presented two models of the nonequilibrium phase transition, a dissipative quench and an external quench.  
In the external quench and at early times, we observed power-law scaling of the length scales of the texture distribution 
in good agreement with the Kibble-Zurek mechanism.  However, by the end 
of the transition the length scales of the texture distribution result from a competition between the 
length scale determined at freeze-out and the ordering dynamics of a textured system.  Therefore a quantitative understanding
of the defect network at the end of the phase transition generally requires an understanding of both critical
dynamics {\it and} the interactions among topological defects. 

Finally, we studied a black hole phase transition in semiclassical gravity.  We reviewed the thermodynamics
of the black hole phase transition, focusing on the calculation of the semiclassical free energy of a black hole
in thermal equilibrium. We determined that the black hole phase transition is entropically driven.
We applied a quantum atomic model to the black hole equilibrium system and showed that the phase transition
occurs as the excitation of a high energy state.  We discussed the nonequilibrium dynamics of the 
black hole phase transition and explored similar examples from the Kosterlitz-Thouless transition in
condensed matter to the Hagedorn transition in string theory.

The power-law scaling of the defect density, observed here in both quantum and classical 
nonequilibrium phase transitions, demonstrates, at least at early times,  the viability of the 
Kibble-Zurek mechanism of topological defect formation.  However, as mentioned in Chapters 3 and 4, the power-law
scaling of the defect density is only an indirect and partial verification of the ideas presented in
the freeze-out scenario.  In fact, the exact details of the dynamical 
process through which the {\it field} correlation length controls the {\it defect} 
separation remain somewhat obscure. Partly this is due to the complicated relationship between the field and
defect degrees of freedom.  To remedy this, one possibility is to seek a dual topological 
defect description of the entire nonequilibrium phase transition.  
At least for certain systems we may be able to phrase the freeze-out 
scenario solely in terms of equilibrium defect distributions and equilibrium relaxation times 
for these defect distributions.  Indeed, some tentative steps have already been made in this direction 
\cite{williams:1999}.

In the context of nonequilibrium phase transitions in quantum field theory, an important issue is the
development of techniques that would allow for an analysis of the entire phase 
transition.  As discussed in Chapter 3, although the two-loop approximation of the 2PI-CTP effective action
is adequate at early times in the phase transition, when long wavelength modes are sampling the spinodal instability,
the approximation is too simplistic to connect to late-time oscillations around the true vacua and the formation 
and evolution of topological defects.  To analyze the entirety of the phase transition in nonequilibrium quantum
field theory we propose a two-stage model of the phase transition.  
In the first stage quantum field fluctuations can be evolved at early times using the known techniques 
of nonequilibrium quantum field theory.  In the second stage, when standard techniques fail, 
we can attempt to match the field fluctuations to an {\it effective} (stochastic) classical Landau-Ginzburg equation. 
The late-time evolution of the Landau-Ginzburg equation allows the field to
settle to the true vacuum.  A similar program was initiated in Ref.~\cite{calzetta:1989} but with neither the focus on
topological defects nor an explicit solution of the late-time dynamics.   The advantage of this 
approach is a picture of the transition that extends from the phase dominated by quantum fluctuations to the phase dominated
by topological defects.  In quantum field theory, such a complete picture has never been given.   
Along these lines we continue our research.

\addcontentsline{toc}{chapter}{Bibliography}
\bibliographystyle{phaip}
\bibliography{thesis}

\spacing{1}

\pagestyle{empty}

\end{document}